\documentclass{article}

\PassOptionsToPackage{numbers}{natbib}
\usepackage[main, final]{neurips_2026}

\usepackage[utf8]{inputenc}
\usepackage[T1]{fontenc}

\usepackage{url}
\usepackage{booktabs}
\usepackage{amsfonts}
\usepackage{nicefrac}
\usepackage{microtype}
\usepackage{xcolor}
\usepackage{colortbl}
\usepackage{xspace}
\usepackage{graphicx}
\usepackage{tikz,graphics,color,float,caption,subcaption}
\usepackage{multirow}

\definecolor{meancol}{RGB}{232,240,252}

\makeatletter
\DeclareRobustCommand\onedot{\futurelet\@let@token\@onedot}
\def\@onedot{\ifx\@let@token.\else.\null\fi\xspace}
\makeatother

\newcommand{\N}{\mathbb{N}}
\newcommand{\Nz}{\mathbb{N}_0}

\newcommand{\R}{\mathbb{R}}

\newcommand{\K}{\mathbb{K}}
\newcommand{\J}{\mathbb{J}}
\newcommand{\I}{\mathbb{I}}
\newcommand{\X}{\mathbb{X}}

\newcommand{\V}{\mathbb{V}}

\def\eg{\emph{e.g}\onedot}
\def\ie{\emph{i.e}\onedot}

\newcommand{\func}[3]{#1:#2\to#3}

\usepackage{xargs}
\newcommandx{\seq}[3][2={},3={}]{(#1)_{#2}^{#3}}
\newcommandx{\set}[3][2={},3={}]{\{#1\}_{#2}^{#3}}

\usepackage{bm}
\newcommand{\vct}[1]{\bm{#1}}
\newcommand{\mtx}[1]{\bm{#1}}
\newcommand{\tns}[1]{\bm{\mathcal{#1}}}

\usepackage{amsmath,amssymb,amsthm}
\newtheorem{theorem}{Theorem}
\newtheorem{definition}{Definition}
\newtheorem{lemma}{Lemma}
\newtheorem{example}{Example}
\newtheorem{proposition}{Proposition}
\newtheorem{corollary}{Corollary}
\theoremstyle{remark}
\newtheorem{remark}{Remark}
\theoremstyle{plain}

\usepackage{enumitem}
\usepackage{placeins}
\newcommand{\cp}[1]{[\![#1]\!]}
\newcommand{\trunc}[1]{\vct{s}_{#1}}
\newcommand{\lowerr}{\underline{\varepsilon}}
\newcommand{\upperr}{\overline{\varepsilon}}

\newcommand{\hadamard}{\odot}
\newcommand{\tenprod}{\otimes}
\newcommand{\dotprod}{\bullet}

\newcommand{\inner}[3][]{\ifx q#1q\langle #2, #3 \rangle\else\langle #2, #3 \rangle_{#1}\fi}
\newcommand{\norm}[2][]{\ifx q#1q\left\lVert#2\right\rVert\else\left\lVert#2\right\rVert_{#1}\fi}

\DeclareMathOperator*{\argmin}{arg\,min}

\usepackage{algorithm,algorithmic}

\usepackage{xcolor}
\usepackage{tikz,pgfplots,graphicx}
\usepackage[normalem]{ulem}
\usetikzlibrary{calc,shapes.arrows,shapes.geometric,arrows.meta}

\makeatletter
\newcommand{\includegraphicswithbbox}[7][\linewidth]{%
  \begingroup
  \sbox0{\includegraphics[width=#1]{#2}}%
  \begin{tikzpicture}
    \node[anchor=south west,inner sep=0] (image) at (0,0) {\usebox0};
    \begin{scope}[x={(image.south east)},y={(image.north west)}]
      \draw[#7, thick] (#3,#4) rectangle (#5,#6);
      \pgfmathsetmacro{\nf@w}{#5 - #3}
      \pgfmathsetmacro{\nf@h}{#6 - #4}
      \pgfmathsetmacro{\nf@s}{0.3/\nf@w}
      \begin{scope}[shift={(0.02,0.02)}, xscale=\nf@s, yscale=\nf@s, transform shape]
        \clip (0,0) rectangle (\nf@w,\nf@h);
        \node[anchor=south west,inner sep=0pt] at (-#3,-#4) {\usebox0};
        \draw[#7, very thick] (0,0) rectangle (\nf@w,\nf@h);
      \end{scope}
    \end{scope}
  \end{tikzpicture}
  \endgroup
}
\makeatother

\makeatletter
\newcommand{\includegraphicsthaistatue}[7][\linewidth]{%
  \begingroup
  \sbox0{\includegraphics[width=#1,trim={25cm 0cm 25cm 5cm},clip]{#2}}%
  \begin{tikzpicture}
    \node[anchor=south west,inner sep=0] (image) at (0,0) {\usebox0};
    \begin{scope}[x={(image.south east)},y={(image.north west)}]
      \draw[#7, thick] (#3,#4) rectangle (#5,#6);
      \pgfmathsetmacro{\nf@w}{#5 - #3}
      \pgfmathsetmacro{\nf@h}{#6 - #4}
      \pgfmathsetmacro{\nf@s}{0.3/\nf@w}
      \begin{scope}[shift={(0.05,0.4)}, xscale=\nf@s, yscale=\nf@s, transform shape]
        \clip (0,0) rectangle (\nf@w,\nf@h);
        \node[anchor=south west,inner sep=0pt] at (-#3,-#4) {\usebox0};
        \draw[#7, very thick] (0,0) rectangle (\nf@w,\nf@h);
      \end{scope}
    \end{scope}
  \end{tikzpicture}
  \endgroup
}
\makeatother

\usepackage{etoolbox}
\usepackage{subcaption}

\AtEndPreamble{
    \usepackage[capitalize]{cleveref}
    \crefname{section}{Sec.}{Secs.}
    \Crefname{section}{Section}{Sections}
    \crefname{table}{Tab.}{Tabs.}
    \Crefname{table}{Table}{Tables}
    \crefname{appendix}{App.}{Apps.}
    \Crefname{appendix}{Appendix}{Appendices}
    \crefname{proposition}{Prop.}{Props.}
    \Crefname{proposition}{Proposition}{Propositions}
    \crefname{corollary}{Cor.}{Cors.}
    \Crefname{corollary}{Corollary}{Corollaries}
    \crefname{theorem}{Thm.}{Thms.}
    \Crefname{theorem}{Theorem}{Theorems}
    \crefname{lemma}{Lem.}{Lems.}
    \Crefname{lemma}{Lemma}{Lemmas}
    \crefname{remark}{Rem.}{Rems.}
    \Crefname{remark}{Remark}{Remarks}
}

\usepackage{hyperref}
\hypersetup{hidelinks, pdftitle={FUTON: Fourier Tensor Networks for Implicit Neural Representations}, pdfauthor={Pooya Ashtari, Pourya Behmandpoor, Nikos Deligiannis, Aleksandra Pižurica}}

\title{FUTON: Fourier Tensor Networks for\texorpdfstring{\\}{ } Implicit Neural Representations}

\author{%
  Pooya Ashtari\textsuperscript{1,}\thanks{Corresponding author: \texttt{pooya.ashtari@ugent.be}}
  \hspace{1em} Pourya Behmandpoor\textsuperscript{2}
  \hspace{1em} Nikos Deligiannis\textsuperscript{2}
  \hspace{1em} Aleksandra Pižurica\textsuperscript{1} \\
  \textsuperscript{1}Ghent University
  \hspace{2em} \textsuperscript{2}Vrije Universiteit Brussel \& imec
}

\begin{document}

\maketitle

\begin{abstract}
    Implicit neural representations (INRs) have emerged as powerful tools for encoding signals, yet dominant MLP-based designs often suffer from slow convergence{, and how their accuracy depends on their size is known only empirically}. We introduce \textbf{FUTON (Fourier Tensor Network)}, which models signals as generalized Fourier series whose coefficients are parameterized by a low-rank tensor decomposition{, read out by a linear or MLP head}. FUTON combines complementary inductive biases: Fourier bases capture smoothness and periodicity, while the low-rank parameterization enforces low-dimensional spectral structure. Beyond universal approximation, we show that with a linear head, the best FUTON's error splits exactly into the truncation of the signal's series and the best low-rank error of its coefficients, which we bracket from the samples before training, with a FUTON that attains the upper bound. On image fitting, occupancy volumes, and novel view synthesis, against twelve INRs of matched size, FUTON achieves the best average quality, reaches the final quality of each task's strongest baseline 4--34$\times$ sooner than that baseline, infers fastest, and trains in the least memory.

    Project page: \href{https://pashtari.github.io/inro/}{\nolinkurl{pashtari.github.io/inro}}
\end{abstract}

\section{Introduction} \label{sec:introduction}

\emph{Implicit neural representations} (INRs), also known as \emph{neural fields}, have emerged as a powerful paradigm for representing signals, including images, volumes, and radiance fields. Unlike grid-based representations that store signals as discrete arrays (\eg, pixels or voxels), INRs encode signals as continuous functions parameterized by neural network weights, mapping coordinates to signal values. This coordinate-based parameterization offers key advantages: (1)~\emph{resolution independence}, enabling super-resolution without retraining; (2)~\emph{memory efficiency} through compact parameter storage; (3)~natural handling of \emph{irregular sampling}; and (4)~\emph{differentiability} with respect to coordinates, facilitating gradient-based optimization for inverse problems.

INRs have demonstrated strong performance across diverse tasks, including 3D reconstruction~\cite{park2019deepsdf,gropp2020implicit}, view synthesis~\cite{mildenhall2020nerf}, and inverse problems~\cite{shen2024nerp,zha2022naf}. Typical INRs employ multilayer perceptrons (MLPs) with specialized activations, \eg, periodic sine functions~\cite{sitzmann2020implicit}, Gaussian nonlinearities~\cite{ramasinghe2022beyond}, or complex Gabor wavelets~\cite{saragadam2023wire}, all of which significantly outperform ReLU-based networks for signal representation~\cite{rahaman2019spectral}.

MLP-based INRs suffer from two key limitations: (1)~\emph{slow convergence}, leading to long training times; and (2)~an \emph{opaque size--accuracy relation}: how width and depth translate into error is known only empirically. These issues arise because MLPs lack inductive biases aligned with the low-dimensional structure of natural signals. Hybrid INRs, which pair a small MLP with feature grids such as hash tables~\cite{muller2022instant} or factorized planes~\cite{chen2022tensorf,fridovich2023kplanes}, converge faster but spend most of their parameters on the grids.

\begin{figure}[t]
    \centering
    \includegraphics{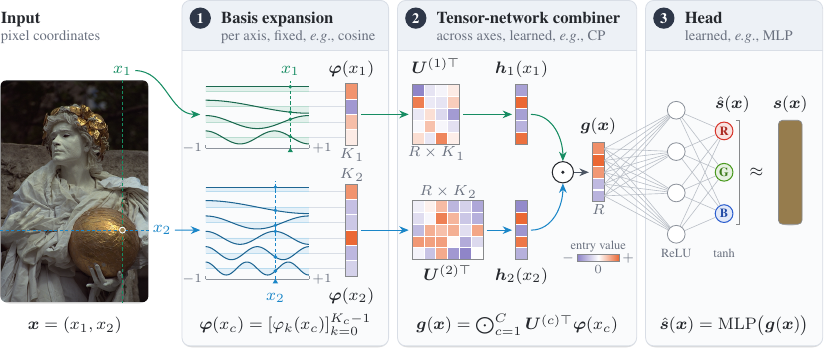}
    \caption{\textbf{FUTON's forward pass} for one pixel of an image ($C = 2$, $D = 3$). \textbf{(1)}~A fixed basis, here cosines, expands each coordinate $x_c$ into $K_c$ functions, at spectral resolution $K = (K_1, K_2)$. \textbf{(2)}~The combiner's factor matrices $\mtx{U}^{(c)}$ project each expansion onto $R$ components, multiplied across axes into $\vct{g}(\vct{x})$: a CP decomposition of the coefficient tensor. \textbf{(3)}~A learned head, here an MLP, maps $\vct{g}(\vct{x})$ to the signal. Arrays hold exact values for random factors, at sizes far below practice.}
    \label{fig:futon_illustration}
\end{figure}

We introduce \emph{Fourier Tensor Network (FUTON)}, an INR that addresses these challenges by expressing signals as \emph{generalized Fourier series} whose coefficients are parameterized via a \emph{low-rank tensor decomposition}. This design embeds complementary inductive biases: Fourier bases capture smoothness and periodicity, while low-rank factorization captures the intrinsic low-dimensional spectral structure of natural signals. As \cref{fig:futon_illustration} shows, a fixed basis expands each coordinate, a tensor-network combiner multiplies the axes' projections, and a small head maps the result to the signal.

FUTON leverages the insight that structured signals exhibit frequency-domain correlations amenable to low-rank approximation. With a Canonical Polyadic (CP) decomposition~\cite{kolda2009tensor}, the construction is classical, a \emph{separated representation}~\cite{beylkin2002numerical} with a fixed basis, used at modest resolution (\cref{sec:related_work}); FUTON brings it to INRs, trained end to end at a spectral resolution per axis comparable to the signal's own.

\paragraph{Contributions.} Our main contributions are:
\begin{itemize}
    \item \textbf{{Architecture}:} We propose FUTON, combining generalized Fourier series with a low-rank tensor decomposition, providing an efficient alternative to MLP-based INRs{, with a spectral resolution per axis, a rank that sets its size, and a linear or MLP head}.
    \item \textbf{Theoretical foundations:} Beyond universal approximation, we show that with a linear head, the best FUTON's error splits exactly into a truncation term and the best rank-$R$ error of the signal's coefficients, bracket it from the samples with a FUTON that attains the upper bound, and verify on 29 signals that training lands in the bracket, to within 0.02\,dB.
    \item \textbf{Computational efficiency:} A coordinate costs $\mathcal{O}(R \sum_c K_c)$ plus the head, linear in each $K_c$ and in $R$; on a grid, the parameter-free basis is cached when the model is built and the per-axis projections are shared, so training and inference cost little beyond the head.
    \item \textbf{{Empirical performance}:} On image fitting, occupancy volumes, and novel view synthesis, against twelve INRs of matched size, FUTON achieves the best average quality, reaches the final quality of the fastest baseline, Instant-NGP~\cite{muller2022instant}, 2.7$\times$ to over 9$\times$ sooner than it does, infers fastest, and trains in the least memory.
\end{itemize}

\paragraph{Notation.} Vectors are denoted by bold lowercase (\eg, $\vct{x}$), matrices by bold uppercase (\eg, $\mtx{U}$), tensors by calligraphic bold uppercase (\eg, $\tns{W}$), and sets by blackboard bold uppercase (\eg, $\X$). Individual elements are indexed by subscripts in non-bold notation: $x_i$, $U_{ij}$, $\mathcal{W}_{ijk}$. Nonnegative integers are denoted by~$\Nz$.

\section{Related Work} \label{sec:related_work}

\paragraph{Implicit Neural Representations.}
NeRF~\cite{mildenhall2020nerf} popularized representing scenes as continuous radiance fields parameterized by MLPs, enabling high-quality novel view synthesis. Subsequent work extended INRs to dynamic scenes~\cite{pumarola2021dnerf,park2021nerfies}, shape representation~\cite{park2019deepsdf,gropp2020implicit}, and generative modeling~\cite{chan2022eg3d}, establishing INRs as a fundamental tool in visual computing.

Most INRs adopt MLP backbones whose inductive bias is strongly shaped by the activation function and input encoding. ReLU MLPs exhibit a low-frequency bias and struggle to fit fine detail~\cite{rahaman2019spectral}. Fourier features mitigate this by mapping coordinates to high-dimensional sinusoidal embeddings~\cite{tancik2020fourier}. SIREN employs sinusoidal activations with principled initialization for stable optimization~\cite{sitzmann2020implicit}. Alternatives include Gaussian activations~\cite{ramasinghe2022beyond} and complex Gabor wavelets (WIRE)~\cite{saragadam2023wire}, which improve space--frequency localization.
FINER~\cite{liu2024finer} introduces variable-periodic activation functions, allowing flexible spectral bias tuning and improved reconstruction of fine details. MFN~\cite{fathony2020multiplicative} replaces depth with multiplicative combinations of sinusoidal or Gabor filters for efficient function approximation in coordinate-based tasks. Recent work speeds up sinusoidal networks by matching their frequency scale to the signal's spectrum (FreSh)~\cite{kania2025fresh}, by scaling their initialization~\cite{yeom2025fast}, or by adapting their frequencies across space~\cite{feng2026sasnet}. Hybrid INRs instead store features on explicit grids and decode them with a small MLP: multiresolution hash tables in Instant-NGP~\cite{muller2022instant}, CP or vector--matrix factors in TensoRF~\cite{chen2022tensorf}, planes in K-Planes~\cite{fridovich2023kplanes}, and lines, planes, and a coarse volume in GA-Planes~\cite{sivgin2024geometric}.

\paragraph{Frequency-Domain Representations.}
Frequency-domain ideas are used to counter the spectral bias of coordinate MLPs. Fourier features and periodic activations inject high frequencies directly into the representation{, and band-limited networks control the frequencies they represent explicitly~\cite{lindell2022bacon}}. More recently, the Fourier reparameterization method~\cite{shi2024improved} learns a coefficient matrix of fixed Fourier bases to compose the weights of MLPs. Unlike these approaches that modulate inputs, nonlinearities, or weights, FUTON parameterizes the Fourier coefficients of the signal directly, via a low-rank tensor decomposition, and uses at most a small MLP after the combination.

\paragraph{Tensor Networks.}
Tensor decompositions and tensor networks provide compact models for multi-dimensional data~\cite{kolda2009tensor}. Canonical Polyadic (CP)~\cite{carroll1970analysis,harshman1970foundations} expresses a tensor as a sum of rank-1 components, while Tucker~\cite{tucker1966some} introduces a low-rank core with factor matrices{, and tensor trains~\cite{oseledets2011tensor} chain three-way cores}. In the INR context, TensoRF factorizes radiance fields into low-rank components; K-Planes extends factorization to dynamic scenes; and F-INR~\cite{vemuri2025f} factorizes an INR into axis-wise sub-networks combined by a tensor decomposition.

Closest to FUTON are \emph{separated representations}~\cite{beylkin2002numerical,beylkin2005algorithms}, sums of products of univariate functions, and their instances with a fixed basis for supervised regression~\cite{kargas2021supervised}, kernel machines~\cite{wesel2021large}, and spectral tensor trains~\cite{bigoni2016spectral}. FUTON is such a model, trained end to end at a high, per-axis spectral resolution, with an MLP head, and an exact error analysis of its linear-head form; unlike TensoRF and F-INR, it factorizes the coefficients of a series in a fixed basis.

\section{Preliminaries} \label{sec:preliminaries}

\subsection{Implicit Neural Representations} \label{sec:inr}

INRs are functions $\func{\bm{f}_{\vct{\theta}}}{\R^C}{\R^D}$ parameterized by neural networks with learnable parameters $\vct{\theta}$ that map continuous coordinates to signal values. For instance, in image representation, $\vct{x} = (x_1, x_2)$ represents pixel coordinates, and $\bm{f}_{\vct{\theta}}(\vct{x}) \in \R^3$ outputs RGB values. Given a ground-truth signal $\vct{s}(\vct{x})$, the goal is to learn parameters $\vct{\theta}$ such that $\vct{s}(\vct{x}) \approx \bm{f}_{\vct{\theta}}(\vct{x})$. This is typically achieved by minimizing the mean squared error over observed coordinate--signal pairs $\set{(\vct{x}_n, \vct{s}_n)}[n=1][N]$, where $\vct{s}_n = \vct{s}(\vct{x}_n)$.

Typical INRs use MLPs{, whose} activation choice critically affects capacity{: a sine $\sin(\omega_0 x)$ (SIREN~\cite{sitzmann2020implicit}), a Gaussian $e^{-(\gamma_0 x)^2}$~\cite{ramasinghe2022beyond}, or a complex Gabor wavelet $e^{j\omega_0 x}e^{-|\gamma_0 x|^2}$ (WIRE~\cite{saragadam2023wire})}.

\subsection{Generalized Fourier Series} \label{sec:gfs}

Fourier series decomposes signals into trigonometric components, enabling compact representation through frequency selectivity. The \emph{generalized Fourier series (GFS)}~\cite{folland2009fourier} extends this idea to arbitrary orthonormal bases and higher-dimensional spaces, offering flexibility for diverse signal classes.

\paragraph{Orthonormal Basis.}
A set of functions $\set{\varphi_k}[k\in\Nz] \subseteq L^2(\X)$  (defined in \cref{sec:l2_space}) is an \textbf{orthonormal basis (ONB)} for $L^2(\X)$ if it satisfies \textbf{orthonormality}, $\inner[L^2]{\varphi_k}{\varphi_\ell} = \delta_{k\ell}$ for all $k, \ell \in \Nz$, with $\delta_{k\ell}$ the Kronecker delta, and \textbf{completeness}: for any function $f \in L^2(\X)$, there exist coefficients $w_k \in \R$ such that the partial sums $s_K = \sum_{k=0}^{K-1} w_k \varphi_k$ converge to $f$ in $L^2(\X)$ as $K \to \infty$ (\cref{def:l2_orthonormal_basis}). As a canonical example, the cosine functions
\begin{equation}
    \label{eq:1d_cosine_basis}
    \varphi_0(x)=1,\qquad \varphi_k(x)= \sqrt{2} \cos(k \pi x) \quad (k \ge 1)
\end{equation}
form an ONB in $L^2([0,1])$, widely studied in Fourier analysis~\cite{stein2011fourier}{; coordinates in $[-1, 1]$, as in our experiments, are mapped affinely onto $[0, 1]$}.

\paragraph{Separable Multi-dimensional Basis.}
Let $\set{\varphi_k}[k\in\Nz]$ be a set of one-dimensional functions in $L^2([0,1])$. We construct a set of \emph{separable} multi-dimensional functions in $L^2([0,1]^C)$ by products of the one-dimensional basis functions:
\begin{equation}
    \label{eq:separable_basis}
    \phi_{\vct{k}}(\vct{x}) \triangleq \prod_{c=1}^C \varphi_{k_c}(x_c),
\end{equation}
where $\vct{k}=(k_1,\ldots,k_C) \in \Nz^C$ is a multi-index, and $\vct{x}=(x_1,\ldots,x_C)\in[0,1]^C$ are coordinates.

\begin{theorem}[Separable Basis Orthonormality{; proof in \cref{sec:proof_separable_basis_orthonormality}}] \label{thm:separable_basis_orthonormality}
    If $\set{\varphi_k}[k\in\Nz]$ is an ONB for $L^2([0,1])$, then $\set{\phi_{\vct{k}}}[\vct{k}\in\Nz^C]$ defined in~\eqref{eq:separable_basis} forms an ONB for $L^2([0,1]^C)$.
\end{theorem}

\paragraph{Generalized Fourier Series.}
For a vector-valued function $\vct{s}(\vct{x})=[s_1(\vct{x}),\ldots,s_D(\vct{x})]^\top$ with $s_d \in L^2([0,1]^C)$ for $d\in\{1,\ldots,D\}$, the partial sum of the \emph{generalized Fourier series (GFS)}~\cite{folland2009fourier} using the ONB $\set{\phi_{\vct{k}}}[\vct{k}\in\Nz^C]$ (defined in~\eqref{eq:separable_basis}) is defined by
\begin{equation}
    \label{eq:gfs}
    {\vct{s}_K}(\vct{x}) \triangleq \sum_{\vct{k}\in\K} \vct{w}^{(\vct{k})} \phi_{\vct{k}}(\vct{x}),
\end{equation}
where $\K=\{0,\ldots,K_1-1\} \times \cdots \times \{0,\ldots,K_C-1\}$ is a finite truncation of all multi-indices, set by the \emph{spectral resolution} $K = (K_1, \ldots, K_C)$, the number of basis functions per axis, and the coefficient vector $\vct{w}^{(\vct{k})}=[w^{(\vct{k})}_1,\ldots,w^{(\vct{k})}_D]^\top \in \R^D$ has components obtained via inner products:
\begin{equation}
    \label{eq:gfs_weights}
    w^{(\vct{k})}_d= \inner[L^2]{s_d}{\phi_{\vct{k}}} = \int_{[0,1]^C} s_d(\vct{x}) \phi_{\vct{k}}(\vct{x}) \, d\vct{x}.
\end{equation}

\begin{theorem}[GFS Convergence{; proof in \cref{sec:proof_gfs_convergence}}]\label{thm:gfs_convergence}
    The partial sum ${\vct{s}_K}$ converges to $\vct{s}$ in $L^2([0,1]^C; \R^D)$, \ie, $\norm[L^2]{s_d - {s_{d,K}}} \to 0$ as $\min_c K_c\to\infty$ for all $d\in\{1,\ldots,D\}$.
\end{theorem}

\paragraph{Tensor Representation.}
The multi-dimensional basis functions in~\eqref{eq:separable_basis} can be expressed as elements of the tensor product (defined in \cref{sec:tensor_operations}):
\begin{equation}
    \label{eq:tensor_product_basis}
    \bm{\Phi}(\vct{x}) = \vct{\varphi}(x_1) \tenprod \vct{\varphi}(x_2) \tenprod \cdots \tenprod \vct{\varphi}(x_C),
\end{equation}
where $\bm{\Phi}(\vct{x})\in\R^{{K_1\times \cdots \times K_C}}$ is a \emph{rank-1 $C$th-order tensor}, and $\vct{\varphi}(x_c) = [\varphi_k(x_c)]_{k=0}^{{K_c}-1} {\in \R^{K_c}}$ with $\set{\varphi_k}[k\in\Nz]$ being the one-dimensional basis from~\eqref{eq:1d_cosine_basis}.

Now let $\tns{W} = [w^{(\vct{k})}_d] \in \R^{{K_1 \times \cdots \times K_C} \times D}$ denote the coefficient tensor of order $C+1$, where $w^{(\vct{k})}_d$ is the coefficient for basis function $\phi_{\vct{k}}$ and output dimension $d$, as defined in~\eqref{eq:gfs_weights}. Using the generalized dot product $\dotprod$, which contracts all modes of its first argument with the leading modes of its second (\cref{def:generalized_dot_product}), the GFS in~\eqref{eq:gfs} can be compactly expressed as
\begin{equation}
    \label{eq:gfs_tensordot}
    {\vct{s}_K}(\vct{x}) =  \bm{\Phi}(\vct{x}) \dotprod \tns{W}.
\end{equation}

\section{Methods} \label{sec:methods}

\subsection{Fourier Tensor Network (FUTON)} \label{sec:futon}

While the GFS provides a theoretically sound foundation, its computational complexity poses a challenge. The tensor contraction in~\eqref{eq:gfs_tensordot} requires $\mathcal{O}({D \prod_c K_c})$ operations, scaling exponentially with input dimension $C$.

\begin{figure}[t]
    \centering
    \captionsetup[subfigure]{font=footnotesize}
    \begin{subfigure}[b]{81bp}\centering\includegraphics{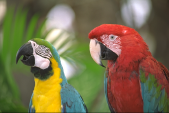}\caption{$\vct{s}$ (image)}\end{subfigure}\hfill
    \begin{subfigure}[b]{62bp}\centering\includegraphics{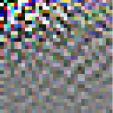}\caption{$\tns{W}^{\star}$ (full rank)}\end{subfigure}\hfill
    \begin{subfigure}[b]{62bp}\centering\includegraphics{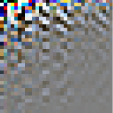}\caption{$\tns{W}$ ($R = 8$)}\end{subfigure}\hfill
    \begin{subfigure}[b]{79.2bp}\centering\includegraphics{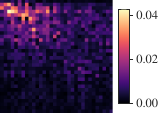}\caption{$\lVert \vct{w}^{\star(\vct{k})} - \vct{w}^{(\vct{k})} \rVert$}\end{subfigure}\hfill
    \begin{subfigure}[b]{106.2bp}\centering\includegraphics{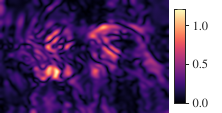}\caption{$\lVert \vct{s}_K(\vct{x}) - \hat{\vct{s}}(\vct{x}) \rVert$}\end{subfigure}
    \caption{\textbf{FUTON's coefficient tensor.} (a)~A Kodak image and its coefficient tensor in the cosine basis at $K_c = 32$, visualized as RGB images with the lowest frequencies at the top left: (b)~in full, 3,072 coefficients (23.79\,dB), and (c)~at rank $R = 8$, the certified FUTON of \cref{prop:error}, built without training, 536 parameters (19.89\,dB). (d,~e)~The norm of their difference per frequency, and per pixel between the images they represent.}
    \label{fig:futon_weights}
\end{figure}

To address this computational bottleneck, we \emph{reparameterize} the coefficient tensor $\tns{W}$ using \emph{low-rank tensor decomposition}, yielding our \emph{Fourier Tensor Network} (FUTON) model. Specifically, we employ the Canonical Polyadic (CP) decomposition format~\cite{kolda2009tensor}, which represents $\tns{W}$ as a sum of $R$ rank-1 tensors:
\begin{equation}\label{eq:cp}
    \tns{W} = \sum_{r=1}^R \vct{u}_r^{(1)} \tenprod \vct{u}_r^{(2)} \tenprod \cdots \tenprod \vct{u}_r^{(C)} \tenprod \vct{v}_r,
\end{equation}
where $\vct{u}_r^{(c)} \in \R^{{K_c}}$, $\vct{v}_r \in \R^{D}$, and $R$ is the CP \emph{rank}. We define factor matrices $\mtx{U}^{(c)} = {[\vct{u}_1^{(c)} \mid \cdots \mid \vct{u}_R^{(c)}]} \in \R^{{K_c} \times R}$ for each input dimension $c$, and $\mtx{V} = {[\vct{v}_1 \mid \cdots \mid \vct{v}_R]} \in \R^{D \times R}$ for the output.
For a Kodak image at $K_c = 32$ (\cref{fig:futon_weights}), rank $R = 8$ keeps 89\% of the energy of $\tns{W}^{\star} \in \R^{32 \times 32 \times 3}$ while reducing parameters from $3{,}072$ to {$536$}. This highlights the inherent low-rank structure of natural signals in the Fourier domain, which FUTON exploits for efficient and compact representation.

Substituting~\eqref{eq:tensor_product_basis} and~\eqref{eq:cp} into~\eqref{eq:gfs_tensordot}, and subsequently rearranging the summations, yields the following compact form{, which \cref{fig:futon_tensor_network} derives in tensor diagram notation}:
\begin{gather}
    \label{eq:futon_cp}
    {\hat{\vct{s}}}(\vct{x}) = \mtx{V} \vct{g}(\vct{x}),\\
    \label{eq:futon_cp_g}
    \vct{g}(\vct{x}) \triangleq \vct{h}_1(x_1) \hadamard \vct{h}_2(x_2) \hadamard \cdots \hadamard \vct{h}_C(x_C), \\
    \label{eq:futon_cp_h}
    \vct{h}_c(x_c) \triangleq \mtx{U}^{(c)\top} \vct{\varphi}(x_c),  \quad c\in\{1,\ldots,C\}
\end{gather}
where $\vct{h}_c(x_c) \in \R^R$ projects basis functions at $x_c$ onto $\mtx{U}^{(c)}$, and the \emph{feature vector} $\vct{g}(\vct{x}) \in \R^R$ is the Hadamard product of projected features.

\paragraph{Basis, Combiner, and Head.}
Equations~\eqref{eq:futon_cp}--\eqref{eq:futon_cp_h} are the three stages of \cref{fig:futon_illustration}: a fixed \emph{basis} expands each coordinate, a learned \emph{combiner}, here CP, projects each expansion onto $R$ components and multiplies them across axes, and a learned \emph{head}, here $\mtx{V}$, maps the feature vector to the signal. With this linear head, FUTON is a GFS whose coefficient tensor has CP rank at most $R$, which makes its error exactly analyzable (\cref{sec:error_analysis}). A nonlinear head adds capacity: a one-hidden-layer MLP of width $M$,
\begin{equation} \label{eq:futon_head}
    \hat{\vct{s}}(\vct{x}) = \vct{\psi}\big(\vct{g}(\vct{x})\big), \qquad \vct{\psi}(\vct{g}) = \mtx{V}_2 \operatorname{ReLU}(\mtx{V}_1 \vct{g} + \vct{b}_1) + \vct{b}_2,
\end{equation}
which contains every linear head when $M \ge 2D$, as $\operatorname{ReLU}(t) - \operatorname{ReLU}(-t) = t$, and adds 7.4\,dB on images (\cref{sec:ablation_details}). An optional output activation, such as $\tanh$, completes the model.

\begin{figure}[t]
    \centering
    \begin{subfigure}[b]{251pt}\centering\includegraphics{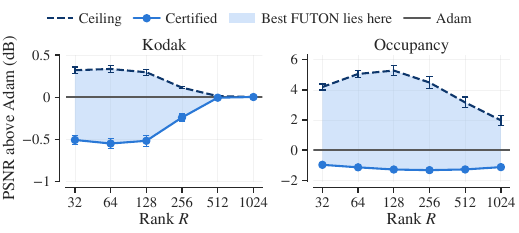}\caption{}\label{fig:error_bound_gap}\end{subfigure}\hfill
    \begin{subfigure}[b]{126pt}\centering\includegraphics{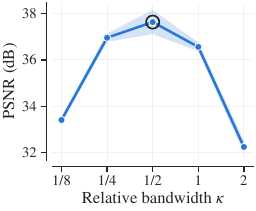}\caption{}\label{fig:bandwidth_main}\end{subfigure}
    \caption{\textbf{The error bound and the bandwidth.} (a)~PSNR relative to Adam's FUTON per rank at $\kappa = 1/2$, averaged over the 24 Kodak images and the five volumes, $\pm$ one standard error: the ceiling and the certified FUTON bracket the best FUTON. (b)~Mean PSNR on Kodak against the relative bandwidth $\kappa$ at 197k parameters, the rank adjusted to keep the size, the default circled.}
    \label{fig:error_bound}
\end{figure}

\paragraph{Spectral Resolution and Rank.}
We set each axis's number of basis functions relative to its number of samples $L_c$, $K = (\kappa L_1, \ldots, \kappa L_C)$, with a \emph{relative bandwidth} $\kappa$. At $\kappa = 1$, the cosines span every signal on the sampling grid, and a smaller $\kappa$ keeps fewer, lower frequencies; $\kappa \in \{1/4, 1/2, 1\}$ work best in practice (\cref{fig:error_bound}b), and we use $\kappa = 1/2$. The rank then sets the size, $R \sum_c K_c$ parameters in the combiner plus $RD$ (and $D$ biases) for a linear head or $RM + M + MD + D$ for an MLP head: we take the largest $R$ within a parameter budget.

Given observed coordinate--signal pairs $\set{(\vct{x}_n, \vct{s}_n)}[n=1][N]$, the FUTON parameters $\bm{\theta} = (\mtx{U}^{(1)}, \allowbreak \ldots, \allowbreak \mtx{U}^{(C)}, \allowbreak {\vct{\psi}})${, with $\vct{\psi}$ the head,} are learned by minimizing the reconstruction loss:
\begin{equation}
    \label{eq:futon_loss}
    \bm{\theta}^* = \argmin_{\bm{\theta}} \frac{1}{N} \sum_{n=1}^{N} \mathcal{L}\left(\vct{s}_n, {\hat{\vct{s}}_{\bm{\theta}}}(\vct{x}_n)\right),
\end{equation}
where $\mathcal{L}$ is a loss function, \eg, mean squared error. Factor matrices are initialized from a uniform distribution (Kaiming-uniform with fan-in $K_c$) and optimized via gradient-based methods (\eg, Adam~\cite{kingma2014adam}). FUTON is a universal approximator (\cref{thm:futon_expressiveness}): it can approximate any function in $L^2([0,1]^C; \R^D)$ to arbitrary precision with a sufficiently large spectral resolution $K$ and rank $R$.

\begin{theorem}[Universal Approximation{; proof in \cref{sec:proof_futon_expressiveness}}]\label{thm:futon_expressiveness}
    For any $\varepsilon > 0$ and vector-valued function $\vct{s} \in L^2([0,1]^C; \R^D)$, there exist $K \in {\N}$ {(the same on every axis)}, rank $R \leq K^{C-1}\min\{K, D\}$, and factor matrices $\seq{\mtx{U}^{(c)}}[c=1][C]$, $\mtx{V}$ such that $\norm[L^2]{{s_d - \hat{s}_d}} < \varepsilon$ for all $d\in\{1,\ldots,D\}$.
\end{theorem}

Universality alone does not distinguish FUTON from MLPs~\cite{cybenko1989approximation,sonoda2017neural}; it certifies that the truncated low-rank parameterization loses no expressivity, also with an MLP head ($M \ge 2D$) or, for targets $\tanh(\tilde{\vct{s}})$, a $\tanh$ output (\cref{rem:nonlinear}).

\subsection{Computational Complexity} \label{sec:futon_complexity}

A key advantage of FUTON is its low time complexity with respect to the input and output dimensions. In the optimal contraction order (\cref{fig:futon_tensor_network,sec:contraction_order}), the forward pass projects each basis vector $\vct{\varphi}(x_c)$ onto the factor matrix $\mtx{U}^{(c)}$~\eqref{eq:futon_cp_h}{, in $\mathcal{O}(R \sum_c K_c)$ operations, multiplies the projections}~\eqref{eq:futon_cp_g}{, in} $\mathcal{O}(CR)$, and maps the feature vector $\vct{g}(\vct{x})$ to the output~\eqref{eq:futon_cp}{, in} $\mathcal{O}(DR)${, or $\mathcal{O}(MR + MD)$ for an MLP head of width $M$}. The total per coordinate, $\mathcal{O}(R \sum_c K_c + CR + DR)$ with a linear head, is linear in each $K_c$, in $R$, and in $C$ and $D$ separately.

\paragraph{Cached Basis and Shared Projections.}
The basis has no parameters, so on a sampling grid FUTON evaluates it once, when the model is built, and caches an $L_c \times K_c$ table per axis: training on grid samples never evaluates the basis again, and inference on the cached grid is particularly fast. Each $\vct{h}_c$ depends on one coordinate, so the projections of the table's rows are shared by all points with that coordinate: on a grid of $N = \prod_c L_c$ points, the combiner costs $\mathcal{O}(R \sum_c L_c K_c + NCR)$ instead of $\mathcal{O}(NR \sum_c K_c)$. For a $768 \times 512$ image at $\kappa = 1/2$ and $R = 280$, that is under $10^3$ operations per pixel instead of $1.8 \times 10^5$, so the head dominates. Off the grid, as along a radiance field's rays, the basis is evaluated per point.

\subsection{Error Analysis} \label{sec:error_analysis}

\begin{figure}[t]
    \centering
    \includegraphics{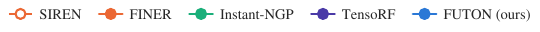}\\[1pt]
    \setlength{\tabcolsep}{0pt}
    \begin{tabular}{@{}c@{\hspace{0.09in}}c@{\hspace{0.09in}}c@{}}
        \footnotesize Images & \footnotesize Occupancy volumes & \footnotesize Radiance fields \\
        \includegraphics{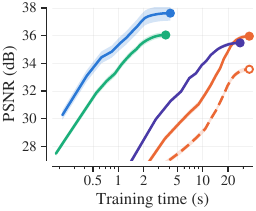} &
        \includegraphics{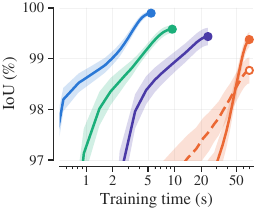} &
        \includegraphics{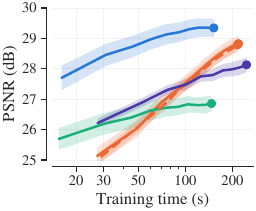}
    \end{tabular}
    \caption{\textbf{Training speed.} Mean quality during training against median training time, with one standard error, for the featured models; radiance fields on validation views. \Cref{fig:tradeoffs,fig:tradeoffs_memory} plot every model's final quality against its costs.}
    \label{fig:learning_speed}
\end{figure}

\begin{table}[t]
    \caption{\textbf{Images}: quality averaged over the 24 Kodak images, and median training time, inference throughput, and peak GPU memory, at about 197k parameters. Best \textbf{bold}, second \underline{underlined}, among the models shown; \cref{tab:image_full} adds standard errors and the other three baselines.}
    \label{tab:image}
    \centering
    \footnotesize
    \setlength{\tabcolsep}{5pt}
    \begin{tabular}{@{}lccccccc@{}}
\toprule
 & \multicolumn{3}{c}{Quality} & \multicolumn{2}{c}{Training} & \multicolumn{2}{c}{Inference} \\
\cmidrule(lr){2-4}\cmidrule(lr){5-6}\cmidrule(l){7-8}
Model & PSNR & SSIM & LPIPS & Time & Memory & Throughput & Memory \\
 & (dB) $\uparrow$ & $\uparrow$ & $\downarrow$ & (s) $\downarrow$ & (GB) $\downarrow$ & (MPix/s) $\uparrow$ & (MB) $\downarrow$ \\
\midrule
RFF~\cite{tancik2020fourier} & 31.28 & 0.846 & 0.271 & 29.8 & 2.08 & 30.7 & 1208 \\
MFN~\cite{fathony2020multiplicative} & \underline{36.72} & \underline{0.938} & 0.129 & 57.3 & 7.67 & 26.8 & 3224 \\
SIREN~\cite{sitzmann2020implicit} & 33.57 & 0.906 & 0.192 & 35.7 & 3.25 & 34.4 & 805 \\
FreSh (SIREN)~\cite{kania2025fresh} & 34.88 & 0.922 & 0.150 & 39.3 & 3.25 & 34.5 & 805 \\
FINER~\cite{liu2024finer} & 35.96 & \underline{0.938} & \underline{0.120} & 35.7 & 3.25 & 34.3 & 805 \\
Instant-NGP~\cite{muller2022instant} & 36.05 & 0.932 & 0.142 & \textbf{3.6} & \underline{0.72} & \underline{198.3} & \underline{503} \\
TensoRF~\cite{chen2022tensorf} & 35.48 & 0.920 & 0.161 & 27.8 & 1.79 & 55.2 & 805 \\
GA-Planes~\cite{sivgin2024geometric} & 32.23 & 0.878 & 0.227 & 19.0 & 1.32 & 68.1 & 805 \\
F-INR (SIREN)~\cite{vemuri2025f} & 29.35 & 0.786 & 0.343 & 7.1 & 2.70 & 136.3 & 1338 \\
\midrule
FUTON (ours) & \textbf{37.63} & \textbf{0.955} & \textbf{0.098} & \underline{4.1} & \textbf{0.36} & \textbf{228.7} & \textbf{201} \\
\bottomrule
\end{tabular}

\end{table}

In practice, the question is what a FUTON of resolution $K$ and rank $R$ loses on a given signal, before training; for the linear head, the answer is exact. Let $\tns{W}^{\star} \in \R^{K_1 \times \cdots \times K_C \times D}$ hold the signal's coefficients~\eqref{eq:gfs_weights}, $\norm{\vct{f}}^2 = \sum_{d} \norm[L^2]{f_d}^2$, $\varepsilon_R(\tns{A}) = \inf\{\norm[F]{\tns{A} - \tns{B}} : \operatorname{rank}\tns{B} \le R\}$ be the best rank-$R$ error of a tensor $\tns{A}$, and $\tns{A}_{\langle \mathbb{S} \rangle}$ be its matricization with the modes in $\mathbb{S}$ as rows.

\begin{proposition}[Error of FUTON with a linear head] \label{prop:error}
    Let $\vct{s} \in L^2([0,1]^C; \R^D)$, $K = (K_1, \ldots, K_C)$, $R \ge 1$, and $\hat{\vct{s}}_{\vct{\theta}}$ the model~\eqref{eq:futon_cp} with parameters $\vct{\theta} = (\mtx{U}^{(1)}, \ldots, \mtx{U}^{(C)}, \mtx{V})$.
    \begin{enumerate}[leftmargin=*, itemsep=1pt, topsep=2pt]
        \item \emph{Every FUTON.} For every $\vct{\theta}$, the error is a truncation plus a CP misfit: $\norm{\vct{s} - \hat{\vct{s}}_{\vct{\theta}}}^2 = \norm{\vct{s} - \trunc{K}}^2 + \norm[F]{\tns{W}^{\star} - \cp{\mtx{U}^{(1)}, \ldots, \mtx{U}^{(C)}, \mtx{V}}}^2$.
        \item \emph{The best FUTON.} $\inf_{\vct{\theta}} \norm{\vct{s} - \hat{\vct{s}}_{\vct{\theta}}}^2 = \norm{\vct{s} - \trunc{K}}^2 + \varepsilon_R(\tns{W}^{\star})^2$.
        \item \emph{Computable bounds.} $\lowerr_R \le \varepsilon_R(\tns{W}^{\star}) \le \upperr_R$, where $\lowerr_R^2 = \max_{\mathbb{S}} \sum_{i > R} \sigma_i^2\big(\tns{W}^{\star}_{\langle \mathbb{S} \rangle}\big)$ over the nonempty proper subsets $\mathbb{S}$ of the $C+1$ modes, and $\upperr_R$ is the misfit of an explicit rank-$R$ tensor built from $\tns{W}^{\star}$ by nested SVDs; the FUTON with its factors, the \emph{certified FUTON}, has squared error exactly $\norm{\vct{s} - \trunc{K}}^2 + \upperr_R^2$.
    \end{enumerate}
\end{proposition}
\Cref{sec:proof_error} proves it for any basis and on sampling grids. For a signal in $[-1, 1]$, the proposition gives three PSNRs at each $(K, R)$: the \emph{truncation} PSNR, which no FUTON of resolution $K$ exceeds; the \emph{ceiling}, from $\lowerr_R$, which no FUTON of that size exceeds; and the \emph{certified} PSNR, from $\upperr_R$, attained without training, with the best FUTON between the last two. They take one pass over the samples and SVDs (\cref{sec:error_algorithm}), on average 0.02 to 3.0\,s per signal for six ranks at once, 16 to 860 times less than training. On the Kodak images and the volumes, in 522 runs over $\kappa$ and $R$ (\cref{sec:error_experiments}), Adam's FUTON always lands between the certified FUTON, to within 0.02\,dB, and the ceiling (\cref{fig:error_bound}): on images, the bracket is a median 0.05\,dB wide, so the bounds predict what training reaches, and on volumes, where the ceiling is loose, the certified FUTON still guarantees a PSNR before training.

The bracket also sizes FUTON before training: for a target PSNR, the smallest FUTON whose ceiling reaches it and the smallest whose certified PSNR reaches it enclose the smallest one Adam trains to it, on every signal and target; on Kodak, the two coincide for 56\% of images and targets, and the certified choice matches Adam's for 76\% (\cref{sec:error_experiments}). \Cref{sec:error_consequences} bounds the truncation by the signal's smoothness, $\norm{\vct{s} - \trunc{K}}^2 \le \sum_c \norm{\partial_c \vct{s}}^2 / (\pi K_c)^2$ for the cosine basis (\cref{cor:rate}), and proves a low-rank sufficiency theorem.

\section{Experiments} \label{sec:experiments}

We compare FUTON with twelve INRs on image fitting, occupancy volumes, and novel view synthesis: MLPs on Fourier features (RFF) or positional encodings (PE-MLP), MFN, MLPs with specialized activations (SIREN, FreSh, Gauss, WIRE, and FINER), the hybrid grid models Instant-NGP, TensoRF, and GA-Planes, and F-INR. The main tables omit PE-MLP, Gauss, and WIRE, which never rank first or second in a task's mean quality or costs (\cref{sec:full_results} reports all twelve).

On each task, every baseline is the smallest of its kind above a common parameter budget, and FUTON's rank keeps it below all of them. Every model trains with Adam~\cite{kingma2014adam} at its best learning rate from one grid, under cosine or exponential decay, on one NVIDIA A100 GPU (\cref{sec:detailed_experimental_setup}). Quality, PSNR, SSIM~\cite{wang2004image}, LPIPS~\cite{zhang2018unreasonable}, or IoU, is a mean over the signals; training time, inference speed on whole signals, and peak GPU memory are medians. FUTON uses the cosine basis~\eqref{eq:1d_cosine_basis} throughout.

\begin{table}[t]
    \caption{\textbf{Occupancy volumes}: IoU on each shape and the mean, and median costs, marked as in \cref{tab:image}, at about 131k parameters. \Cref{tab:occupancy_full} adds standard errors and the other baselines.}
    \label{tab:occupancy}
    \centering
    \footnotesize
    \setlength{\tabcolsep}{3.5pt}
    \begin{tabular}{@{}lccccc>{\columncolor{meancol}}ccccc@{}}
\toprule
 & \multicolumn{6}{c}{IoU (\%) $\uparrow$} & \multicolumn{2}{c}{Training} & \multicolumn{2}{c}{Inference} \\
\cmidrule(lr){2-7}\cmidrule(lr){8-9}\cmidrule(l){10-11}
Model & Armadillo & Dragon & Happy & Lucy & Thai & Mean & Time & Memory & Throughput & Memory \\
 & & & Buddha & & Statue & & (s) $\downarrow$ & (GB) $\downarrow$ & (MVox/s) $\uparrow$ & (GB) $\downarrow$ \\
\midrule
RFF & \underline{99.71} & 99.65 & 99.59 & 99.78 & 99.28 & 99.60 & 46.5 & 4.15 & 48.9 & 3.33 \\
MFN & 99.23 & 97.47 & 97.50 & 99.10 & 98.74 & 98.41 & 108.5 & 15.07 & 38.9 & 5.49 \\
SIREN & 98.78 & 99.18 & 99.41 & 98.53 & 97.94 & 98.77 & 70.0 & 6.48 & 51.0 & 2.26 \\
FreSh (SIREN) & 99.34 & 99.67 & 99.65 & 99.15 & 98.83 & 99.33 & 263.6 & 6.48 & 51.3 & 2.26 \\
FINER & 99.45 & 99.59 & 99.70 & 99.29 & 98.85 & 99.38 & 69.9 & 6.48 & 51.3 & 2.26 \\
Instant-NGP & 99.50 & 99.59 & 99.69 & 99.78 & \underline{99.35} & 99.58 & \underline{9.4} & 2.11 & 162.6 & 2.93 \\
TensoRF & 99.53 & 99.71 & 99.78 & 99.34 & 98.82 & 99.44 & 23.8 & \underline{1.75} & \underline{202.3} & \underline{1.19} \\
GA-Planes & 99.58 & \underline{99.79} & \underline{99.93} & \underline{99.79} & 99.34 & \underline{99.69} & 62.2 & 2.31 & 111.9 & 1.37 \\
F-INR (SIREN) & 97.75 & 98.25 & 98.08 & 96.27 & 96.42 & 97.35 & 20.9 & 6.59 & 179.2 & 3.39 \\
\midrule
FUTON (ours) & \textbf{99.86} & \textbf{99.94} & \textbf{99.99} & \textbf{99.96} & \textbf{99.74} & \textbf{99.90} & \textbf{5.4} & \textbf{0.28} & \textbf{537.4} & \textbf{0.61} \\
\bottomrule
\end{tabular}

\end{table}

\begin{figure}[t]
    \centering
    \includegraphics{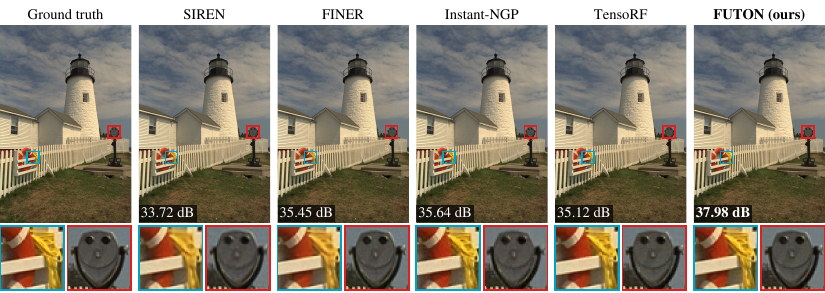}
    \caption{\textbf{Image fitting} on kodim19, with PSNR and two regions magnified; \cref{fig:qualitative_kodim01} shows more.}
    \label{fig:qualitative_image}
\end{figure}

\subsection{Image Fitting} \label{sec:image_representation}

On the 24 Kodak images~\cite{kodak1993kodak}, fitted whole at every step for 1,000 steps at about 197k parameters, FUTON ($\kappa = 1/2$, $R = 280$, MLP head of width $M = 64$) reaches 37.63\,dB PSNR on average (\cref{tab:image}), 0.9\,dB above MFN, the strongest baseline, with the best SSIM and LPIPS, and is the most accurate model on 21 of the 24 images, kodim19 among them (\cref{fig:qualitative_image}).

FUTON converges fastest (\cref{fig:learning_speed}), helped by being shallow, with one layer of factor matrices and a one-hidden-layer head: it passes 35\,dB in 0.9\,s, against 1.6\,s for Instant-NGP and 14.6\,s for MFN, and reaches MFN's final PSNR, which MFN needs 57\,s to reach, in 1.7\,s (\cref{tab:time_to_target}). Its 4.1\,s of training are second only to Instant-NGP's 3.6\,s, whose final PSNR FUTON reaches in 1.3\,s. FUTON also infers fastest and uses the least memory, half of Instant-NGP's in training and 40\% in inference: on the pixel grid, its cached basis and shared projections leave only the products across axes and the head to compute per pixel (\cref{sec:futon_complexity}), and, being shallow, it keeps few activations in training.

\begin{figure}[t]
    \centering
    \includegraphics{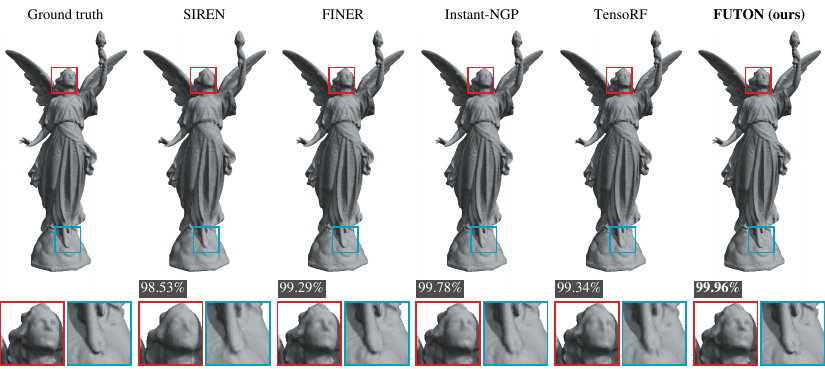}
    \caption{\textbf{Occupancy volumes} on Lucy, with IoU and two regions magnified; \cref{fig:qualitative_volumes_a} shows more.}
    \label{fig:qualitative_occupancy}
\end{figure}

\begin{table}[t]
    \caption{\textbf{Radiance fields}: PSNR on each Blender scene's test views and the mean, and median costs, marked as in \cref{tab:image}, speed in frames per second, at about 74k parameters. \Cref{tab:nerf_means} adds SSIM, LPIPS, standard errors, and the other baselines.}
    \label{tab:nerf}
    \centering
    \footnotesize
    \setlength{\tabcolsep}{2.5pt}
    \begin{tabular}{@{}lcccccccc>{\columncolor{meancol}}ccccc@{}}
\toprule
 & \multicolumn{9}{c}{PSNR (dB) $\uparrow$} & \multicolumn{2}{c}{Training} & \multicolumn{2}{c}{Inference} \\
\cmidrule(lr){2-10}\cmidrule(lr){11-12}\cmidrule(l){13-14}
Model & Chair & Drums & Ficus & Hotdog & Lego & Materials & Mic & Ship & Mean & Time & Mem. & Speed & Mem. \\
 & & & & & & & & & & (s) $\downarrow$ & (GB) $\downarrow$ & (FPS) $\uparrow$ & (GB) $\downarrow$ \\
\midrule
RFF & \underline{33.10} & 22.91 & 27.47 & 32.76 & 28.48 & 25.62 & 33.37 & 20.88 & 28.07 & 179.3 & 2.91 & 38.7 & 2.12 \\
MFN & 32.94 & 23.77 & 27.28 & 32.56 & 28.52 & 25.60 & 33.04 & 22.50 & 28.28 & 275.6 & 9.97 & 35.5 & 3.61 \\
SIREN & 32.54 & \underline{24.58} & 27.61 & 32.40 & 29.03 & 26.53 & 33.25 & 22.29 & 28.53 & 215.5 & 5.08 & 38.4 & 2.14 \\
FreSh (SIREN) & 32.38 & 24.19 & 27.60 & 32.81 & 29.62 & 25.18 & 32.58 & 22.32 & 28.34 & 217.2 & 5.03 & 39.4 & 2.13 \\
FINER & 32.85 & \textbf{24.61} & \textbf{27.74} & 32.30 & 29.46 & 26.11 & \textbf{33.58} & 22.27 & \underline{28.62} & 218.8 & 5.02 & 39.1 & 2.15 \\
Instant-NGP & 31.35 & 23.03 & 22.64 & 30.87 & 27.29 & 24.57 & 31.92 & 21.16 & 26.60 & \textbf{147.0} & 2.80 & 44.0 & 3.26 \\
TensoRF & 31.60 & 23.70 & 25.74 & 32.73 & 28.60 & 26.31 & 32.43 & 22.78 & 27.99 & 246.2 & 2.14 & \underline{47.2} & \textbf{2.04} \\
GA-Planes & 32.58 & 23.75 & 25.77 & \underline{33.06} & \underline{29.63} & 26.09 & 32.29 & \underline{22.79} & 28.24 & 271.6 & \underline{2.10} & 45.7 & \textbf{2.04} \\
F-INR (SIREN) & 32.81 & 21.42 & 24.23 & 32.39 & 29.22 & \underline{27.37} & 31.78 & 20.93 & 27.52 & 292.8 & 16.18 & 33.7 & 6.69 \\
\midrule
FUTON (ours) & \textbf{33.18} & 23.86 & \underline{27.70} & \textbf{33.20} & \textbf{31.28} & \textbf{27.65} & \underline{33.40} & \textbf{23.48} & \textbf{29.22} & \underline{152.0} & \textbf{1.88} & \textbf{49.6} & \textbf{2.04} \\
\bottomrule
\end{tabular}

\end{table}

\subsection{Occupancy Volumes} \label{sec:occupancy_volumes}

Five Stanford scans~\cite{stanford3dscan}, normalized as in BACON~\cite{lindell2022bacon}, are labeled $\pm 1$ at each voxel center at 512 voxels per unit length, and fitted from $2^{20}$ random voxels per step for 1,000 steps at about 131k parameters. With just a linear head ($\kappa = 1/2$, $R = 215$ to $310$), FUTON has the best IoU on every shape (\cref{tab:occupancy,fig:qualitative_occupancy}), 99.90\% on average, against 99.69\% for GA-Planes, the strongest baseline, and at most 99.60\% for the MLP-based INRs. Its margin is largest on Thai Statue, 0.39 points, the shape on which most models score lowest. FUTON trains fastest, reaching 99.5\% IoU in 3.2\,s against 7.2\,s for Instant-NGP and 44.5\,s for GA-Planes, infers 2.7 times faster than TensoRF, the next fastest, and uses the least memory: a sixth of TensoRF's in training and half in inference.

\begin{figure}[t]
    \centering
    \includegraphics{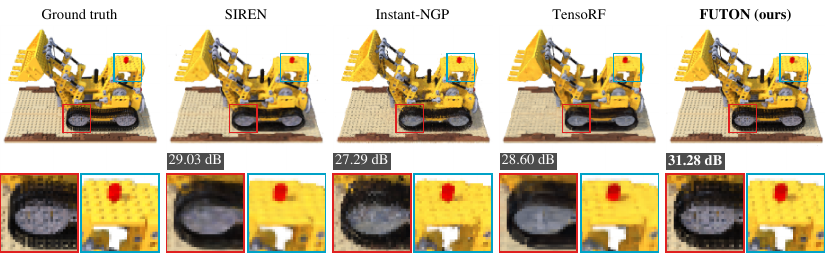}
    \caption{\textbf{Novel view synthesis} on Lego, with mean test-view PSNR and two regions magnified; \cref{fig:qualitative_nerf_more} shows more.}
    \label{fig:qualitative_nerf}
\end{figure}

\subsection{Novel View Synthesis} \label{sec:view_synthesis}

On the eight Blender scenes~\cite{mildenhall2020nerf}, under FINER's protocol~\cite{liu2024finer}, the compared INR is a radiance field's density network, with a shared color network, at about 74k parameters, trained for 25,000 steps; FUTON has $K_c = 100$, $R = 180$, and an MLP head of width $M = 64$. It achieves the best mean PSNR (\cref{tab:nerf}), 29.22\,dB against 28.62\,dB for FINER, with the best mean SSIM and LPIPS, 0.943 and 0.065 against 0.936 and 0.071 for GA-Planes (\cref{tab:nerf_means}), and the best PSNR on five of the eight scenes, Lego by 1.65\,dB (\cref{fig:qualitative_nerf}); FINER leads on Drums by 0.75\,dB and on Ficus and Mic by under 0.2\,dB.

FUTON trains in 152\,s, close to Instant-NGP's 147\,s, reaches FINER's final validation PSNR 4.2 times sooner, and trains in the least memory. It renders fastest too, but by little (49.6 against 47.2\,FPS): along rays, samples lie off the grid, so the basis is evaluated per point, and the shared renderer keeps every model between 33.7 and 49.6\,FPS.

\subsection{Ablations} \label{sec:ablations}

Five ablations on Kodak (\cref{sec:ablation_details}) motivate FUTON's defaults. The relative bandwidth trades truncation against rank at a fixed size (\cref{fig:error_bound}b): $\kappa = 1/8$ truncates the images and $\kappa = 2$ leaves too small a rank, 4.2 and 5.4\,dB below $\kappa = 1/2$, while $\kappa = 1/4$ and $1$ stay within 1.1\,dB; training time falls as $\kappa$ grows, with the rank.

The MLP head adds 7.4\,dB over a linear head, and wider heads at most 0.5\,dB, at 2--3 times the training time; at about equal size, CP leads a tensor ring and Tucker by 0.4 and 2.5\,dB, though Tucker trains in half the time, and cosines lead five other bases by 0.2 to 1.6\,dB. PSNR grows by 3.4 to 5.0\,dB per doubling of $R$, with no plateau up to $R = 1{,}120$, while training time grows from 2.2 to 10.9\,s: a smooth trade of size for quality.

\section{Conclusion} \label{sec:conclusion}

We presented FUTON, a structured INR that unifies generalized Fourier series with a low-rank CP decomposition of the coefficient tensor{, read out by a linear or MLP head}. This design addresses common limitations of MLP-based INRs {(slow convergence, an opaque size--accuracy relation): its cost is linear in each axis's resolution and in the rank, and with a linear head its best error splits exactly into terms bounded before training, bounds that also size it for a target quality}. Against twelve INRs of matched size on three tasks, FUTON delivered the best average quality, reached every quality target first, inferred fastest, and trained in the least memory. These findings position FUTON as a practical alternative to MLPs for signal representation.

\paragraph{Limitations.} FUTON's separable basis favors axis-aligned structure: an oblique edge needs a rank that grows with its sharpness (\cref{tab:edge_test}). Signals whose coefficients lack low-rank structure, such as noise, need a large rank, which \cref{prop:error} reveals before training. The exact error analysis covers the linear head on fully sampled signals: only its upper bound transfers to the MLP head, radiance fields lie outside it, and on volumes its bracket is wide (\cref{sec:limitations}).

\clearpage
\begin{ack}
This work is funded by the Flanders AI Research Program, under grant no. 174B09119. The computational resources (Stevin Supercomputer Infrastructure) and services used in this work were provided by the VSC (Flemish Supercomputer Center), funded by Ghent University, FWO, and the Flemish Government -- department EWI. The authors declare no competing interests. Pourya Behmandpoor and Nikos Deligiannis acknowledge the support of the Innoviris WatchBOT Project (Project 2025-JRDIC-3b) and the Francqui Research Professorship (N. Deligiannis). They also acknowledge the support of the ERC Consolidator Grant IONIAN (No. 101171240, DOI: 10.3030/101171240). Funded by the European Union. Views and opinions expressed are however those of the author(s) only and do not necessarily reflect those of the European Union or the European Research Council Executive Agency. Neither the European Union nor the granting authority can be held responsible for them.
\end{ack}

\bibliographystyle{unsrt}
\bibliography{ref}

\appendix
\clearpage
\makeatletter\setlength{\@fptop}{0pt}\setlength{\@fpsep}{12pt}\makeatother
\section*{Supplementary Material}

This supplementary material is organized as follows. \Cref{sec:l2_space,sec:tensor_networks} provide background on $L^2$ spaces and tensor networks, respectively, which underpin the theoretical framework of FUTON. \Cref{sec:proof_separable_basis_orthonormality,sec:proof_gfs_convergence,sec:proof_futon_expressiveness} contain complete proofs of the theorems stated in the main paper. \Cref{sec:proof_error} proves the error analysis of \cref{sec:error_analysis} for any basis and on sampling grids, draws its consequences (a truncation rate, a low-rank sufficiency theorem, and an edge test), gives its algorithm, and reports its experiments. \Cref{sec:detailed_experimental_setup} provides detailed experimental configurations for reproducibility{, \cref{sec:full_results} the full results of every task, and \cref{sec:ablation_details} the ablation studies}. Finally, \cref{sec:limitations} discusses limitations.

\FloatBarrier
\section{\texorpdfstring{$L^2$}{L2} Space} \label{sec:l2_space}

FUTON's theoretical foundation relies on generalized Fourier series in $L^2$ spaces (\cref{sec:gfs}). We provide formal definitions of the $L^2$ space and related concepts.

\begin{definition}[$L^2$ Space]\label{def:l2_space}
    Let $\X \subseteq \R^C$ for some $C\in\N$. The Hilbert space $L^2(\X)$ consists of \textbf{square-integrable} functions on $\X$, where $f \in L^2(\X)$ if and only if
    \begin{equation} \label{eq:square-integrable}
        \int_{\X} \lvert f(\vct{x}) \rvert^2 \, d\vct{x} < \infty.
    \end{equation}
    For instance, $L^2([0,1]^C)$ denotes the space of square-integrable functions on the $C$-dimensional unit hypercube.
\end{definition}

\begin{definition}[$L^2$ Inner Product]\label{def:l2_inner_product}
    For $f, g \in L^2(\X)$, the inner product is defined as
    \begin{equation} \label{eq:l2_inner_product}
        \inner[L^2]{f}{g} \triangleq \int_{\X} f(\vct{x}) g(\vct{x}) \, d\vct{x}.
    \end{equation}
\end{definition}

\begin{definition}[$L^2$ Norm]\label{def:l2_norm}
    For $f \in L^2(\X)$, the norm is defined as
    \begin{equation} \label{eq:l2_norm}
        \norm[L^2]{f} \triangleq \sqrt{\inner[L^2]{f}{f}}.
    \end{equation}
\end{definition}

\begin{definition}[$L^2$ Convergence]\label{def:l2_convergence}
    A sequence $\seq{f_k}[k\in\Nz]$ in $L^2(\X)$ is said to converge in $L^2(\X)$ to a function $f \in L^2(\X)$ if $\lim_{k\to\infty} \norm[L^2]{f_k - f} = 0$, denoted as $f_k \xrightarrow{L^2} f$.
\end{definition}

\begin{definition}[$L^2$ Orthonormality]\label{def:l2_orthonormality}
    A set of functions $\set{\varphi_k}[k\in\I] \subseteq L^2(\X)$ is \textbf{orthonormal} in $L^2(\X)$ if $\inner[L^2]{\varphi_k}{\varphi_\ell} = \delta_{k\ell}$ for all $k, \ell \in \I$, where $\delta_{k\ell}$ is the Kronecker delta ($1$ if $k=\ell$, $0$ otherwise).
\end{definition}

\begin{definition}[$L^2$ Density]\label{def:l2_density}
    A subset $\mathbb{D} \subseteq L^2(\X)$ is said to be \textbf{dense} in $L^2(\X)$ if for every function $f \in L^2(\X)$ there exists a sequence $\seq{f_k}[k\in\Nz]$ in $\mathbb{D}$ such that $f_k \xrightarrow{L^2} f$. In other words, any function in $L^2(\X)$ can be approximated arbitrarily closely by functions from $\mathbb{D}$.
\end{definition}

\begin{definition}[$L^2$ Completeness] \label{def:l2_completeness}
    A set of functions is said to be \textbf{complete} in $L^2(\X)$ if its \textbf{span} is \textbf{dense} in $L^2(\X)$. Formally, $\set{\varphi_k}[k\in\I] \subseteq L^2(\X)$ is complete if for every function $f \in L^2(\X)$ and every $\varepsilon > 0$, some finite linear combination $g$ of the $\varphi_k$ satisfies $\norm[L^2]{f - g} < \varepsilon$. In other words, any function in $L^2(\X)$ can be approximated arbitrarily closely by a finite linear combination of functions from $\set{\varphi_k}[k\in\I]$.
\end{definition}

\begin{definition}[$L^2$ Orthonormal Basis] \label{def:l2_orthonormal_basis}
    A set of functions $\set{\varphi_k}[k\in\Nz] \subseteq L^2(\X)$ is said to be an \textbf{orthonormal basis (ONB)} for $L^2(\X)$ if it is both \textbf{orthonormal} and \textbf{complete} in $L^2(\X)$.
\end{definition}

\FloatBarrier
\section{Tensor Networks} \label{sec:tensor_networks}

\subsection{Tensor Diagram Notation} \label{sec:tensor_diagram_notation}

A tensor is a multi-dimensional array of numbers. A $T$th-order tensor $\tns{A} \in \R^{I_1 \times I_2 \times \cdots \times I_T}$ is a multi-dimensional array where the $m$th \emph{mode} (dimension or index) has size $I_m$. Special cases include scalars ($T=0$), vectors ($T=1$), and matrices ($T=2$).

\paragraph{Index Contraction.}
An \emph{index contraction} sums over all values of shared indices between tensors. For example, matrix multiplication
\begin{equation} \label{eq:matrix_product}
    c_{ij} = \sum_{k} a_{ik} b_{kj}
\end{equation}
contracts the shared index $k$. More complex contractions involve multiple tensors and indices:
\begin{equation} \label{eq:complex_contraction}
    e_{ijk} = \sum_{\ell, m, n, p, q=1}^{I} a_{i \ell m} b_{j \ell n p} c_{k p q} d_{m n q}.
\end{equation}
Indices that remain after contraction ($i$, $j$, $k$) are called \emph{open indices}. A \emph{tensor network} is a set of tensors with indices contracted according to a specified pattern.

\begin{figure}[t]
    \centering
    \begin{subfigure}{.2\linewidth}
        \centering
        \scalebox{.9}{\begin{tikzpicture}[scale=1, transform shape]
\tikzstyle{nodeStyle} = [circle, shade, draw=red!50!black, inner color=red!30, outer color=red!50, line width=1pt, minimum size=10pt]

\node[nodeStyle, label=$a$] (v) at (0, 0) {};

\end{tikzpicture}}
        \caption{scalar}
        \label{fig:scalar}
    \end{subfigure}%
    \hfill
    \begin{subfigure}{.22\linewidth}
        \centering
        \scalebox{.9}{\begin{tikzpicture}[scale=1, transform shape]
\tikzstyle{nodeStyle} = [circle, shade, draw=red!50!black, inner color=red!30, outer color=red!50, line width=1pt, minimum size=10pt]
\tikzstyle{edgeStyle} = [line width = 1pt]

\node[nodeStyle, label=$\bm{a}$] (v) at (0, 0) {};

\node (i) at (1, 0) {$ i $};

\draw[edgeStyle] (v) -- (i);

\end{tikzpicture}}
        \caption{vector}
        \label{fig:vector}
    \end{subfigure}%
    \hfill
    \begin{subfigure}{.27\linewidth}
        \centering
        \scalebox{.9}{\begin{tikzpicture}[scale=1, transform shape]
\tikzstyle{nodeStyle} = [circle, shade, draw=red!50!black, inner color=red!30, outer color=red!50, line width=1pt, minimum size=10pt]
\tikzstyle{edgeStyle} = [line width = 1pt]

\node[nodeStyle, label=$\mtx{A}$] (A) at (0, 0) {};
\node (i) at (-1, 0) {$ i $};
\node (j) at (1, 0) {$ j $};

\draw[edgeStyle] (A) -- (i);
\draw[edgeStyle] (A) -- (j);

\end{tikzpicture}}
        \caption{matrix}
        \label{fig:matrix}
    \end{subfigure}%
    \hfill
    \begin{subfigure}{.27\linewidth}
        \centering
        \scalebox{.9}{\begin{tikzpicture}[scale=1, transform shape]
    \tikzstyle{nodeStyle} = [circle, shade, draw=red!50!black, inner color=red!30, outer color=red!50, line width=1pt, minimum size=10pt]
    \tikzstyle{edgeStyle} = [line width = 1pt]

    \node[nodeStyle, label=north east:$\tns{A}$] (A) at (0, 0) {};

    \node (i) at (-1, 0) {$ i $};
    \node (j) at (1, 0) {$ j $};
    \node (k) at (0, 1) {$ k $};

    \draw[edgeStyle] (A) -- (i);
    \draw[edgeStyle] (A) -- (j);
    \draw[edgeStyle] (A) -- (k);

\end{tikzpicture}}
        \caption{third-order tensor}
        \label{fig:3rd_order_tensor}
    \end{subfigure}

    \medskip

    \begin{subfigure}{.49\linewidth}
        \centering
        \scalebox{1}{\begin{tikzpicture}[scale=1, transform shape]
    \tikzstyle{nodeStyle} = [circle, shade, draw=red!50!black, inner color=red!30, outer color=red!50, line width=1pt, minimum size=10pt]
    \tikzstyle{edgeStyle} = [line width = 1pt]
    \tikzstyle{backStyle} = [shade, inner color= orange!10, outer color=orange!40, rounded corners]

    \begin{scope}
        \node[backStyle, diamond, minimum width = 3.8cm, minimum height = 3.8cm, rounded corners=4ex, label=above left:$\tns{E}$] (E) at (0,0) {};

        \node[nodeStyle, label={[label distance=-3pt]below left:$\tns{A}$}] (A) at (-1,0) {};
        \node[nodeStyle, label={[label distance=-3pt]above right:$\tns{B}$}] (B) at (0,1) {};
        \node[nodeStyle, label={[label distance=-3pt]below right:$\tns{C}$}] (C) at (1,0) {};
        \node[nodeStyle, label={[label distance=-3pt]below right:$\tns{D}$}] (D) at (0,-1) {};

        \node (i) at (-2, 0) {$i$};
        \node (j) at (0, 2) {$j$};
        \node (k) at (2, 0) {$k$};

        \draw[edgeStyle] (A) -- (i);
        \draw[edgeStyle] (B) -- (j);
        \draw[edgeStyle] (C) -- (k);
        \path[edgeStyle] (A) edge[above left]  node {$\ell$} (B);
        \path[edgeStyle] (A) edge[below left]  node {$m$} (D);
        \path[edgeStyle] (B) edge[left]  node {$n$} (D);
        \path[edgeStyle] (B) edge[above right]  node {$p$} (C);
        \path[edgeStyle] (C) edge[below right]  node {$q$} (D);

    \end{scope}

\end{tikzpicture}}
        \caption{$e_{ijk} = \sum_{\ell, m, n, p, q=1}^{I} a_{i \ell m} b_{j \ell n p} c_{k p q} d_{m n q}$}
        \label{fig:complex}
    \end{subfigure}%
    \hfill
    \begin{subfigure}{.49\linewidth}
        \centering
        \scalebox{1}{\begin{tikzpicture}[scale=1, transform shape]
\tikzstyle{nodeStyle} = [circle, shade, draw=red!50!black, inner color=red!30, outer color=red!50, line width=1pt, minimum size=10pt]
\tikzstyle{edgeStyle} = [line width = 1pt]
\tikzstyle{backStyle} = [shade, inner color= orange!10, outer color=orange!40, rounded corners]

\begin{scope}
\node[backStyle, circle, minimum width = 3.6cm, minimum height = 3.6cm, rounded corners=5ex] (g) at (0,0) {};

\draw[edgeStyle] (0,0) circle (1.00);

\node[nodeStyle, label={[label distance=-3pt]left:$\mtx{A}$}] (A) at (180:1.00) {};
\node[nodeStyle, label={[label distance=-3pt]above left:$\mtx{B}$}] (B) at (120:1.00) {};
\node[nodeStyle, label={[label distance=-3pt]above right:$\mtx{C}$}] (C) at (60:1.00) {};
\node[nodeStyle, label={[label distance=-3pt]right:$\mtx{D}$}] (D) at (0:1.00) {};
\node[nodeStyle, label={[label distance=-3pt]below right:$\mtx{E}$}] (E) at (-60:1.00) {};
\node[nodeStyle, label={[label distance=-3pt]below left:$\mtx{F}$}] (F) at (-120:1.00) {};

\end{scope}

\end{tikzpicture}}
        \caption{$\operatorname{tr}(\mtx{A}\mtx{B}\mtx{C}\mtx{D}\mtx{E}\mtx{F})$}
        \label{fig:ring}
    \end{subfigure}
    \caption{\textbf{Tensor diagram notation.} Graphical representation of tensors with different orders: (a)~scalar (0th-order), (b)~vector (1st-order), (c)~matrix (2nd-order), and (d)~third-order tensor. In tensor diagrams, tensors are represented as nodes (shapes) with edges (lines) denoting indices. Dangling edges represent open (uncontracted) indices. (e)~Complex tensor network from~\eqref{eq:complex_contraction} with multiple contracted indices. Edges connecting different nodes represent shared indices that are summed over. (f)~Trace of matrix product $\operatorname{tr}(\mtx{A}\mtx{B}\mtx{C}\mtx{D}\mtx{E}\mtx{F})$ represented as a ring diagram, where the cyclic property becomes visually apparent, \ie, any rotation of the ring yields the same value.}
    \label{fig:basic_tensor_diagrams}\label{fig:complex_diagram_examples}
\end{figure}

\paragraph{Tensor Diagrams.}
\emph{Tensor diagrams} provide powerful graphical notation for tensor operations. As shown in \cref{fig:basic_tensor_diagrams}, tensors are represented as nodes with edges corresponding to indices. Edges connecting tensors represent contracted indices, while dangling edges represent open indices. This notation simplifies understanding of complex tensor operations and is used {for FUTON in \cref{fig:futon_tensor_network}}.

\Cref{fig:complex} illustrates tensor diagrams for the complex contraction in~\eqref{eq:complex_contraction}. As another example of complex calculations, the trace of the product of six matrices can be represented by a simple ring-like diagram as seen in \cref{fig:ring}. From this tensor diagram, the cyclic property of the trace becomes clear. This is a simple example of why tensor diagrams are advantageous. Tensor diagrams make structural properties visually apparent and are particularly useful for understanding operations like those in FUTON (see {\cref{fig:futon_tensor_network}}).

\begin{figure}[t]
    \centering
    \begin{subfigure}{.33\linewidth}
        \centering
        \scalebox{.9}{\begin{tikzpicture}[scale=1, transform shape]
\tikzstyle{nodeStyle} = [circle, shade, draw=red!50!black, inner color=red!30, outer color=red!50, line width=1pt, minimum size=10pt]
\tikzstyle{edgeStyle} = [line width = 1pt]

\node[nodeStyle] (A) at (0, 0) {};

\draw[edgeStyle] (A) edge[out=150,in=30,looseness=9] (A);

\end{tikzpicture}}
        \caption{$\text{tr}(\mtx{A})$}
        \label{fig:matrix_trace}
    \end{subfigure}%
    \hfill
    \begin{subfigure}{.33\linewidth}
        \centering
        \scalebox{.9}{\begin{tikzpicture}[scale=1, transform shape]
\tikzstyle{nodeStyle} = [circle, shade, draw=red!50!black, inner color=red!30, outer color=red!50, line width=1pt, minimum size=10pt]
\tikzstyle{edgeStyle} = [line width = 1pt]

\node[nodeStyle, label=left:$\vct{b}$] (b) at (0, 0) {};
\node[nodeStyle, label=left:$\vct{a}$] (a) at (0, 1) {};

\node[circle, fill = black, inner sep=0pt, minimum size=3pt] (jj) at (.5, .5) {};

\node (j) at (1, .5) {};

\draw[edgeStyle] (b) edge[in=-90, out=0] (jj);
\draw[edgeStyle] (a) edge[in=90, out=0] (jj);

\draw[edgeStyle] (jj) -- (j);

\end{tikzpicture}}
        \caption{$\vct{a} \odot \vct{b}$}
        \label{fig:hadamard_product}
    \end{subfigure}%
    \hfill
    \begin{subfigure}{.33\linewidth}
        \centering
        \scalebox{.9}{\definecolor{top}{RGB}{248,206,204}
\definecolor{bottom}{RGB}{234,107,102}
\definecolor{border}{RGB}{184,84,80}

\begin{tikzpicture}[scale=1, transform shape]
    \tikzstyle{nodeStyle} = [circle, shade, draw = border, top color= top, bottom color=bottom, line width = 1pt, minimum size=10pt]
    \tikzstyle{edgeStyle} = [line width = 1pt]

    \node[nodeStyle, label=below:$\tns{A}$] (A) at (0, 0) {};
    \node[nodeStyle, label=below:$\mtx{U}$] (U) at (1, 0) {};

    \node (i) at (-.75, 0) {};
    \node (j) at (0, .75) {};
    \node (l) at (1.75, 0) {};

    \draw[edgeStyle] (A) -- (i);
    \draw[edgeStyle] (A) -- (j);
    \draw[edgeStyle] (A) -- (U);
    \draw[edgeStyle] (U) -- (l);

\end{tikzpicture}}
        \caption{$\tns{A} \times_1 \mtx{U}$}
        \label{fig:mode-n_product}
    \end{subfigure}%

    \begin{subfigure}{.33\linewidth}
        \centering
        \scalebox{.9}{\begin{tikzpicture}[scale=1, transform shape]
\tikzstyle{nodeStyle} = [circle, shade, draw=red!50!black, inner color=red!30, outer color=red!50, line width=1pt, minimum size=10pt]
\tikzstyle{edgeStyle} = [line width = 1pt]

\node[nodeStyle, label=left:$\vct{a}$] (a) at (0, .75) {};
\node[nodeStyle, label=left:$\vct{b}$] (b) at (0, 0) {};

\node (i) at (.75, .75) {};
\node (j) at (.75, 0) {};

\draw[edgeStyle] (a) -- (i);
\draw[edgeStyle] (b) -- (j);

\end{tikzpicture}}
        \caption{$\vct{a} \otimes \vct{b}$}
        \label{fig:tensor_product}
    \end{subfigure}%
    \hfill
    \begin{subfigure}{.33\linewidth}
        \centering
        \scalebox{.9}{\begin{tikzpicture}[scale=1, transform shape]
\tikzstyle{nodeStyle} = [circle, shade, draw=red!50!black, inner color=red!30, outer color=red!50, line width=1pt, minimum size=10pt]
\tikzstyle{edgeStyle} = [line width = 1pt]

\node[nodeStyle, label=above:$\vct{v}$] (v) at (0, 0) {};
\node[nodeStyle, label=above:$\mtx{A}$] (A) at (1, 0) {};

\node (j) at (1.75, 0) {};

\draw[edgeStyle] (A) -- (j);
\draw[edgeStyle] (v) -- (A);

\end{tikzpicture}}
        \caption{$\mtx{A}\vct{v}$}
        \label{fig:matrix_product}
    \end{subfigure}%
    \hfill
    \begin{subfigure}{.33\linewidth}
        \centering
        \scalebox{.9}{\begin{tikzpicture}[scale=1, transform shape]
    \tikzstyle{nodeStyle} = [circle, shade, draw=red!50!black, inner color=red!30, outer color=red!50, line width=1pt, minimum size=10pt]
    \tikzstyle{edgeStyle} = [line width = 1pt]

    \node[nodeStyle, label=left:$\mtx{A}$] (A) at (0, 0) {};
    \node[nodeStyle, label=left:$\tns{B}$] (B) at (1.25, 0) {};

    \node (k) at (2, 0) {};

    \draw[edgeStyle] (A) edge[in=120, out=60] (B);
    \draw[edgeStyle] (A) edge[in=-120, out=-60] (B);
    \draw[edgeStyle] (B) -- (k);

\end{tikzpicture}}
        \caption{$\mtx{A} \dotprod \tns{B}$}
        \label{fig:generalized_dot_product}
    \end{subfigure}

    \caption{\textbf{Tensor diagrams of basic operations.} (a)~matrix trace; (b)~Hadamard product, used in~\eqref{eq:futon_cp_g}; (c)~mode-$1$ product; (d)~tensor product, used in~\eqref{eq:tensor_product_basis} and~\eqref{eq:cp}; (e)~matrix-vector multiplication, used in~\eqref{eq:futon_cp_h} and~\eqref{eq:futon_cp}; and (f)~generalized dot product, used in~\eqref{eq:gfs_tensordot}.}
    \label{fig:basic_tensor_operations}
\end{figure}

Many matrix and tensor operations perform some form of tensor network contraction, and hence, can be represented graphically using tensor diagrams. \Cref{fig:basic_tensor_operations} illustrates the tensor diagrams of some basic operations.

\subsection{Basic Tensor Operations} \label{sec:tensor_operations}

\begin{definition}[Fiber] \label{def:fiber}
    A mode-$m$ fiber of a tensor $\tns{A} \in \R^{I_1 \times I_2 \times \cdots \times I_T}$ is a vector obtained by fixing every index except the $m$th, i.e., $\tns{A}_{i_1 \cdots i_{m-1} :i_{m+1} \cdots i_T}$ for fixed indices $i_j \in \{1, \ldots, I_j\}$ where $j \neq m$.
\end{definition}

\begin{definition}[Matricization] \label{def:matricization}
    The mode-$m$ matricization (or unfolding) of a tensor $\tns{A} \in \R^{I_1 \times I_2 \times \cdots \times I_T}$ is defined as the matrix $\mtx{A}_{(m)} \in \R^{I_m \times (I_1 \cdots I_{m-1} I_{m+1} \cdots I_T)}$ obtained by stacking all mode-$m$ fibers of $\tns{A}$ along the columns, arranged in lexicographical order.
\end{definition}

\begin{definition}[Hadamard Product] \label{def:hadamard_product}
    The Hadamard product $\tns{C} = \tns{A} \hadamard \tns{B}$ of equally sized tensors $\tns{A}, \tns{B} \in \R^{I_1 \times \cdots \times I_T}$ performs elementwise multiplication:
    \begin{equation} \label{eq:hadamard_product}
        c_{i_1 \cdots i_T} = a_{i_1 \cdots i_T} b_{i_1 \cdots i_T}.
    \end{equation}
\end{definition}

\Cref{fig:hadamard_product} illustrates the tensor diagram for the Hadamard product of two vectors.

\begin{definition}[Generalized Dot Product]\label{def:generalized_dot_product}
    For tensors $\tns{A} \in \R^{I_1 \times \cdots \times I_T}$ and $\tns{B} \in \R^{I_1 \times \cdots \times I_T \times J_1 \times \cdots \times J_P}$, the \emph{generalized dot product} $\tns{C} = \tns{A} \dotprod \tns{B} \in \R^{J_1 \times \cdots \times J_P}$ contracts over all $T$ dimensions of $\tns{A}$:
    \begin{equation}
        c_{j_1,\ldots,j_P} = \sum_{i_1=1}^{I_1} \cdots \sum_{i_T=1}^{I_T} a_{i_1,\ldots,i_T} \, b_{i_1,\ldots,i_T,j_1,\ldots,j_P}.
    \end{equation}
\end{definition}

\begin{definition}[Inner Product] \label{def:inner_product}
    The inner product (or dot product) of equally sized tensors $\tns{A}, \tns{B} \in \R^{I_1 \times \cdots \times I_T}$ is
    \begin{equation} \label{eq:inner_product}
        \inner{\tns{A}}{\tns{B}} \triangleq \sum_{i_1, \ldots, i_T} a_{i_1 \cdots i_T} b_{i_1 \cdots i_T}.
    \end{equation}
\end{definition}

The inner product is a special case of the generalized dot product (\cref{def:generalized_dot_product}) where both tensors have identical shapes.

\begin{definition}[Tensor Product] \label{def:tensor_product}
    The tensor product $\tns{C} = \tns{A} \tenprod \tns{B}$ of tensors $\tns{A} \in \R^{I_1 \times \cdots \times I_T}$ and $\tns{B} \in \R^{J_1 \times \cdots \times J_P}$ yields $\tns{C} \in \R^{I_1 \times \cdots \times I_T \times J_1 \times \cdots \times J_P}$ with elements
    \begin{equation} \label{eq:tensor_product}
        c_{i_1 \cdots i_T j_1 \cdots j_P} = a_{i_1 \cdots i_T} b_{j_1 \cdots j_P}.
    \end{equation}
\end{definition}

\Cref{fig:tensor_product} illustrates the tensor diagram for the tensor product of two vectors.

\begin{definition}[Tensor-Matrix Product] \label{def:mode-n_product}
    The mode-$m$ product of a tensor $\tns{A} \in \R^{I_1 \times I_2 \times \cdots \times I_T}$ and a matrix $\mtx{U} \in \R^{J \times I_m}$ is the tensor
    \begin{equation} \label{eq:mode-n_product}
        \tns{C} = \tns{A} \times_m \mtx{U} \in \R^{I_1 \times \cdots \times I_{m-1} \times J \times I_{m+1} \times \cdots \times I_T}
    \end{equation}
    with elements
    \begin{equation} \label{eq:mode-n_product_elements}
        c_{i_1 \cdots i_{m-1} j i_{m+1} \cdots i_T} = \sum_{i_m=1}^{I_m} a_{i_1 \cdots i_T} U_{j i_m}.
    \end{equation}
\end{definition}

\Cref{fig:mode-n_product} illustrates the tensor diagram of the mode-1 product of a third-order tensor and a matrix.

An equivalent matrix representation of the mode-$m$ product is $\mtx{C}_{(m)} = \mtx{U} \mtx{A}_{(m)}$, which allows us to employ established fast matrix multiplications when dealing with very large-scale tensors.

\subsection{Contraction Order} \label{sec:contraction_order}

The computational cost of tensor network contractions depends critically on the order in which indices are contracted. \Cref{fig:contraction_order} illustrates this: contracting $\mtx{A}$ and $\tns{C}$ first (\cref{fig:contraction_order_1}) requires $\mathcal{O}(I^4)$ operations, while contracting $\mtx{A}$ and $\mtx{B}$ first (\cref{fig:contraction_order_2}) requires only $\mathcal{O}(I^3)$ operations. This principle is crucial for FUTON's linear complexity (see \cref{sec:futon_complexity}). Its panels (c)--(f) apply it to FUTON with a linear head.

\begin{figure}[t]
    \centering
    \begin{subfigure}{\linewidth}
        \centering
        \scalebox{.9}{\begin{tikzpicture}[scale=1, transform shape]
    \tikzstyle{nodeStyle} = [circle, shade, draw=red!50!black, inner color=red!30, outer color=red!50, line width=1pt, minimum size=10pt]
    \tikzstyle{edgeStyle} = [line width = 1pt, font=\footnotesize]
    \tikzstyle{arrowStyle} = [single arrow, draw = gray, fill = gray!30, minimum width=10, minimum height=25, inner sep=.5, single arrow head extend=1]
    \tikzstyle{backStyle} = [shade, inner color= orange!10, outer color=orange!40, rounded corners]

    \begin{scope}
        \path[backStyle, rounded corners=2ex] (-.4,-.5) rectangle (1.4,.5);

        \node[nodeStyle, label=below:$\mtx{A}$] (A) at (0, 0) {};
        \node[nodeStyle, label=above:$\mtx{B}$] (B) at (0, 1) {};
        \node[nodeStyle, label=below:$\tns{C}$] (C) at (1, 0) {};

        \node (i) at (1.75, 0) {};

        \path[edgeStyle] (A) edge[left] node {$r_1$} (B);
        \path[edgeStyle] (A) edge[below] node {$r_2$} (C);
        \path[edgeStyle] (B) edge[right] node {$r_3$} (C);
        \path[edgeStyle] (C) edge[above] node {$r_4$} (i);
    \end{scope}

    \begin{scope}[xshift = 2.75cm]

        \node[arrowStyle, label=above:\footnotesize $\mathcal{O}(I^4)$] at (0, .5) {};

    \end{scope}

    \begin{scope}[xshift = 4.25cm]

        \node[nodeStyle] (B) at (0, 1) {};

        \node (i) at (1.75, 0) {};

        \path[edgeStyle] (0,0) edge[below] node {} (B);
        \path[edgeStyle] (1,0) edge[right] node {} (B);
        \path[edgeStyle] (1,0) edge[above] node {} (i);

        \path[nodeStyle, rounded corners] (-.2,-.2) rectangle (1.2,.2);
    \end{scope}

    \begin{scope}[xshift = 7cm]

        \node[arrowStyle,label=above:\footnotesize $\mathcal{O}(I^3)$] at (0, .5) {};

    \end{scope}

    \begin{scope}[xshift = 8.25cm]

        \node[nodeStyle] (A) at (0, 0) {};
        \node (q) at (.75, 0) {};

        \path[edgeStyle] (A) edge[above]  node {} (q);

    \end{scope}

\end{tikzpicture}}
        \caption{}
        \label{fig:contraction_order_1}
    \end{subfigure}

    \begin{subfigure}{\linewidth}
        \centering
        \scalebox{.9}{\begin{tikzpicture}[scale=1, transform shape]
    \tikzstyle{nodeStyle} = [circle, shade, draw=red!50!black, inner color=red!30, outer color=red!50, line width=1pt, minimum size=10pt]
    \tikzstyle{edgeStyle} = [line width = 1pt, font=\footnotesize]
    \tikzstyle{arrowStyle} = [single arrow, draw = gray, fill = gray!30, minimum width=10, minimum height=25, inner sep=.5, single arrow head extend=1]
    \tikzstyle{backStyle} = [shade, inner color= orange!10, outer color=orange!40, rounded corners]

    \begin{scope}
        \path[backStyle, rounded corners=2ex] (-.5,-.75) rectangle (.4,1.75);

        \node[nodeStyle, label=below:$\mtx{A}$] (A) at (0, 0) {};
        \node[nodeStyle, label=above:$\mtx{B}$] (B) at (0, 1) {};
        \node[nodeStyle, label=below:$\tns{C}$] (C) at (1, 0) {};

        \node (i) at (1.75, 0) {};

        \path[edgeStyle] (A) edge[left] node {$r_1$} (B);
        \path[edgeStyle] (A) edge[below] node {$r_2$} (C);
        \path[edgeStyle] (B) edge[right] node {$r_3$} (C);
        \path[edgeStyle] (C) edge[above] node {$r_4$} (i);
    \end{scope}

    \begin{scope}[xshift = 2.75cm]

        \node[arrowStyle, label=above:\footnotesize $\mathcal{O}(I^3)$] at (0, .5) {};

    \end{scope}

    \begin{scope}[xshift = 4.25cm]

        \node[nodeStyle] (C) at (1, 0) {};

        \node (i) at (1.75, 0) {};

        \path[edgeStyle] (0,0) edge[below] node {} (C);
        \path[edgeStyle] (0,1) edge[right] node {} (C);
        \path[edgeStyle] (C) edge[above] node {} (i);

        \path[nodeStyle, rounded corners] (-.2,-.2) rectangle (.2,1.2);
    \end{scope}

    \begin{scope}[xshift = 7cm]

        \node[arrowStyle,label=above:\footnotesize $\mathcal{O}(I^3)$] at (0, .5) {};

    \end{scope}

    \begin{scope}[xshift = 8.25cm]

        \node[nodeStyle] (A) at (0, 0) {};
        \node (q) at (.75, 0) {};

        \path[edgeStyle] (A) edge[above]  node {} (q);

    \end{scope}

\end{tikzpicture}}
        \caption{}
        \label{fig:contraction_order_2}
    \end{subfigure}

    \medskip

    \captionsetup[subfigure]{skip=5pt}
    \begin{minipage}[b]{.4\linewidth}
        \begin{subfigure}[b]{\linewidth}
            \centering
            \scalebox{.67}{\begin{tikzpicture}[scale=1, transform shape]
    \tikzstyle{nodeStyle1} = [circle, shade, draw=blue!50!black, inner color=blue!30, outer color=blue!50, line width=1pt, minimum size=10pt]
    \tikzstyle{nodeStyle2} = [circle, shade, draw=red!50!black, inner color=red!30, outer color=red!50, line width=1pt, minimum size=10pt]
    \tikzstyle{intersectionStyle} = [circle, fill=black, inner sep=0pt, minimum size=3pt]
    \tikzstyle{edgeStyle} = [line width=1pt, font=\footnotesize]
    \tikzstyle{inputBackStyle} = [shade, inner color=blue!3, outer color=blue!15, rounded corners]
    \tikzstyle{weightBackStyle} = [shade, inner color=red!3, outer color=red!15, rounded corners]

    \path[inputBackStyle] (-0.7,-4) rectangle (0.7, 2);
    \path[weightBackStyle] (1.3,-4) rectangle (4.7, 2);
    \node (Phi) at (0, 1.5) {\large$\color{blue!50!black} \bm{\Phi}(\vct{x})$};
    \node (W) at (3, 1.5) {\large$\color{red!50!black} \tns{W}$};
    \path[fill=black] (0, -1.8) circle (1pt) (0, -2.1) circle (1pt) (0, -2.4) circle (1pt);
    \path[fill=black] (2, -1.8) circle (1pt) (2, -2.1) circle (1pt) (2, -2.4) circle (1pt);

    \node[nodeStyle1, label=above:$\vct{\varphi}(x_1)$] (phi1) at (0, 0) {};
    \node[nodeStyle1, label=above:$\vct{\varphi}(x_2)$] (phi2) at (0, -1.3) {};
    \node[nodeStyle1, label=above:$\vct{\varphi}(x_C)$] (phiC) at (0, -3.5) {};

    \node[nodeStyle2, label=above:$\mtx{U}^{(1)}$] (U1) at (2, 0) {};
    \node[nodeStyle2, label=above:$\mtx{U}^{(2)}$] (U2) at (2, -1.3) {};
    \node[nodeStyle2, label=above:$\mtx{U}^{(C)}$] (UC) at (2, -3.5) {};
    \node[nodeStyle2, label=above:$\mtx{V}$] (V) at (4, -1.75) {};

    \node[intersectionStyle, label={[label distance=-2pt]45:$r$}] (r) at (3, -1.75) {};
    \node[intersectionStyle, label={[label distance=-2pt]45:$d$}, minimum size=0pt] (d) at (5, -1.75) {};

    \path[edgeStyle] (phi1) edge[below] node {$k_1$} (U1);
    \path[edgeStyle] (phi2) edge[below] node {$k_2$} (U2);
    \path[edgeStyle] (phiC) edge[below] node {$k_C$} (UC);
    \draw[edgeStyle] (U1) to[out=0, in=90] (r);
    \draw[edgeStyle] (U2) to[out=0, in=135] (r);
    \draw[edgeStyle] (UC) to[out=0, in=-90] (r);
    \draw[edgeStyle] (r) -- (V);
    \draw[edgeStyle] (V) -- (d);

\end{tikzpicture}}
            \caption{FUTON's forward pass}
            \label{fig:futon_cp}
        \end{subfigure}
    \end{minipage}%
    \hfill
    \begin{minipage}[b]{.33\linewidth}
        \begin{subfigure}[b]{\linewidth}
            \centering
            \scalebox{.76}{\begin{tikzpicture}[scale=1, transform shape]
    \tikzstyle{nodeStyle1} = [circle, shade, draw=blue!50!black, inner color=blue!20, outer color=blue!50, line width=1pt, minimum size=10pt]
    \tikzstyle{nodeStyle2} = [circle, shade, draw=red!50!black, inner color=red!20, outer color=red!50, line width=1pt, minimum size=10pt]
    \tikzstyle{nodeStyle3} = [circle, shade, draw = violet!50!black, inner color=black!20, outer color=violet!50, line width = 1pt, minimum size=10pt]
    \tikzstyle{nodeStyle4} = [circle, shade, draw = teal!50!black, inner color=teal!20, outer color=teal!50, line width = 1pt, minimum size=10pt]
    \tikzstyle{nodeStyle5} = [circle, shade, draw = darkgray!50!black, inner color=darkgray!20, outer color=darkgray!50, line width = 1pt, minimum size=10pt]
    \tikzstyle{intersectionStyle} = [circle, fill=black, inner sep=0pt, minimum size=3pt]
    \tikzstyle{edgeStyle} = [line width=1pt, font=\footnotesize]
    \tikzstyle{arrowStyle} = [single arrow, draw = gray, fill = gray!30, minimum width=10, minimum height=30, inner sep=.5, single arrow head extend=1]
    \tikzstyle{backStyle} = [shade, inner color= orange!10, outer color=orange!40, rounded corners]

    \path[backStyle] (-0.7,-0.4) rectangle (2, 0.7);
    \path[backStyle] (-0.7,-1.7) rectangle (2, -0.6);
    \path[backStyle] (-0.7,-4) rectangle (2, -2.8);
    \path[fill=black] (.75, -2.1) circle (1pt) (.75, -2.4) circle (1pt) (.75, -2.7) circle (1pt);

    \node[nodeStyle1, label=above:$\vct{\varphi}(x_1)$] (phi1) at (0, 0) {};
    \node[nodeStyle1, label=above:$\vct{\varphi}(x_2)$] (phi2) at (0, -1.3) {};
    \node[nodeStyle1, label=above:$\vct{\varphi}(x_C)$] (phiC) at (0, -3.5) {};

    \node[nodeStyle2, label=above:$\mtx{U}^{(1)}$] (U1) at (1.5, 0) {};
    \node[nodeStyle2, label=above:$\mtx{U}^{(2)}$] (U2) at (1.5, -1.3) {};
    \node[nodeStyle2, label=above:$\mtx{U}^{(C)}$] (UC) at (1.5, -3.5) {};
    \node[nodeStyle2, label=above:$\mtx{V}$] (V) at (3.5, -1.75) {};

    \node[intersectionStyle, label=-60:$r$] (r) at (2.5, -1.75) {};
    \node[intersectionStyle, label=below:$d$, minimum size=0pt] (d) at (4.25, -1.75) {};

    \path[edgeStyle] (phi1) edge[below] node {$k_1$} (U1);
    \path[edgeStyle] (phi2) edge[below] node {$k_2$} (U2);
    \path[edgeStyle] (phiC) edge[below] node {$k_C$} (UC);
    \draw[edgeStyle] (U1) to[out=0, in=90] (r);
    \draw[edgeStyle] (U2) to[out=0, in=135] (r);
    \draw[edgeStyle] (UC) to[out=0, in=-90] (r);
    \draw[edgeStyle] (r) -- (V);
    \draw[edgeStyle] (V) -- (d);

\end{tikzpicture}}
            \caption{Matrix-vector products}
            \label{fig:futon_basis_projection}
        \end{subfigure}
    \end{minipage}%
    \hfill
    \begin{minipage}[b]{.27\linewidth}
        \begin{subfigure}[b]{\linewidth}
            \centering
            \scalebox{.55}{\begin{tikzpicture}[scale=1, transform shape]
    \tikzstyle{nodeStyle1} = [circle, shade, draw=blue!50!black, inner color=blue!20, outer color=blue!50, line width=1pt, minimum size=10pt]
    \tikzstyle{nodeStyle2} = [circle, shade, draw=red!50!black, inner color=red!20, outer color=red!50, line width=1pt, minimum size=10pt]
    \tikzstyle{nodeStyle3} = [circle, shade, draw = violet!50!black, inner color=black!20, outer color=violet!50, line width = 1pt, minimum size=10pt]
    \tikzstyle{nodeStyle4} = [circle, shade, draw = teal!50!black, inner color=teal!20, outer color=teal!50, line width = 1pt, minimum size=10pt]
    \tikzstyle{nodeStyle5} = [circle, shade, draw = darkgray!50!black, inner color=darkgray!20, outer color=darkgray!50, line width = 1pt, minimum size=10pt]
    \tikzstyle{intersectionStyle} = [circle, fill=black, inner sep=0pt, minimum size=3pt]
    \tikzstyle{edgeStyle} = [line width=1pt, font=\footnotesize]
    \tikzstyle{arrowStyle} = [single arrow, draw = gray, fill = gray!30, minimum width=10, minimum height=30, inner sep=.5, single arrow head extend=1]
    \tikzstyle{backStyle} = [shade, inner color= orange!10, outer color=orange!40, rounded corners]

    \path[backStyle] (-0.7, 0.8) rectangle (1.5, -3.9);
    \path[fill=black] (0, -1.9) circle (1pt) (0, -2.2) circle (1pt) (0, -2.5) circle (1pt);

    \node[nodeStyle3, label=above:$\vct{h}_1(x_1)$] (h1) at (0, 0) {};
    \node[nodeStyle3, label=above:$\vct{h}_2(x_2)$] (h2) at (0, -1.3) {};
    \node[nodeStyle3, label=above:$\vct{h}_C(x_C)$] (h3) at (0, -3.5) {};
    \node[nodeStyle2, label=above:$\mtx{V}$] (V) at (2, -1.75) {};

    \node[intersectionStyle, label=-60:$r$] (r) at (1, -1.75) {};
    \node[intersectionStyle, label=below:$d$, minimum size=0pt] (d) at (2.75, -1.75) {};

    \draw[edgeStyle] (h1) to[out=0, in=90] (r);
    \draw[edgeStyle] (h2) to[out=0, in=135] (r);
    \draw[edgeStyle] (h3) to[out=0, in=-90] (r);
    \draw[edgeStyle] (r) -- (V);
    \draw[edgeStyle] (V) -- (d);

\end{tikzpicture}}
            \caption{Hadamard products}
            \label{fig:futon_hadamard_product}
        \end{subfigure}

        \medskip

        \begin{subfigure}[b]{\linewidth}
            \centering
            \scalebox{.55}{\begin{tikzpicture}[scale=1, transform shape]
    \tikzstyle{nodeStyle1} = [circle, shade, draw=blue!50!black, inner color=blue!20, outer color=blue!50, line width=1pt, minimum size=10pt]
    \tikzstyle{nodeStyle2} = [circle, shade, draw=red!50!black, inner color=red!20, outer color=red!50, line width=1pt, minimum size=10pt]
    \tikzstyle{nodeStyle3} = [circle, shade, draw = violet!50!black, inner color=black!20, outer color=violet!50, line width = 1pt, minimum size=10pt]
    \tikzstyle{nodeStyle4} = [circle, shade, draw = teal!50!black, inner color=teal!20, outer color=teal!50, line width = 1pt, minimum size=10pt]
    \tikzstyle{nodeStyle5} = [circle, shade, draw = darkgray!50!black, inner color=darkgray!20, outer color=darkgray!50, line width = 1pt, minimum size=10pt]
    \tikzstyle{intersectionStyle} = [circle, fill=black, inner sep=0pt, minimum size=3pt]
    \tikzstyle{edgeStyle} = [line width=1pt, font=\footnotesize]
    \tikzstyle{arrowStyle} = [single arrow, draw = gray, fill = gray!30, minimum width=10, minimum height=30, inner sep=.5, single arrow head extend=1]
    \tikzstyle{backStyle} = [shade, inner color= orange!10, outer color=orange!40, rounded corners]

    \path[backStyle] (-0.5, -2.3) rectangle (1.5, -1);

    \node[nodeStyle4, label=above:$\vct{g}(\vct{x})$] (g) at (0, -1.75) {};
    \node[nodeStyle2, label=above:$\mtx{V}$] (V) at (1, -1.75) {};
    \node[intersectionStyle, label=below:$d$, minimum size=0pt] (d) at (1.75, -1.75) {};

    \path[edgeStyle] (g) edge[below] node {$r$} (V);
    \draw[edgeStyle] (V) -- (d);

\end{tikzpicture}}
            \caption{Matrix-vector product}
            \label{fig:futon_feature_mapping}
        \end{subfigure}
    \end{minipage}
    \caption{\textbf{Contraction order, and FUTON's tensor network} {(linear head)}. (a,~b)~Sequential contraction of a tensor network whose modes are all of size $I$. In (a), the matrix $\mtx{A}$ and the tensor $\tns{C}$ are first contracted (mode-$m$ product), then the resulting tensor is contracted with the matrix $\mtx{B}$, requiring $\mathcal{O}(I^4)$ operations in total. In (b), the matrices $\mtx{A}$ and $\mtx{B}$ are first contracted (multiplied), then the resulting matrix is contracted with the tensor $\tns{C}$, requiring $\mathcal{O}(I^3)$ operations in total. (c)~Tensor diagram of FUTON's forward pass, where nodes represent tensors, connected edges represent contracted indices (summed over), and dangling edges correspond to open indices. See \cref{sec:tensor_diagram_notation} for tensor diagram notation. (d--f)~Optimal contraction order achieving linear complexity {$\mathcal{O}(R \sum_c K_c + CR + DR)$}. Orange boxes indicate tensors being contracted at each step: (d)~project each basis vector $\vct{\varphi}(x_c)$ onto its factor matrix $\mtx{U}^{(c)}$ via matrix-vector multiplication~\eqref{eq:futon_cp_h}, incurring $\mathcal{O}(R \sum_c K_c)$ operations; (e)~combine the projected features $\vct{h}_c(x_c)$ via Hadamard product~\eqref{eq:futon_cp_g}, incurring $\mathcal{O}(CR)$ operations; and (f)~map the resulting vector $\vct{g}(\vct{x})$ to the output via matrix-vector multiplication with $\mtx{V}$~\eqref{eq:futon_cp}, incurring $\mathcal{O}(DR)$ operations. This contrasts with the naive approach of directly computing $\bm{\Phi}(\vct{x}) \dotprod \tns{W}$, which requires $\mathcal{O}(D \prod_c K_c)$ operations and scales exponentially with the input dimension $C$.}
    \label{fig:futon_tensor_network}\label{fig:contraction_order}
\end{figure}

\subsection{Canonical Polyadic Decomposition} \label{sec:cp_decomposition}

\begin{definition}[Rank-1 Tensor] \label{def:rank-1_tensor}
    A rank-1 tensor $\tns{A} \in \R^{I_1 \times I_2 \times \cdots \times I_T}$ is a tensor that can be expressed as the tensor product of nonzero vectors $\{\vct{u}^{(m)} \in \R^{I_m}\}_{m=1}^T$:
    \begin{equation} \label{eq:rank-1_tensor}
        \tns{A} = \vct{u}^{(1)} \tenprod \vct{u}^{(2)} \tenprod \cdots \tenprod \vct{u}^{(T)}.
    \end{equation}
    Equivalently, the elements of $\tns{A}$ satisfy
    \begin{equation} \label{eq:rank-1_tensor_elements}
        a_{i_1 \cdots i_T} = u^{(1)}_{i_1} u^{(2)}_{i_2} \cdots u^{(T)}_{i_T}.
    \end{equation}
\end{definition}

\begin{definition}[Rank] \label{def:rank}
    The rank of a tensor $\tns{A}$, denoted by $\operatorname{rank}(\tns{A})$, is the minimal number of rank-1 tensors that sum to $\tns{A}$.
\end{definition}

\begin{definition}[CP Decomposition] \label{def:cp_decomposition}
    Canonical polyadic (CP) decomposition~\cite{carroll1970analysis, harshman1970foundations} represents a tensor $\tns{A} \in \R^{I_1 \times \cdots \times I_T}$ as a sum of $R$ rank-1 tensors:
    \begin{equation} \label{eq:cp_decomposition}
        \tns{A} = \sum_{r=1}^R \vct{u}_r^{(1)} \tenprod \vct{u}_r^{(2)} \tenprod \cdots \tenprod \vct{u}_r^{(T)},
    \end{equation}
    where $\vct{u}_r^{(m)} \in \R^{I_m}$ are factor vectors, and $R$ is the CP rank. Factor matrices $\mtx{U}^{(m)} = [\vct{u}_1^{(m)} \mid \cdots \mid \vct{u}_R^{(m)}] \in \R^{I_m \times R}$ collect the vectors for mode $m$.
\end{definition}

\begin{figure}[t]
    \centering
    \scalebox{.9}{\begin{tikzpicture}[scale=1, transform shape]
\tikzstyle{nodeStyle} = [circle, shade, draw=red!50!black, inner color=red!30, outer color=red!50, line width=1pt, minimum size=10pt]
\tikzstyle{edgeStyle} = [line width = 1pt, font=\footnotesize]

\begin{scope}

\node[nodeStyle] (X) at (0, 0) {};
\node[label={[label distance=-3pt]above:$i_1$}] (i1) at (0, 1) {};
\node[label={[label distance=-3pt]left:$i_2$}] (i2) at (-1, 0) {};
\node[label={[label distance=-3pt]below:$i_T$}] (iM) at (-45:1) {};

\node[circle, fill = black, inner sep=0pt, minimum size=2pt] at (-102.5:.75) {};
\node[circle, fill = black, inner sep=0pt, minimum size=2pt] at (-112.5:.75) {};
\node[circle, fill = black, inner sep=0pt, minimum size=2pt] at (-122.5:.75) {};

\draw[edgeStyle] (X) -- (i1);
\draw[edgeStyle] (X) -- (i2);
\draw[edgeStyle] (X) -- (iM);

\end{scope}

\begin{scope}[xshift = 1.5cm]
\node (eq) at (0, 0) {$\approx$};
\end{scope}

\begin{scope}[xshift = 5cm]

\node[nodeStyle, label={[label distance=-2pt]above right:$\mtx{U}^{(1)}$}] (U1) at (0, 1) {};
\node[nodeStyle, label={[label distance=-2pt]above:$\mtx{U}^{(2)}$}] (U2) at (-1, 0) {};
\node[nodeStyle, label={[label distance=0pt]right:$\mtx{U}^{(T)}$}] (UM) at (-45:1) {};

\node[label={[label distance=-3pt]above:$i_1$}] (i1) at (0, 2) {};
\node[label={[label distance=-3pt]left:$i_2$}] (i2) at (-2, 0) {};
\node[label={[label distance=-3pt]below:$i_T$}] (iM) at (-45:2) {};
\node[circle, fill = black, inner sep=0pt, minimum size=3pt, label={[label distance=-2pt]above right: $r$}] (r) at (0, 0) {};

\node[circle, fill = black, inner sep=0pt, minimum size=2pt] at (-102.5:1) {};
\node[circle, fill = black, inner sep=0pt, minimum size=2pt] at (-112.5:1) {};
\node[circle, fill = black, inner sep=0pt, minimum size=2pt] at (-122.5:1) {};

\draw[edgeStyle] (U1) -- (i1);
\draw[edgeStyle] (U2) -- (i2);
\draw[edgeStyle] (UM) -- (iM);
\draw[edgeStyle] (U1) -- (r);
\draw[edgeStyle] (U2) -- (r);
\draw[edgeStyle] (UM) -- (r);

\end{scope}

\end{tikzpicture}}
    \caption{\textbf{Tensor diagram of CP decomposition.}}
    \label{fig:cp_decomposition_tensor}
\end{figure}

\Cref{fig:cp_decomposition_tensor} shows the tensor diagram for CP decomposition, which can be understood by expressing~\eqref{eq:cp_decomposition} elementwise:
\begin{equation} \label{eq:cp_decomposition_elementwise}
    a_{i_1 \cdots i_T} = \sum_{r=1}^{R} \prod_{m=1}^{T} u_{i_m r}^{(m)}.
\end{equation}
The CP decomposition~\eqref{eq:cp_decomposition} can be written in terms of matricization as:
\begin{equation} \label{eq:cp_decomposition_matricization}
    \mtx{A}_{(m)} = \mtx{U}^{(m)} \mtx{Y}^{{(m)}^\top},
\end{equation}
where $\mtx{Y}^{(m)} = [\vct{y}_1^{(m)} \mid \cdots \mid \vct{y}_R^{(m)}] \in \R^{(\prod_{j \neq m} I_j) \times R}$, and $\vct{y}_r^{(m)} = \text{vec}\big(\bigotimes_{j \neq m} \vct{u}_r^{(j)}\big)$ for $r \in \{1, \ldots, R\}$.

It is important to distinguish between the \emph{CP rank} $R$ in~\eqref{eq:cp_decomposition} and the \emph{(exact) rank} $\operatorname{rank}(\tns{A})$ from \cref{def:rank}. The CP rank specifies the number of rank-1 components in a decomposition~\eqref{eq:cp_decomposition}, exact or approximate, while the (exact) rank is the minimal number required for exact representation. When $R = \operatorname{rank}(\tns{A})$ in~\eqref{eq:cp_decomposition}, the result is called a \emph{tensor rank decomposition} or \emph{minimal CP decomposition}. However, determining the (exact) rank is NP-hard in general~\cite{hillar2013most}. In practice, approximate low-rank CP decompositions are computed using iterative methods such as gradient-based optimization~\cite{kolda2009tensor}, where $R$ is chosen based on computational constraints and approximation accuracy.

\subsection{Tensor Networks as a Computational Tool} \label{sec:tensor_networks_computational_tool}

Low-rank tensor representations enable efficient computation with high-dimensional data. Storing a $T$th-order tensor with each mode of size $I$ requires $I^T$ elements, which grows exponentially in $T$. In contrast, CP decomposition with rank $R$ requires only $\mathcal{O}(TIR)$ parameters, reducing storage and computational costs dramatically.

\begin{example}[Rank-1 Inner Product] \label{ex:rank-1_inner_product}
    Consider computing the inner product of two $T$th-order tensors $\tns{A}, \tns{B} \in \R^{I \times I \times \cdots \times I}$. Direct computation using \eqref{eq:inner_product} requires $\mathcal{O}(I^T)$ operations, which becomes intractable for large $T$. However, when $\tns{A}$ and $\tns{B}$ are rank-1 tensors, \ie, $\tns{A} = \vct{a}^{(1)} \tenprod \vct{a}^{(2)} \tenprod \cdots \tenprod \vct{a}^{(T)}$ and $\tns{B} = \vct{b}^{(1)} \tenprod \vct{b}^{(2)} \tenprod \cdots \tenprod \vct{b}^{(T)}$, the inner product decomposes as
    \begin{align*}
        \inner{\tns{A}}{\tns{B}}
         & = \sum_{i_1, \ldots, i_T=1}^{I} a_{i_1 \cdots i_T} b_{i_1 \cdots i_T}                                                  \\
         & = \sum_{i_1, \ldots, i_T=1}^{I} \left[\prod_{m=1}^{T} a_{i_m}^{(m)}\right]  \left[\prod_{m=1}^{T} b_{i_m}^{(m)}\right] \\
         & = \prod_{m=1}^{T}  \left[\sum_{i_m = 1}^{I} a_{i_m}^{(m)} b_{i_m}^{(m)}\right]                                         \\
         & = \prod_{m=1}^{T}  \inner{\vct{a}^{(m)}}{\vct{b}^{(m)}}.
    \end{align*}
    This requires only $\mathcal{O}(TI)$ operations by computing the mode-wise inner products $\inner{\vct{a}^{(m)}}{\vct{b}^{(m)}}$ first, then multiplying the results---an exponential speedup. More generally, representing high-order tensors with low-rank tensor networks (\eg, CP decomposition) and executing contractions in optimal order (as detailed in \cref{sec:contraction_order}) dramatically reduces computational complexity in numerous machine learning applications.
\end{example}

\FloatBarrier
\section{Proof of \texorpdfstring{\cref{thm:separable_basis_orthonormality}}{Thm. 1}} \label{sec:proof_separable_basis_orthonormality}

\paragraph{Orthonormality.}
Let $\vct{k}, \vct{\ell} \in \Nz^C$. By Fubini's theorem and the separable definition of $\phi_{\vct{k}}$ in~\eqref{eq:separable_basis}, the inner product decomposes over dimensions:
\begin{align}
    \inner[L^2]{\phi_{\vct{k}}}{\phi_{\vct{\ell}}}
     & = \int_{[0,1]^C} \phi_{\vct{k}}(\vct{x}) \phi_{\vct{\ell}}(\vct{x}) \, d\vct{x}                                                                                               \nonumber  \\
     & = \int_{[0,1]^C} \left[\prod_{c=1}^C \varphi_{k_c}(x_c)\right] \left[\prod_{c=1}^C \varphi_{\ell_c}(x_c)\right] \, d\vct{x}                                           \nonumber          \\
     & = \prod_{c=1}^C \int_0^1 \varphi_{k_c}(x_c) \varphi_{\ell_c}(x_c) \, dx_c                                                                                                      \nonumber \\
     & = \prod_{c=1}^C \delta_{k_c \ell_c} = \delta_{\vct{k}\vct{\ell}}. \label{eq:orthonormality_multidim}
\end{align}
The penultimate equality follows from the orthonormality of the one-dimensional basis $\set{\varphi_k}[k\in\Nz]$.

\paragraph{Completeness.}
We prove the completeness by the \emph{orthogonal complement} criterion. Let $f \in L^2([0,1]^C)$ satisfy $\inner[L^2]{f}{\phi_{\vct{k}}} = 0$ for all $\vct{k} \in \Nz^C$. We show $f=0$ almost everywhere (\ie, $\norm[L^2]{f}=0$).

We argue by induction on $C$. The case $C=1$ holds because $\set{\varphi_k}[k\in\Nz]$ is an ONB in $L^2([0,1])$ by assumption. Assume the claim holds for dimension $C-1$. For $k_1\in\Nz$, define
\begin{equation} \label{eq:h_k1}
    h_{k_1}(\vct{x}_{-1}) \triangleq \int_0^1 f(x_1,\vct{x}_{-1})\, \varphi_{k_1}(x_1)\, dx_1,
\end{equation}
where $\vct{x}_{-1}=(x_2,\ldots,x_C)$ is the vector of all coordinates except $x_1$. By the Cauchy--Schwarz inequality applied in the $x_1$-coordinate, for almost every $\vct{x}_{-1}\in[0,1]^{C-1}$,
\begin{align}
    \lvert h_{k_1}(\vct{x}_{-1}) \rvert^2
     & = \left\lvert \int_0^1 f(x_1,\vct{x}_{-1})\, \varphi_{k_1}(x_1)\, dx_1 \right\rvert^2 \nonumber \\
     & \le \left( \int_0^1 \lvert f(\vct{x}) \rvert^2 dx_1 \right)
    \left( \int_0^1 \lvert \varphi_{k_1}(x_1) \rvert^2 dx_1 \right) \nonumber                          \\
     & = \int_0^1 \lvert f(\vct{x}) \rvert^2 dx_1, \label{eq:h_k1_squared}
\end{align}
since $\{\varphi_k\}_{k\in\Nz}$ is orthonormal in $L^2([0,1])$. Integrating over $\vct{x}_{-1}$ and applying Fubini's theorem yields
\begin{align}
    \norm[L^2]{h_{k_1}}^2
     & = \int_{[0,1]^{C-1}} \lvert h_{k_1}(\vct{x}_{-1}) \rvert^2 d\vct{x}_{-1} \nonumber    \\
     & \le \int_{[0,1]^C} \lvert f(\vct{x}) \rvert^2 d\vct{x} < \infty, \label{eq:h_k1_norm}
\end{align}
hence $h_{k_1}\in L^2([0,1]^{C-1})$. For any $\vct{k}_{-1}=(k_2,\ldots,k_C)\in\Nz^{C-1}$,
\begin{align}
    0 & = \inner[L^2]{f}{\phi_{(k_1,\vct{k}_{-1})}} \nonumber                                                                      \\
      & = \int_{[0,1]^{C-1}} h_{k_1}(\vct{x}_{-1}) \prod_{c=2}^C \varphi_{k_c}(x_c)\, d\vct{x}_{-1}. \label{eq:h_k1_orthogonality}
\end{align}
Hence $h_{k_1}$ is orthogonal in $L^2([0,1]^{C-1})$ to the set $\big\{\prod_{c=2}^C \varphi_{k_c}\big\}_{\vct{k}_{-1}\in\Nz^{C-1}}$, which is an ONB by the induction hypothesis; therefore $h_{k_1}=0$ almost everywhere for every $k_1\in\Nz$.

As the exceptional null sets, one per $k_1$, are countably many, their union is null, so for almost every $\vct{x}_{-1}$, the slice $x_1\mapsto f(x_1,\vct{x}_{-1})$ has all one-dimensional Fourier coefficients zero:
\begin{equation} \label{eq:orthogonality_h_k1}
    \int_0^1 f(x_1,\vct{x}_{-1})\, \varphi_{k_1}(x_1)\, dx_1 = 0, \quad \forall\, k_1\in\Nz.
\end{equation}
By completeness of $\{\varphi_k\}_{k\in\Nz}$ in $L^2([0,1])$, it follows that $f(\cdot,\vct{x}_{-1})=0$ in $L^2([0,1])$ for almost every $\vct{x}_{-1}$, and thus $f=0$ almost everywhere on $[0,1]^C$. \qed

\FloatBarrier
\section{Proof of \texorpdfstring{\cref{thm:gfs_convergence}}{Thm. 2}} \label{sec:proof_gfs_convergence}

It suffices to prove convergence for each component $s_d$ independently. Fix $d \in \{1,\ldots,D\}$ and let $s_d \in L^2([0,1]^C)$ be arbitrary. By \cref{thm:separable_basis_orthonormality}, the separable family $\{\phi_{\vct{k}}\}_{\vct{k}\in\Nz^C}$ from~\eqref{eq:separable_basis} forms an ONB for $L^2([0,1]^C)$.

For $K = (K_1, \ldots, K_C) \in \N^C$, recall $\K=\{0,\ldots,K_1-1\} \times \cdots \times \{0,\ldots,K_C-1\}$ and define the finite-dimensional subspace
\begin{equation} \label{eq:finite_dimensional_subspace}
    \V_K \triangleq \operatorname{span}\{\phi_{\vct{k}}:\vct{k}\in\K\}.
\end{equation}
Clearly $\V_K \subseteq \V_{K'}$ {whenever $K_c \le K'_c$ for every $c$}, and by completeness of the ONB, $\bigcup_{K} \V_K$ is dense in $L^2([0,1]^C)$.

Let $P_K\colon L^2([0,1]^C) \to \V_K$ denote the orthogonal projection onto $\V_K$. Since $\{\phi_{\vct{k}}\}_{\vct{k}\in\K}$ is an orthonormal set, the projection has the Fourier expansion
\begin{equation} \label{eq:projection_fourier}
    P_K(s_d) = \sum_{\vct{k}\in\K} \inner[L^2]{s_d}{\phi_{\vct{k}}} \phi_{\vct{k}}.
\end{equation}
Comparing with~\eqref{eq:gfs} and~\eqref{eq:gfs_weights}, we identify $P_K(s_d) = {s_{d,K}}$.

To establish $L^2$ convergence, fix $\varepsilon>0$. By density, there exists a finite linear combination $f = \sum_{\vct{k}\in\J} a_{\vct{k}} \phi_{\vct{k}}$ for some finite index set $\J \subset \Nz^C$ such that $\norm[L^2]{s_d - f} < \varepsilon/2$. Since $\J$ is finite, there exists $K_0 \in \N$ sufficiently large that $\J \subseteq \K$ {whenever $\min_c K_c \ge K_0$}; hence $f \in \V_K$ for all {such $K$}.

For {$\min_c K_c \ge K_0$}, the best approximation property of orthogonal projections (\ie, $P_K(s_d)$ minimizes $\norm[L^2]{s_d - v}$ over all $v \in \V_K$) yields
\begin{equation*}
    \norm[L^2]{s_d - {s_{d,K}}} = \norm[L^2]{s_d - P_K(s_d)} \le \norm[L^2]{s_d - f} < \frac{\varepsilon}{2} < \varepsilon,
\end{equation*}
where the inequality holds since $f \in \V_K$. Since $\varepsilon>0$ was arbitrary, we conclude $\norm[L^2]{s_d - {s_{d,K}}} \to 0$ as {$\min_c K_c\to\infty$}.

Since the choice of $d \in \{1,\ldots,D\}$ was arbitrary, the convergence holds for all components, yielding ${\vct{s}_K} \xrightarrow{L^2} \vct{s}$ as {$\min_c K_c \to \infty$}. \qed

\FloatBarrier
\section{Proof of \texorpdfstring{\cref{thm:futon_expressiveness}}{Thm. 3}} \label{sec:proof_futon_expressiveness}

\begin{lemma}[Rank Upper Bound] \label{lem:rank_upperbound}
    Let $\tns{A} \in \R^{I_1 \times \cdots \times I_T}$ be a $T$th-order tensor. Then
    \begin{equation} \label{eq:rank_upperbounds}
        \operatorname{rank}(\tns{A}) \le \min_{1 \le m \le T} \prod_{\substack{j = 1 \\ j \ne m}}^T I_j.
    \end{equation}
    \begin{proof} \label{proof:rank_upperbound}
        Fix $m \in \{1,\ldots,T\}$. We construct an explicit rank-1 decomposition with at most $\prod_{j \ne m} I_j$ terms by expanding in canonical bases.

        Let $\vct{e}^{(j)}_{i_j} \in \R^{I_j}$ denote the $i_j$th standard basis vector, \ie, $[\vct{e}^{(j)}_{i_j}]_k = \delta_{i_j k}$. Any tensor $\tns{A}$ admits the expansion
        \begin{equation} \label{eq:full_canonical_expansion}
            \tns{A} = \sum_{i_1=1}^{I_1} \cdots \sum_{i_T=1}^{I_T} a_{i_1 \cdots i_T} \, \vct{e}^{(1)}_{i_1} \tenprod \cdots \tenprod \vct{e}^{(T)}_{i_T}.
        \end{equation}
        Define the multi-index $\vct{i}_{-m} \triangleq (i_1,\ldots,i_{m-1},i_{m+1},\ldots,i_T) \in \prod_{j\ne m} \{1,\ldots,I_j\}$. Reordering the summations in~\eqref{eq:full_canonical_expansion} and factoring out the mode-$m$ sum yields
        \begin{align}
            \tns{A}
             & = \sum_{\vct{i}_{-m}} \sum_{i_m=1}^{I_m} a_{i_1 \cdots i_T} \, \vct{e}^{(1)}_{i_1} \tenprod \cdots \tenprod \vct{e}^{(m)}_{i_m} \tenprod \cdots \tenprod \vct{e}^{(T)}_{i_T} \nonumber                                                                                \\
             & = \sum_{\vct{i}_{-m}} \vct{e}^{(1)}_{i_1} \tenprod \cdots \tenprod \underbrace{\left(\sum_{i_m=1}^{I_m} a_{i_1 \cdots i_T} \, \vct{e}^{(m)}_{i_m}\right)}_{\vct{a}_{\vct{i}_{-m}}^{(m)}} \tenprod \cdots \tenprod \vct{e}^{(T)}_{i_T}. \label{eq:rank1_decomposition}
        \end{align}
        Here $\vct{a}_{\vct{i}_{-m}}^{(m)} \in \R^{I_m}$ is the mode-$m$ fiber of $\tns{A}$ at index $\vct{i}_{-m}$ (\cref{def:fiber}), with components $[\vct{a}_{\vct{i}_{-m}}^{(m)}]_{i_m} = a_{i_1 \cdots i_T}$. Each summand in~\eqref{eq:rank1_decomposition} with a nonzero fiber is rank-1 (\cref{def:rank-1_tensor}), as the tensor product of $T$ nonzero vectors; the others vanish. The sum runs over $\prod_{j \ne m} I_j$ distinct values of $\vct{i}_{-m}$, so~\eqref{eq:rank1_decomposition} expresses $\tns{A}$ as a sum of at most $\prod_{j \ne m} I_j$ rank-1 tensors. By \cref{def:rank}, $\operatorname{rank}(\tns{A}) \le \prod_{j \ne m} I_j$. Since $m\in\{1,\ldots,T\}$ was arbitrary, taking the minimum over all modes yields~\eqref{eq:rank_upperbounds}.
    \end{proof}
\end{lemma}

\begin{proof}[Proof of \cref{thm:futon_expressiveness}]
    Fix $\varepsilon>0$. By \cref{thm:gfs_convergence}, taking the same number $K$ of functions on every axis, for each component $d\in\{1,\ldots,D\}$ there exists $\tilde{K}_d\in\N$ such that for all $K\ge \tilde{K}_d$, the truncated generalized Fourier series ${s_{d,K}}$ in~\eqref{eq:gfs} satisfies $\norm[L^2]{s_d-{s_{d,K}}}<\varepsilon$. Let $K_0\triangleq \max_d \tilde{K}_d$ and fix any $K\ge K_0$. Then $\norm[L^2]{s_d-{s_{d,K}}}<\varepsilon$ holds simultaneously for all $d$.

    For this $K$, collect the Fourier coefficients $w_d^{(\vct{k})}$ from~\eqref{eq:gfs_weights} into the $(C{+}1)$st-order coefficient tensor $\tns{W}\in\R^{K^{\times C}\times D}$ so that ${\vct{s}_K}(\vct{x})=\bm{\Phi}(\vct{x})\dotprod \tns{W}$ by~\eqref{eq:gfs_tensordot}, with $\bm{\Phi}$ defined in~\eqref{eq:tensor_product_basis}. Applying \cref{lem:rank_upperbound} to $\tns{W}$ (of sizes $I_1=\cdots=I_C=K$ and $I_{C+1}=D$) yields
    \begin{equation} \label{eq:futon_rank_bound}
        \operatorname{rank}(\tns{W})\le \min\{K^C,\,K^{C-1}D\}=K^{C-1}\min\{K,D\}.
    \end{equation}
    Hence $\tns{W}$ admits a CP decomposition of the form~\eqref{eq:cp} with some rank $R\le K^{C-1}\min\{K,D\}$ and factor matrices $\{\mtx{U}^{(c)}\}_{c=1}^C,\,\mtx{V}$. Substituting this decomposition together with~\eqref{eq:tensor_product_basis} into~\eqref{eq:gfs_tensordot} and regrouping terms gives exactly the FUTON computation~\eqref{eq:futon_cp}--\eqref{eq:futon_cp_h}. Therefore ${\hat{\vct{s}}}(\vct{x})\equiv {\vct{s}_K}(\vct{x})$, and for the chosen $K$ and corresponding $R$ and factors
    \begin{equation} \label{eq:futon_convergence}
        \norm[L^2]{s_d- {\hat{s}_d}}=\norm[L^2]{s_d-{s_{d,K}}}<\varepsilon,\quad \forall d\in\{1,\ldots,D\}.
    \end{equation}
    This establishes the theorem.
\end{proof}

\FloatBarrier
\section{Error Analysis of FUTON with a Linear Head} \label{sec:proof_error}

\paragraph{Objective and takeaway.}
How accurate can a FUTON of spectral resolution $K$ and rank $R$ be on a given signal, before it is trained? With a linear head, the answer is exact (\cref{thm:error_general}). The squared error is the \emph{truncation}, the signal's energy beyond the first $K_c$ basis functions per axis, which only $K$ controls, plus the \emph{CP misfit}, the distance from the signal's coefficient tensor to FUTON's rank-$R$ tensor, which $R$ limits and training reduces. The truncation follows from the samples, and SVDs bracket the best misfit: below by the \emph{ceiling}, a PSNR no FUTON of that size passes, and above by a \emph{certified FUTON}, built without training. Trained FUTONs land in the bracket, to within 0.02\,dB; on images it is a median 0.05\,dB wide, so it predicts training, at 16 to 860 times less cost (\cref{sec:error_experiments}). We prove this for any basis, on the continuous domain and on sampling grids, then ask what sets each part (\cref{sec:error_consequences}) and how the bounds are computed (\cref{sec:error_algorithm}).

\paragraph{Setting.}
Let $\X = \X_1 \times \cdots \times \X_C$ carry a product measure $\mu = \mu_1 \otimes \cdots \otimes \mu_C$, with $\inner{\vct{f}}{\vct{f}'} = \sum_{d} \int_{\X} f_d f'_d \, d\mu$ and $\norm{\vct{f}}^2 = \inner{\vct{f}}{\vct{f}}$ for signals $\func{\vct{f}, \vct{f}'}{\X}{\R^D}$. Either $\X_c = [0, 1]$ with the Lebesgue measure, as in the main text, or $\X_c$ holds the $L_c$ sample positions of axis $c$ with the uniform measure; then $\norm{\vct{f}}^2 / D$ is the mean squared error over the $N = \prod_c L_c$ grid points, on which FUTON is trained and evaluated. Sampled cosines are not orthonormal on FUTON's grid, which includes both ends, so axis $c$ may use any $K_c$ square-integrable functions $\vct{\varphi}(x_c) \in \R^{K_c}$.

One change of basis does the work. The functions of axis $c$ span a space with an orthonormal basis $\vct{\xi}(x_c) \in \R^{K'_c}$, $K'_c \le K_c$, and $\vct{\varphi} = \mtx{B}_c^\top \vct{\xi}$ for a matrix $\mtx{B}_c$ of full row rank (\cref{lem:span_basis}); for an orthonormal basis, $\vct{\xi} = \vct{\varphi}$ and $\mtx{B}_c = \mtx{I}$. The products $\prod_c \xi_{j_c}(x_c)$ are orthonormal, by Fubini as in \cref{thm:separable_basis_orthonormality}, and span the space $\V_K$ of the products $\phi_{\vct{k}}$, $\vct{k} \in \K$; with $P_K$ the orthogonal projection onto $\V_K$, per output dimension, $\trunc{K} = P_K \vct{s}$ is the best approximation of $\vct{s}$ in $\V_K$, the partial sum~\eqref{eq:gfs} for an orthonormal basis. The signal's coefficients in the product basis form
\begin{equation} \label{eq:coefficients}
    \tns{Z} \in \R^{K'_1 \times \cdots \times K'_C \times D}, \qquad z^{(\vct{j})}_d = \inner{s_d}{\textstyle\prod_{c} \xi_{j_c}},
\end{equation}
which is $\tns{W}^{\star}$ for an orthonormal basis. With $\bm{\Xi}(\vct{x}) = \vct{\xi}(x_1) \tenprod \cdots \tenprod \vct{\xi}(x_C)$, $\trunc{K} = \bm{\Xi} \dotprod \tns{Z}$, and Parseval reads $\norm{\bm{\Xi} \dotprod \tns{A}} = \norm[F]{\tns{A}}$. FUTON is $\hat{\vct{s}}_{\vct{\theta}} = \bm{\Phi} \dotprod \tns{W}$ with $\tns{W} = \cp{\mtx{U}^{(1)}, \ldots, \mtx{U}^{(C)}, \mtx{V}}$, by~\eqref{eq:gfs_tensordot} and~\eqref{eq:cp}, with no bias or output activation; $\varepsilon_R$ and $\tns{A}_{\langle \mathbb{S} \rangle}$ are as in \cref{sec:error_analysis}.

\begin{theorem}[\cref{prop:error} for any basis] \label{thm:error_general}
    Let $\vct{s}$ be square-integrable, $K = (K_1, \ldots, K_C)$ and $R \ge 1$.
    \begin{enumerate}[leftmargin=*, itemsep=1pt, topsep=2pt]
        \item \emph{Every FUTON.} For every $\vct{\theta}$, $\norm{\vct{s} - \hat{\vct{s}}_{\vct{\theta}}}^2 = \norm{\vct{s} - \trunc{K}}^2 + \norm[F]{\tns{Z} - \cp{\mtx{B}_1 \mtx{U}^{(1)}, \ldots, \mtx{B}_C \mtx{U}^{(C)}, \mtx{V}}}^2$, and the truncation is $\norm{\vct{s} - \trunc{K}}^2 = \norm{\vct{s}}^2 - \norm[F]{\tns{Z}}^2$.
        \item \emph{The best FUTON.} $\inf_{\vct{\theta}} \norm{\vct{s} - \hat{\vct{s}}_{\vct{\theta}}}^2 = \norm{\vct{s} - \trunc{K}}^2 + \varepsilon_R(\tns{Z})^2$.
        \item \emph{Computable bounds.} $\lowerr_R \le \varepsilon_R(\tns{Z}) \le \upperr_R$, where $\lowerr_R^2 = \max_{\mathbb{S}} \sum_{i > R} \sigma_i^2(\tns{Z}_{\langle \mathbb{S} \rangle})$ over the nonempty proper subsets $\mathbb{S}$ of the modes, and $\upperr_R^2$ is the squared misfit of the $R$ heaviest terms of an orthonormal rank-1 expansion of $\tns{Z}$ by nested SVDs (\cref{lem:expansion}). The FUTON built from these terms, the \emph{certified FUTON}, has squared error exactly $\norm{\vct{s} - \trunc{K}}^2 + \upperr_R^2$.
        \item \emph{When the bounds meet.} If at most two modes of $\tns{Z}$ have more than one entry, then $\lowerr_R = \varepsilon_R(\tns{Z}) = \upperr_R$, and the infimum in (2) is attained; with three or more such modes, it need not be.
    \end{enumerate}
    All four quantities depend on the basis only through the span of $\vct{\varphi}$ on each axis.
\end{theorem}

Item (1) separates what $K$ decides from what $R$ and training decide; (2) reduces the best FUTON to the best rank-$R$ approximation of $\tns{Z}$, which (3) brackets and (4) pins down for two-axis signals with one output. Two lemmas supply the change of basis and the expansion.

\begin{lemma}[Orthonormal basis of the span] \label{lem:span_basis}
    Factor the Gram matrix $\mtx{G}_c = \int_{\X_c} \vct{\varphi} \vct{\varphi}^\top d\mu_c = \mtx{B}_c^\top \mtx{B}_c$ with $\mtx{B}_c \in \R^{K'_c \times K_c}$ of full row rank $K'_c = \operatorname{rank} \mtx{G}_c$. The entries of $\vct{\xi} = (\mtx{B}_c^{+})^\top \vct{\varphi}$ are orthonormal in $L^2(\X_c, \mu_c)$, and $\vct{\varphi} = \mtx{B}_c^\top \vct{\xi}$ almost everywhere, so $\vct{\xi}$ is an orthonormal basis of the span of $\vct{\varphi}$.
    \begin{proof}
        Full row rank gives $\mtx{B}_c \mtx{B}_c^{+} = \mtx{I}$, hence $\int \vct{\xi} \vct{\xi}^\top d\mu_c = (\mtx{B}_c^{+})^\top \mtx{B}_c^\top \mtx{B}_c \mtx{B}_c^{+} = \mtx{I}$. Next, $\mtx{B}_c^\top \vct{\xi} = \mtx{B}_c^{+} \mtx{B}_c \vct{\varphi}$, as $\mtx{B}_c^{+} \mtx{B}_c$ is symmetric, and the rest of $\vct{\varphi}$ vanishes almost everywhere: $\int \norm[2]{(\mtx{I} - \mtx{B}_c^{+} \mtx{B}_c) \vct{\varphi}}^2 d\mu_c = \norm[F]{\mtx{B}_c - \mtx{B}_c \mtx{B}_c^{+} \mtx{B}_c}^2 = 0$. So each of $\vct{\varphi}$ and $\vct{\xi}$ is a linear map of the other, and they span the same space.
    \end{proof}
\end{lemma}

\begin{lemma}[Orthonormal rank-1 expansion] \label{lem:expansion}
    For any ordering of its modes, nested SVDs write a tensor $\tns{A}$ with two or more modes as $\tns{A} = \sum_\ell \lambda_\ell \tns{T}_\ell$, with orthonormal rank-1 tensors $\tns{T}_\ell$ and weights $\lambda_\ell \ge 0$. A sum of $R$ of the terms has rank at most $R$, and its squared distance to $\tns{A}$ is $\norm[F]{\tns{A}}^2$ minus their squared weights, smallest for the $R$ heaviest.
    \begin{proof}
        By induction on the number of modes; with two, the SVD is such an expansion. Otherwise, an SVD of the unfolding along the first mode of the ordering gives $\tns{A} = \sum_i \sigma_i \, \vct{a}_i \tenprod \tns{G}_i$, modes in place, with orthonormal vectors $\vct{a}_i$ and orthonormal tensors $\tns{G}_i$ of the other modes. Expanding each $\tns{G}_i$ by induction gives terms $\vct{a}_i \tenprod \tns{T}'_{i\ell}$, weighted by $\sigma_i$ times the inner weights and orthonormal, since $\inner{\vct{a}_i \tenprod \tns{T}'_{i\ell}}{\vct{a}_{i'} \tenprod \tns{T}'_{i'\ell'}} = \inner{\vct{a}_i}{\vct{a}_{i'}} \inner{\tns{T}'_{i\ell}}{\tns{T}'_{i'\ell'}}$. The residual of a sum of terms is the sum of the others, whose squared norm is the sum of their squared weights.
    \end{proof}
\end{lemma}

\begin{proof}[Proof of \cref{thm:error_general}]
    (1) By \cref{lem:span_basis}, $\bm{\Phi}(\vct{x}) = \bigotimes_c \mtx{B}_c^\top \vct{\xi}(x_c)$ almost everywhere, so $\hat{\vct{s}}_{\vct{\theta}} = \bm{\Xi} \dotprod \tns{Y}_{\vct{\theta}}$ with $\tns{Y}_{\vct{\theta}} = \tns{W} \times_1 \mtx{B}_1 \cdots \times_C \mtx{B}_C$. A mode product with $\mtx{B}_c$ replaces $\vct{u}_r^{(c)}$ by $\mtx{B}_c \vct{u}_r^{(c)}$ in every rank-1 term (\cref{def:mode-n_product}), so $\tns{Y}_{\vct{\theta}} = \cp{\mtx{B}_1 \mtx{U}^{(1)}, \ldots, \mtx{B}_C \mtx{U}^{(C)}, \mtx{V}}$. As $\vct{s} - \hat{\vct{s}}_{\vct{\theta}} = (\vct{s} - \trunc{K}) + \bm{\Xi} \dotprod (\tns{Z} - \tns{Y}_{\vct{\theta}})$, the first term orthogonal to $\V_K$ and the second in it, Pythagoras and Parseval give the identity, and $\tns{Y}_{\vct{\theta}} = \bm{0}$ the truncation.

    (2) The tensors $\tns{Y}_{\vct{\theta}}$ are exactly those of rank at most $R$: each is a sum of $R$ rank-1 terms, and any such sum $\cp{\tilde{\mtx{U}}^{(1)}, \ldots, \tilde{\mtx{U}}^{(C)}, \tilde{\mtx{V}}}$ is $\tns{Y}_{\vct{\theta}}$ for $\vct{\theta} = (\mtx{B}_1^{+} \tilde{\mtx{U}}^{(1)}, \ldots, \mtx{B}_C^{+} \tilde{\mtx{U}}^{(C)}, \tilde{\mtx{V}})$, since $\mtx{B}_c \mtx{B}_c^{+} = \mtx{I}$. So (1) is minimized over that set.

    (3) \emph{Lower bound.} Every rank-1 term of a tensor $\tns{Y}$ of rank at most $R$ unfolds to a rank-1 matrix, as~\eqref{eq:cp_decomposition_matricization} shows for one mode, so $\operatorname{rank} \tns{Y}_{\langle \mathbb{S} \rangle} \le R$; unfolding keeps the Frobenius norm, and the Eckart--Young--Mirsky theorem gives $\norm[F]{\tns{Z} - \tns{Y}}^2 \ge \sum_{i > R} \sigma_i^2(\tns{Z}_{\langle \mathbb{S} \rangle})$ for every $\mathbb{S}$. \emph{Upper bound.} By \cref{lem:expansion}, the $R$ heaviest terms sum to a tensor of rank at most $R$ at squared distance $\upperr_R^2$ from $\tns{Z}$, which is $\tns{Y}_{\vct{\theta}}$ for the $\vct{\theta}$ of (2); (1) gives that FUTON's error. Any tensor of rank at most $R$ gives an upper bound in the same way.

    (4) With two modes of more than one entry, rank means matrix rank in those modes, and the expansion of \cref{lem:expansion} is the SVD, whose $R$ heaviest terms attain the Eckart--Young--Mirsky optimum, so $\upperr_R = \lowerr_R$. With three or more, the tensors of rank at most $R$ need not form a closed set, and a best approximation need not exist~\cite{desilva2008tensor}.

    Last, any orthonormal basis of the same spans is $\mtx{O}_c \vct{\xi}$ with $\mtx{O}_c$ orthogonal, which rotates $\tns{Z}$ mode by mode and changes none of the four quantities.
\end{proof}

\subsection{What Sets the Two Terms} \label{sec:error_consequences}

The truncation falls with smoothness (\cref{cor:rate}); the misfit falls with structure (\cref{prop:separation,thm:sufficiency}), but no bound on the rank holds for all signals (\cref{prop:typical_rank}), so \cref{prop:error} measures the signal at hand. The edge test shows both effects, and \cref{rem:nonlinear} what carries over to the benchmarks' heads. \Cref{cor:rate,thm:sufficiency} use the cosine basis on $[0,1]^C$ of \cref{sec:gfs}.

\begin{corollary}[Truncation rate] \label{cor:rate}
    For the cosine basis~\eqref{eq:1d_cosine_basis} and $\vct{s} \in L^2([0,1]^C; \R^D)$ whose weak partial derivatives $\partial_c \vct{s}$ are square-integrable,
    \begin{equation} \label{eq:rate}
        \norm{\vct{s} - \trunc{K}}^2 \le \sum_{c=1}^{C} \frac{\norm{\partial_c \vct{s}}^2}{(\pi K_c)^2},
    \end{equation}
    to which the best FUTON adds $\varepsilon_R(\tns{W}^{\star})^2$. If, along every axis, the odd derivatives of $\vct{s}$ of order below $m$ vanish at $0$ and $1$, the bound improves to $\sum_c \norm{\partial_c^m \vct{s}}^2 / (\pi K_c)^{2m}$.
    \begin{proof}
        Take one axis and $f \in L^2(0, 1)$ with $f' \in L^2(0, 1)$, and let $w_k = \inner{f}{\varphi_k}$. The functions $\chi_k(x) = \sqrt{2} \sin(k \pi x)$, $k \ge 1$, are orthonormal, and integrating by parts, $\inner{f'}{\chi_k} = [f \chi_k]_0^1 - k\pi \inner{f}{\varphi_k} = -k \pi w_k$, since $\chi_k(0) = \chi_k(1) = 0$. By Bessel's inequality, $\sum_{k \ge 1} (k\pi)^2 w_k^2 \le \norm{f'}^2$, hence $\sum_{k \ge K} w_k^2 \le \norm{f'}^2 / (\pi K)^2$. In $C$ dimensions, a multi-index outside $\K$ has $k_c \ge K_c$ for some $c$, so by Parseval $\norm{\vct{s} - \trunc{K}}^2 \le \sum_c \sum_{\vct{k} : k_c \ge K_c} \norm{\vct{w}^{(\vct{k})}}^2$. For fixed indices of the other axes, $\vct{w}^{(\vct{k})}$ is the $k_c$-th cosine coefficient of the integral of $\vct{s}$ against the other axes' basis functions, whose derivative is the same integral of $\partial_c \vct{s}$; the one-axis bound, summed with Bessel's inequality, bounds the inner sum by $\norm{\partial_c \vct{s}}^2 / (\pi K_c)^2$. Under the boundary conditions, $m$ integrations by parts leave no boundary terms and give the rate $(\pi K_c)^{-2m}$.
    \end{proof}
\end{corollary}

\begin{proposition}[Separation rank] \label{prop:separation}
    If $\vct{s}$ is within $\varepsilon$ of a sum of $R$ separable terms, $\norm{\vct{s} - \sum_{r=1}^{R} q^{(1)}_r(x_1) \cdots q^{(C)}_r(x_C) \, \vct{v}_r} \le \varepsilon$ with $q^{(c)}_r \in L^2(\X_c, \mu_c)$ and $\vct{v}_r \in \R^D$, then, for every basis, the best FUTON of rank $R$ at resolution $K$ has squared error at most $\norm{\vct{s} - \trunc{K}}^2 + \varepsilon^2$: a fixed basis costs only the truncation.
    \begin{proof}
        The projection onto $\V_K$ factorizes over the axes, $P_K = P_{K_1} \otimes \cdots \otimes P_{K_C}$, so it maps each separable term to a separable term, and $P_K \sum_r q^{(1)}_r \cdots q^{(C)}_r \vct{v}_r = \bm{\Xi} \dotprod \tns{Y}$ with $\tns{Y}$ of rank at most $R$. By \cref{thm:error_general}~(1) and~(2), the best FUTON's squared error is at most $\norm{\vct{s} - \trunc{K}}^2 + \norm[F]{\tns{Z} - \tns{Y}}^2$, and $\norm[F]{\tns{Z} - \tns{Y}} = \norm{P_K(\vct{s} - \sum_r q^{(1)}_r \cdots q^{(C)}_r \vct{v}_r)} \le \varepsilon$, as projections are nonexpansive.
    \end{proof}
\end{proposition}

\begin{lemma}[Box covers bound the rank] \label{lem:box}
    Let $\mathcal{F}$ be a finite family of boxes $\mathbb{I} = \mathbb{I}_1 \times \cdots \times \mathbb{I}_C \subseteq \K$, $\mathbb{I}_c \subseteq \{0, \ldots, K_c - 1\}$, and
    \begin{equation} \label{eq:box_cost}
        \operatorname{cost}(\mathcal{F}) = \sum_{\mathbb{I} \in \mathcal{F}} \frac{D \prod_c |\mathbb{I}_c|}{\max\{|\mathbb{I}_1|, \ldots, |\mathbb{I}_C|, D\}}.
    \end{equation}
    Every tensor in $\R^{K_1 \times \cdots \times K_C \times D}$ that vanishes outside $\bigcup \mathcal{F}$ has rank at most $\operatorname{cost}(\mathcal{F})$.
    \begin{proof}
        Split the tensor into its parts on each box minus the earlier boxes. Each is a tensor of size $|\mathbb{I}_1| \times \cdots \times |\mathbb{I}_C| \times D$ padded with zeros, of rank at most the product of its sizes but the largest (\cref{lem:rank_upperbound}), its term in~\eqref{eq:box_cost}; rank is subadditive.
    \end{proof}
\end{lemma}

\begin{theorem}[Low-rank sufficiency] \label{thm:sufficiency}
    Let $\vct{s} \in L^2([0,1]^C; \R^D)$ have cosine coefficients $\vct{w}^{(\vct{k})}$, and let a family of boxes $\mathcal{F}$ cover a set $\mathbb{A} \subseteq \K$ of multi-indices. Some FUTON of resolution $K$ and rank at most $\operatorname{cost}(\mathcal{F})$, with a linear head, has squared error at most $\sum_{\vct{k} \notin \mathbb{A}} \norm{\vct{w}^{(\vct{k})}}^2$, the energy of $\vct{s}$ outside $\mathbb{A}$.
    \begin{proof}
        Keep $\tns{W}^{\star}$ on $\bigcup \mathcal{F} \supseteq \mathbb{A}$ and set it to zero elsewhere. By \cref{lem:box} the result has rank at most $\operatorname{cost}(\mathcal{F})$, so it is $\cp{\mtx{U}^{(1)}, \ldots, \mtx{U}^{(C)}, \mtx{V}}$ for factors of that rank; by \cref{prop:error}~(1) and Parseval, that FUTON's squared error is the energy outside $\bigcup \mathcal{F}$, at most the energy outside $\mathbb{A}$.
    \end{proof}
\end{theorem}

\Cref{thm:futon_expressiveness} is the case $\mathbb{A} = \K$. A \emph{sub-box} $\{0, \ldots, q - 1\}^C$ costs $q^{C-1} \min\{q, D\}$ and, by \cref{cor:rate}, leaves at most $\norm{\nabla \vct{s}}^2 / (\pi q)^2$ outside, so rank $q^{C-1} \min\{q, D\}$ with $q = \lceil \norm{\nabla \vct{s}} / (\pi \varepsilon) \rceil$ suffices for squared error $\varepsilon^2$, set by the accuracy, not by $K$. A \emph{hyperbolic cross} $\{\vct{k} : (k_1 + 1)(k_2 + 1) \le \nu\}$, where signals of dominating mixed smoothness concentrate~\cite{bungartz2004sparse}, lies in two strips of width $\lceil \sqrt{\nu} \rceil$, at cost $2D \lceil \sqrt{\nu} \rceil$: 138 at $\nu = K_c = 512$ and $D = 3$, against 1,536 for the full grid. Both bound what FUTONs can represent, not what training finds.

\begin{proposition}[No rank bound holds for all signals] \label{prop:typical_rank}
    If $R(\sum_c K'_c + D - C) < D \prod_c K'_c$, the tensors of rank at most $R$ in $\R^{K'_1 \times \cdots \times K'_C \times D}$, with their limits, form a set of Lebesgue measure zero. So $\varepsilon_R(\tns{Z}) = 0$ only for exceptional coefficient tensors; almost every $K \times K \times K$ tensor has rank at least $\lceil K^3 / (3K - 2) \rceil$, 21,903 at $K = 256$.
    \begin{proof}
        The map from factors to $\cp{\mtx{U}^{(1)}, \ldots, \mtx{U}^{(C)}, \mtx{V}}$ is polynomial, and rescaling a term's $C + 1$ factors by numbers of product one leaves it unchanged, so the tensors of rank at most $R$ form a semialgebraic set of dimension at most $R(\sum_c K'_c + D - C)$, as does its closure, which lies in the $R$-th secant variety of a Segre variety~\cite{landsberg2012tensors}; such a set has measure zero when its dimension is below the ambient one. For $K \times K \times K$ tensors the condition reads $R(3K - 2) < K^3$, and the bound is the typical rank over the complex numbers~\cite{lickteig1985typical}.
    \end{proof}
\end{proposition}

So typical tensors attain the order of \cref{thm:futon_expressiveness}'s bound, and only structure in the signal brings the rank down.

\paragraph{The edge test.}
A soft edge of width $\delta$ through the center of $[0,1]^2$, $s(\vct{x}) = \tanh(\vct{a}^\top (\vct{x} - \frac{1}{2}\vct{1}) / \delta)$ with a unit direction $\vct{a}$, tests these predictions. Along an axis, it is separable, of rank one at any width. Obliquely, a profile within $\varepsilon$ of a polynomial of degree $p$ gives a polynomial of total degree $p$ in $(x_1, x_2)$, of rank at most $p + 1$, with $p$ growing as $(1/\delta) \log(1/\varepsilon)$ for $\tanh(t / \delta)$. With $C = 2$ and $D = 1$, \cref{thm:error_general}~(4) gives the exact rank each accuracy needs (\cref{tab:edge_test}): one for the axis-aligned edge at every width, and about $(1/\delta)^{0.6}$ to $(1/\delta)^{0.8}$ for oblique edges, within the predicted $\mathcal{O}(1/\delta)$; orientation costs rank only through sharpness.

\begin{table}[t]
    \caption{\textbf{The edge test.} Smallest rank $R$ at which an edge's cosine coefficients ($K_c = 256$) have relative squared error at most $10^{-3}$, $10^{-4}$, and $10^{-5}$, per width $\delta$ and direction, exactly from the singular values ($4096 \times 4096$ midpoint quadrature).}
    \label{tab:edge_test}
    \centering
    \small
    \setlength{\tabcolsep}{4pt}
    \begin{tabular}{@{}l ccc ccc ccc@{}}
        \toprule
        & \multicolumn{3}{c}{Axis-aligned} & \multicolumn{3}{c}{Oblique, $30^\circ$} & \multicolumn{3}{c}{Diagonal, $45^\circ$} \\
        \cmidrule(lr){2-4} \cmidrule(lr){5-7} \cmidrule(l){8-10}
        Width $\delta$ & $10^{-3}$ & $10^{-4}$ & $10^{-5}$ & $10^{-3}$ & $10^{-4}$ & $10^{-5}$ & $10^{-3}$ & $10^{-4}$ & $10^{-5}$ \\
        \midrule
        0.2   & 1 & 1 & 1 & 4  & 4  & 5  & 4  & 4  & 6   \\
        0.1   & 1 & 1 & 1 & 5  & 6  & 8  & 6  & 8  & 10  \\
        0.05  & 1 & 1 & 1 & 7  & 10 & 12 & 9  & 13 & 16  \\
        0.02  & 1 & 1 & 1 & 13 & 19 & 25 & 18 & 26 & 35  \\
        0.01  & 1 & 1 & 1 & 21 & 33 & 45 & 30 & 47 & 63  \\
        0.005 & 1 & 1 & 1 & 34 & 57 & 80 & 49 & 82 & 115 \\
        \bottomrule
    \end{tabular}
\end{table}

\begin{remark}[Nonlinear heads and the output activation] \label{rem:nonlinear}
    The benchmarks' FUTON adds an MLP head of width $M = 64$ on images and radiance fields, or a bias on volumes, and a $\tanh$ output on images and volumes, which the analysis does not cover. The upper bound transfers to both heads, since a bias can be zero and an MLP head of width $M \ge 2D$ contains every linear head, as $\operatorname{ReLU}(t) - \operatorname{ReLU}(-t) = t$: the best such FUTON at $(K, R)$ is at least as good as the certified one, with more parameters. The lower bound does not, as biases and a hidden layer fit what no linear head can. As $\tanh$ is 1-Lipschitz, for a target $\vct{s} = \tanh(\tilde{\vct{s}})$ with $\tilde{\vct{s}}$ square-integrable (on a grid, any target with $|\vct{s}| < 1$), the error of $\tanh \circ \hat{\vct{s}}$ against $\vct{s}$ is at most that of $\hat{\vct{s}}$ against $\tilde{\vct{s}}$, so the analysis of $\tilde{\vct{s}}$ certifies the $\tanh$ model; a target that reaches $\pm 1$, as occupancy and the Kodak images do, has no such $\tilde{\vct{s}}$.
\end{remark}

\paragraph{Related results.}
The ingredients are classical; new is their combination for this model, where every quantity is computed from the signal and the upper bound is attained. The truncated SVD is optimal for matrices~\cite{eckart1936approximation,mirsky1960symmetric}, item (4), but with three or more modes a best approximation need not exist~\cite{desilva2008tensor}, hence (2) and (3). The lower bound is the matricization tail that also bounds the HOSVD~\cite{delathauwer2000multilinear} and TT-SVD~\cite{oseledets2011tensor} from below, and \cref{lem:expansion} is a tree of SVDs, one per branch, with its heaviest terms kept, for two modes the Kronecker-product SVD~\cite{vanloan1993kronecker}. Separated representations~\cite{beylkin2002numerical,beylkin2005algorithms} keep their separation rank in a fixed basis up to the truncation (\cref{prop:separation}), and CP models of Fourier coefficients~\cite{kargas2021supervised,wesel2021large} come with bounds in expectation or asymptotically, where \cref{prop:error} is an identity for the signal at hand; for MLP-based INRs, no analogue of item (2) is known~\cite{yuce2022structured}.

\subsection{Algorithm} \label{sec:error_algorithm}

On the grid, axis $c$'s basis is the matrix $\bm{\Phi}_c \in \R^{L_c \times K_c}$ of its values at the sample positions, and the signal is the array $\tns{S} \in \R^{L_1 \times \cdots \times L_C \times D}$ of its samples. A thin SVD $\bm{\Phi}_c / \sqrt{L_c} = \mtx{Q}_c \bm{\Sigma}_c \mtx{E}_c^\top$, truncated at the numerical rank $K'_c$, orthonormalizes the axis: $\mtx{B}_c = \bm{\Sigma}_c \mtx{E}_c^\top$, and the values of $\vct{\xi}$ are the rows of $\sqrt{L_c} \, \mtx{Q}_c$. \Cref{tab:error_algorithm} lists the steps, all in double precision. Only Steps 2 and 3 read the signal, one matrix product per axis and a sum of squares; the rest works on $\tns{Z}$, whose size is set by $K$, not by the signal, which keeps the bound cheap. Dividing by $D$ turns the outputs into mean squared errors.

\begin{table}[t]
    \caption{The bounds from samples. Steps 5 and 6 give the certified FUTON.}
    \label{tab:error_algorithm}
    \centering
    \small
    \begin{tabular}{@{}cp{7.3cm}l@{}}
        \toprule
        Step & Computes & Cost \\
        \midrule
        1 & $\mtx{Q}_c, \bm{\Sigma}_c, \mtx{E}_c$ from the thin SVD of $\bm{\Phi}_c / \sqrt{L_c}$ & $\mathcal{O}(L_c K_c^2)$ per axis \\
        2 & $\tns{Z} = \tns{S} \times_1 \mtx{Q}_1^\top \cdots \times_C \mtx{Q}_C^\top / \sqrt{N}$ & $\mathcal{O}(N D \sum_c K_c)$ \\
        3 & $\norm{\vct{s} - \trunc{K}}^2 = \norm[F]{\tns{S}}^2 / N - \norm[F]{\tns{Z}}^2$, exact & $\mathcal{O}(N D)$ \\
        4 & $\lowerr_R^2$: the singular values of every matricization of $\tns{Z}$, tails beyond each $R$ & one eigendecomposition per split \\
        5 & \cref{lem:expansion} for every mode ordering, keeping the ordering with the most weight in its $R$ heaviest terms; $\upperr_R^2 = \norm[F]{\tns{Z} - \tns{Y}}^2$ for their sum $\tns{Y} = \cp{\tilde{\mtx{U}}^{(1)}, \ldots, \tilde{\mtx{U}}^{(C)}, \tilde{\mtx{V}}}$ & nested SVDs of $\tns{Z}$ \\
        6 & $\hat{\mtx{U}}^{(c)} = \mtx{E}_c \bm{\Sigma}_c^{-1} \tilde{\mtx{U}}^{(c)}$, $\hat{\mtx{V}} = \tilde{\mtx{V}}$ & $\mathcal{O}(R \sum_c K_c K'_c)$ \\
        \bottomrule
    \end{tabular}
\end{table}

\subsection{Experiments} \label{sec:error_experiments}

The experiments test what \cref{prop:error} tells before training: whether its guarantees hold, where training lands, how well it sizes a FUTON, and what it costs. In short: the bracket always holds; on images it is narrow and locates training, on volumes it is wide and guarantees a PSNR; and it costs a fraction of training.

\paragraph{Setup.}
The signals are the 24 Kodak images ($C = 2$, $D = 3$, values in $[-1, 1]$) and the five occupancy volumes of \cref{sec:occupancy_volumes} ($C = 3$, $D = 1$, values $\pm 1$), in the cosine basis at $K_c = \kappa L_c$ with $\kappa \in \{1/8, 1/4, 1/2\}$ and $R \in \{32, 64, \ldots, 1024\}$, around the benchmark's ranks: 18 settings, 522 runs, and $\kappa \in \{1, 2\}$ for the cost alone. For each $K$, the algorithm of \cref{sec:error_algorithm} gives the truncation, the ceiling and the certified FUTON at all six ranks in one timed call; the same linear-head model is trained by Adam under the protocol of \cref{sec:experiments}, at learning rate $10^{-1}$ on images and $3 \cdot 10^{-2}$ on volumes. Errors are mean squared errors over the whole grid, and $\mathrm{PSNR} = 10 \log_{10}(4 / \mathrm{MSE})$. The bound runs mostly on the CPU, in double precision, so its timings depend on the CPU's load; the volumes' were timed in a run of their own. \Cref{fig:error_bound} compares the bounds with training at $\kappa = 1/2$, and \cref{tab:error_means_images} lists the means at every setting.

\paragraph{The bracket holds.}
No trained FUTON passes the ceiling beyond rounding, in any of the 522 runs: in 15 of them, all on Kodak at full rank, where the ceiling is the truncation and the full-rank FUTON reaches it, Adam's error lies below it by a relative $2.3 \cdot 10^{-8}$ at most, the single-precision rounding of its forward pass. The certified side is a construction, not a limit on training, yet training respects it too: Adam's FUTON never ends more than 0.02\,dB below the certified one. Training thus lands in the bracket in every run, to within 0.02\,dB. Unit tests check that every certified FUTON attains its upper bound, to a relative $10^{-9}$, as well as the Eckart--Young case, the invariance within a span, and a DCT cross-check.

\begin{table}[t]
    \caption{\textbf{Means per setting} over the 24 Kodak images and over the five occupancy volumes. PSNR (dB), $\pm$ one within-signal standard error: the truncation, the certified FUTON, the ceiling and Adam's FUTON; seconds on an A100 for the bound at all six ranks of a $K$ and for one Adam run.}
    \label{tab:error_means_images}\label{tab:error_means_volumes}
    \centering
    {\small\setlength{\tabcolsep}{3pt}%
    \begin{tabular}{@{}lrrrrrrr@{}}
    \toprule
    \multicolumn{6}{c}{} & \multicolumn{2}{c}{Seconds} \\
    \cmidrule(lr){7-8}
    $\kappa$ & \multicolumn{1}{c}{$R$} & \multicolumn{1}{c}{Truncation} & \multicolumn{1}{c}{Certified} & \multicolumn{1}{c}{Ceiling} & \multicolumn{1}{c}{Adam} & \multicolumn{1}{c}{Bound} & \multicolumn{1}{c}{Adam} \\
    \midrule
    \multicolumn{8}{@{}l}{\textit{Images}} \\
    $1/8$ & 32 & 23.65{\scriptsize$\pm$0.06} & 22.85{\scriptsize$\pm$0.06} & 23.27{\scriptsize$\pm$0.01} & 23.13{\scriptsize$\pm$0.02} & 0.02 & 1.7 \\
    $1/8$ & 64 & 23.65{\scriptsize$\pm$0.01} & 23.45{\scriptsize$\pm$0.03} & 23.62{\scriptsize$\pm$0.01} & 23.57{\scriptsize$\pm$0.00} & 0.02 & 1.7 \\
    $1/8$ & 128 & 23.65{\scriptsize$\pm$0.00} & 23.64{\scriptsize$\pm$0.00} & 23.65{\scriptsize$\pm$0.00} & 23.65{\scriptsize$\pm$0.00} & 0.02 & 1.7 \\
    $1/8$ & 256 & 23.65{\scriptsize$\pm$0.00} & 23.65{\scriptsize$\pm$0.00} & 23.65{\scriptsize$\pm$0.00} & 23.65{\scriptsize$\pm$0.00} & 0.02 & 1.7 \\
    $1/8$ & 512 & 23.65{\scriptsize$\pm$0.00} & 23.65{\scriptsize$\pm$0.00} & 23.65{\scriptsize$\pm$0.00} & 23.65{\scriptsize$\pm$0.00} & 0.02 & 1.7 \\
    $1/8$ & 1024 & 23.65{\scriptsize$\pm$0.00} & 23.65{\scriptsize$\pm$0.00} & 23.65{\scriptsize$\pm$0.00} & 23.65{\scriptsize$\pm$0.00} & 0.02 & 1.7 \\
    $1/4$ & 32 & 25.88{\scriptsize$\pm$0.15} & 23.70{\scriptsize$\pm$0.10} & 24.35{\scriptsize$\pm$0.03} & 24.10{\scriptsize$\pm$0.05} & 0.07 & 1.7 \\
    $1/4$ & 64 & 25.88{\scriptsize$\pm$0.07} & 24.90{\scriptsize$\pm$0.07} & 25.41{\scriptsize$\pm$0.02} & 25.23{\scriptsize$\pm$0.02} & 0.07 & 1.7 \\
    $1/4$ & 128 & 25.88{\scriptsize$\pm$0.02} & 25.64{\scriptsize$\pm$0.03} & 25.85{\scriptsize$\pm$0.01} & 25.79{\scriptsize$\pm$0.00} & 0.07 & 1.7 \\
    $1/4$ & 256 & 25.88{\scriptsize$\pm$0.00} & 25.87{\scriptsize$\pm$0.00} & 25.88{\scriptsize$\pm$0.00} & 25.88{\scriptsize$\pm$0.00} & 0.07 & 1.7 \\
    $1/4$ & 512 & 25.88{\scriptsize$\pm$0.00} & 25.88{\scriptsize$\pm$0.00} & 25.88{\scriptsize$\pm$0.00} & 25.88{\scriptsize$\pm$0.00} & 0.07 & 1.7 \\
    $1/4$ & 1024 & 25.88{\scriptsize$\pm$0.00} & 25.88{\scriptsize$\pm$0.00} & 25.88{\scriptsize$\pm$0.00} & 25.88{\scriptsize$\pm$0.00} & 0.07 & 1.7 \\
    $1/2$ & 32 & 29.86{\scriptsize$\pm$0.25} & 24.34{\scriptsize$\pm$0.13} & 25.17{\scriptsize$\pm$0.06} & 24.85{\scriptsize$\pm$0.08} & 0.17 & 1.7 \\
    $1/2$ & 64 & 29.86{\scriptsize$\pm$0.19} & 26.22{\scriptsize$\pm$0.11} & 27.11{\scriptsize$\pm$0.04} & 26.77{\scriptsize$\pm$0.06} & 0.17 & 1.7 \\
    $1/2$ & 128 & 29.86{\scriptsize$\pm$0.11} & 28.13{\scriptsize$\pm$0.09} & 28.94{\scriptsize$\pm$0.03} & 28.65{\scriptsize$\pm$0.03} & 0.17 & 1.7 \\
    $1/2$ & 256 & 29.86{\scriptsize$\pm$0.02} & 29.46{\scriptsize$\pm$0.04} & 29.82{\scriptsize$\pm$0.02} & 29.70{\scriptsize$\pm$0.01} & 0.17 & 1.7 \\
    $1/2$ & 512 & 29.86{\scriptsize$\pm$0.00} & 29.84{\scriptsize$\pm$0.00} & 29.86{\scriptsize$\pm$0.00} & 29.85{\scriptsize$\pm$0.00} & 0.17 & 1.7 \\
    $1/2$ & 1024 & 29.86{\scriptsize$\pm$0.00} & 29.86{\scriptsize$\pm$0.00} & 29.86{\scriptsize$\pm$0.00} & 29.86{\scriptsize$\pm$0.00} & 0.17 & 1.7 \\
    \midrule
    \multicolumn{8}{@{}l}{\textit{Occupancy volumes}} \\
    $1/8$ & 32 & 20.62{\scriptsize$\pm$0.26} & 16.94{\scriptsize$\pm$0.25} & 20.22{\scriptsize$\pm$0.17} & 17.92{\scriptsize$\pm$0.18} & 0.05 & 3.5 \\
    $1/8$ & 64 & 20.62{\scriptsize$\pm$0.19} & 18.08{\scriptsize$\pm$0.24} & 20.62{\scriptsize$\pm$0.19} & 19.08{\scriptsize$\pm$0.14} & 0.05 & 3.8 \\
    $1/8$ & 128 & 20.62{\scriptsize$\pm$0.15} & 19.13{\scriptsize$\pm$0.21} & 20.62{\scriptsize$\pm$0.15} & 19.99{\scriptsize$\pm$0.09} & 0.05 & 4.7 \\
    $1/8$ & 256 & 20.62{\scriptsize$\pm$0.09} & 20.02{\scriptsize$\pm$0.18} & 20.62{\scriptsize$\pm$0.09} & 20.47{\scriptsize$\pm$0.01} & 0.05 & 6.4 \\
    $1/8$ & 512 & 20.62{\scriptsize$\pm$0.03} & 20.49{\scriptsize$\pm$0.09} & 20.62{\scriptsize$\pm$0.03} & 20.60{\scriptsize$\pm$0.02} & 0.05 & 9.8 \\
    $1/8$ & 1024 & 20.62{\scriptsize$\pm$0.00} & 20.61{\scriptsize$\pm$0.01} & 20.62{\scriptsize$\pm$0.00} & 20.62{\scriptsize$\pm$0.00} & 0.05 & 17.6 \\
    $1/4$ & 32 & 23.64{\scriptsize$\pm$0.33} & 17.10{\scriptsize$\pm$0.25} & 21.69{\scriptsize$\pm$0.10} & 18.10{\scriptsize$\pm$0.18} & 0.44 & 3.5 \\
    $1/4$ & 64 & 23.64{\scriptsize$\pm$0.25} & 18.34{\scriptsize$\pm$0.25} & 23.16{\scriptsize$\pm$0.17} & 19.48{\scriptsize$\pm$0.17} & 0.44 & 3.8 \\
    $1/4$ & 128 & 23.64{\scriptsize$\pm$0.21} & 19.57{\scriptsize$\pm$0.25} & 23.64{\scriptsize$\pm$0.21} & 20.80{\scriptsize$\pm$0.18} & 0.44 & 4.6 \\
    $1/4$ & 256 & 23.64{\scriptsize$\pm$0.19} & 20.81{\scriptsize$\pm$0.23} & 23.64{\scriptsize$\pm$0.19} & 21.99{\scriptsize$\pm$0.15} & 0.44 & 6.5 \\
    $1/4$ & 512 & 23.64{\scriptsize$\pm$0.16} & 22.00{\scriptsize$\pm$0.22} & 23.64{\scriptsize$\pm$0.16} & 22.92{\scriptsize$\pm$0.09} & 0.44 & 9.9 \\
    $1/4$ & 1024 & 23.64{\scriptsize$\pm$0.10} & 22.97{\scriptsize$\pm$0.19} & 23.64{\scriptsize$\pm$0.10} & 23.43{\scriptsize$\pm$0.02} & 0.44 & 17.6 \\
    $1/2$ & 32 & 26.94{\scriptsize$\pm$0.39} & 17.20{\scriptsize$\pm$0.24} & 22.37{\scriptsize$\pm$0.05} & 18.18{\scriptsize$\pm$0.18} & 2.95 & 3.4 \\
    $1/2$ & 64 & 26.94{\scriptsize$\pm$0.32} & 18.48{\scriptsize$\pm$0.24} & 24.68{\scriptsize$\pm$0.10} & 19.62{\scriptsize$\pm$0.18} & 2.95 & 3.8 \\
    $1/2$ & 128 & 26.94{\scriptsize$\pm$0.27} & 19.77{\scriptsize$\pm$0.25} & 26.35{\scriptsize$\pm$0.18} & 21.06{\scriptsize$\pm$0.20} & 2.95 & 4.7 \\
    $1/2$ & 256 & 26.94{\scriptsize$\pm$0.24} & 21.12{\scriptsize$\pm$0.26} & 26.94{\scriptsize$\pm$0.24} & 22.45{\scriptsize$\pm$0.21} & 2.95 & 6.5 \\
    $1/2$ & 512 & 26.94{\scriptsize$\pm$0.22} & 22.49{\scriptsize$\pm$0.26} & 26.94{\scriptsize$\pm$0.22} & 23.78{\scriptsize$\pm$0.19} & 2.95 & 9.9 \\
    $1/2$ & 1024 & 26.94{\scriptsize$\pm$0.21} & 23.83{\scriptsize$\pm$0.25} & 26.94{\scriptsize$\pm$0.21} & 24.96{\scriptsize$\pm$0.17} & 2.95 & 17.6 \\
    \bottomrule
    \end{tabular}}
\end{table}

\paragraph{Images: training, located.}
For an image, the bracket is narrow: a median 0.05\,dB over images and settings, at most 0.1\,dB in 56\% of them and at most 0.5\,dB in 78\%. It widens with $K$, to medians of 0.00, 0.05 and 0.37\,dB at $\kappa = 1/8$, $1/4$ and $1/2$, and closes as the certified FUTON reaches the truncation, which it does on every image from $R = 256$ on at $\kappa = 1/8$, from 512 on at $1/4$, and at 1024 at $1/2$. Adam lands at a steady place inside it, a median 0.66 of the way from the certified FUTON to the ceiling, and 0.60 to 0.70 in half the settings where it is open; at $\kappa = 1/2$, a median 0.24\,dB above the certified FUTON and 0.11\,dB below the ceiling. Before training, \cref{prop:error} thus says what Adam will reach to within a tenth of a decibel in over half the settings, and within a few tenths at the largest $K$.

\paragraph{Volumes: a wider bracket.}
With three axes and one output, every matricization of $\tns{Z}$ has rank at most the largest $K_c$, so once $R$ passes it the ceiling falls back to the truncation, although the misfit need not vanish; and the certified FUTON trails training further than on images. The bracket is a median 0.98, 3.43 and 5.37\,dB wide at $\kappa = 1/8$, $1/4$ and $1/2$, with Adam a median 0.67, 1.07 and 1.19\,dB above the certified FUTON and, at $\kappa = 1/4$ and $1/2$, the ceiling the looser side. At $\kappa = 1/8$, the truncation caps the volumes at 20.6\,dB on average, which the certified FUTON reaches to within 3.69\,dB at $R = 32$, 0.60\,dB at 256 and 0.01\,dB at 1024. On volumes the bracket is a guarantee more than a forecast: a PSNR that a FUTON of the size reaches without training, and one that none passes.

\begin{figure}[t]
    \centering
    \begin{subfigure}[b]{0.64\linewidth}\centering\includegraphics{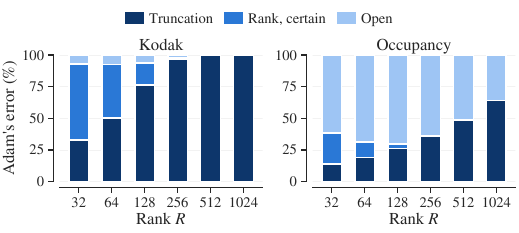}\caption{Where Adam's error goes}\label{fig:error_decomposition}\end{subfigure}\hfill
    \begin{subfigure}[b]{0.34\linewidth}\centering\includegraphics{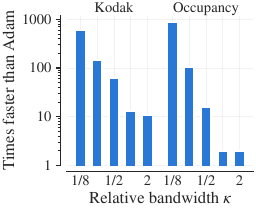}\caption{Cost of the bound}\label{fig:error_cost}\end{subfigure}
    \caption{\textbf{Where Adam's error goes, and what the bound costs.} (a)~Adam's error at $\kappa = 1/2$, in percent, as \cref{prop:error} splits it, averaged over the 24 Kodak images and the five occupancy volumes, per rank: the truncation, which $K$ forces; the part of the misfit up to the ceiling, which $R$ forces with certainty; and the rest, which the bounds leave open. (b)~How many times faster the bound, certified FUTON included, is than the six Adam runs it stands in for, per relative bandwidth $\kappa$, from the seconds per signal averaged over each dataset.}
\end{figure}

\paragraph{Where Adam's error goes.}
\Cref{prop:error} splits every FUTON's error into the truncation, which $K$ forces, and the misfit, which the ceiling splits again: up to the ceiling it is forced by $R$, and above it is open, between the rank and training (\cref{fig:error_decomposition}). At $\kappa = 1/2$ on Kodak, the truncation is 33\% of Adam's error at $R = 32$ and all but 0.1\% at 1024, the part that $R$ forces falls from 60\% to nothing, and at most 7\% stays open: Adam's error is almost entirely accounted for before training. On the volumes, the truncation is 14\% at 32 and 64\% at 1024, $R$ forces 25\% at 32 and nothing from 256 on, where the ceiling meets the truncation, and the open part, 36 to 70\%, is the largest share up to $R = 512$.

\paragraph{How close is training to the best FUTON?}
The ceiling also grades training: a trained FUTON's distance to it is the most a better optimizer could add. \Cref{tab:error_checks} (top) trains every signal at $\kappa = 1/2$ and the benchmark's rank with Adam for 1000 and 5000 steps. On Kodak the best FUTON of that size lies between 29.80\,dB, Adam's at 5000 steps, and the ceiling's 29.84\,dB: the benchmark's 1000 steps end within 0.09\,dB of the ceiling on average and within 0.18\,dB on every image. On the volumes it lies between 22.94\,dB, Adam's at 5000 steps, and the loose ceiling's 26.94\,dB. Every trained FUTON passes the certified one on every signal, by up to 0.65\,dB on the images and by 1.22 to 2.11\,dB on the volumes.

\begin{table}[t]
    \caption{\textbf{Training against the bound.} Top: mean PSNR (dB) over each dataset, $\pm$ one within-signal standard error, at $\kappa = 1/2$ and the benchmark's rank, for the certified FUTON and the ceiling against Adam at the benchmark's 1000 steps and at 5000. Bottom: seconds per signal for the six ranks of a $K$, from $\kappa = 1/8$ to $2$, averaged over each dataset, for the bound, certified FUTON included, and the six Adam runs; and how far the certified FUTON stays below Adam's, in dB, over the signals and ranks.}
    \label{tab:error_checks}\label{tab:error_cost}
    \centering
    {\small\setlength{\tabcolsep}{6pt}%
    \begin{tabular}{@{}lrrrrr@{}}
    \toprule
    \multicolumn{4}{c}{} & \multicolumn{2}{c}{Adam} \\
    \cmidrule(l){5-6}
    Dataset & \multicolumn{1}{c}{$R$} & \multicolumn{1}{c}{Certified} & \multicolumn{1}{c}{Ceiling} & \multicolumn{1}{c}{1000 steps} & \multicolumn{1}{c}{5000 steps} \\
    \midrule
    Kodak & 280 & 29.56{\scriptsize$\pm$0.03} & 29.84{\scriptsize$\pm$0.01} & 29.75{\scriptsize$\pm$0.00} & 29.80{\scriptsize$\pm$0.01} \\
    Occupancy & 215--310 & 21.22{\scriptsize$\pm$0.16} & 26.94{\scriptsize$\pm$0.43} & 22.55{\scriptsize$\pm$0.10} & 22.94{\scriptsize$\pm$0.07} \\
    \bottomrule
    \end{tabular}}

    \medskip
    {\small\setlength{\tabcolsep}{5pt}%
    \begin{tabular}{@{}lrrrrrrrr@{}}
    \toprule
    & \multicolumn{4}{c}{Kodak images} & \multicolumn{4}{c}{Occupancy volumes} \\
    \cmidrule(lr){2-5}\cmidrule(l){6-9}
    & \multicolumn{2}{c}{Seconds} & \multicolumn{2}{c}{Behind Adam (dB)} & \multicolumn{2}{c}{Seconds} & \multicolumn{2}{c}{Behind Adam (dB)} \\
    \cmidrule(lr){2-3}\cmidrule(lr){4-5}\cmidrule(lr){6-7}\cmidrule(l){8-9}
    $\kappa$ & \multicolumn{1}{c}{Bound} & \multicolumn{1}{c}{Adam} & \multicolumn{1}{c}{Median} & \multicolumn{1}{c}{Max} & \multicolumn{1}{c}{Bound} & \multicolumn{1}{c}{Adam} & \multicolumn{1}{c}{Median} & \multicolumn{1}{c}{Max} \\
    \midrule
    $1/8$ & 0.02 & 10 & 0.00 & 0.93 & 0.05 & 46 & 0.67 & 1.29 \\
    $1/4$ & 0.07 & 10 & 0.03 & 1.11 & 0.44 & 46 & 1.07 & 1.47 \\
    $1/2$ & 0.17 & 10 & 0.24 & 1.28 & 2.95 & 46 & 1.19 & 1.56 \\
    $1$ & 0.79 & 10 & 0.85 & 5.83 & 24.19 & 46 & 1.28 & 1.57 \\
    $2$ & 0.97 & 10 & 0.81 & 5.25 & 23.82 & 47 & 1.31 & 1.66 \\
    \bottomrule
    \end{tabular}}
\end{table}

\paragraph{Sizing a FUTON for a target PSNR.}
The bracket on PSNR is also one on size: among the candidate settings, no FUTON smaller than the smallest whose ceiling reaches a target reaches it, and the smallest whose certified PSNR reaches it does, with its model. \Cref{tab:error_design} sets both against an oracle that trains every setting and takes the smallest that Adam reaches. The oracle's pick lies between the two on every signal and target, and the certified pick is never smaller. On Kodak the bracket on size is tight: the two picks coincide for 56\% of the images and targets, so the size a target needs is known before training, and the certified pick is the oracle's for 76\% and at most twice it otherwise. On the volumes, the ceiling's pick is a median eight times smaller than the certified one, which is safe but generous, at most twice the oracle's parameters at the median of every target; at 23\,dB it reaches four of the five shapes that Adam does. Read along a budget, the certified mean picks the spectral resolution that Adam's results pick at every parameter count on Kodak ($\kappa = 1/8$ at about 5k, $1/4$ at 10k and 21k, $1/2$ from 41k on), and at six of eight on the volumes.

\begin{table}[t]
    \caption{\textbf{The smallest setting for a target PSNR.} Reached: the signals of each dataset for which some setting reaches the target by Adam, by the certified FUTON, and by the ceiling (possible). Pick vs.\ oracle: over the signals both reach, how often the certified pick has the parameter count of the oracle's, which trains every setting and takes the smallest that Adam reaches, or a larger one; it is never smaller. \#Params: the median parameter count of the ceiling's pick, below which no candidate reaches the target, of the oracle's, and of the certified pick.}
    \label{tab:error_design}
    \centering
    {\small\setlength{\tabcolsep}{3pt}%
    \begin{tabular}{@{}lrrrrrrrr@{}}
    \toprule
    \multicolumn{1}{c}{} & \multicolumn{3}{c}{Reached} & \multicolumn{2}{c}{Pick vs. oracle} & \multicolumn{3}{c}{\#Params (k)} \\
    \cmidrule(lr){2-4}\cmidrule(lr){5-6}\cmidrule(lr){7-9}
    Target (dB) & \multicolumn{1}{c}{Adam} & \multicolumn{1}{c}{Certified} & \multicolumn{1}{c}{Possible} & \multicolumn{1}{c}{Same} & \multicolumn{1}{c}{Larger} & \multicolumn{1}{c}{Ceiling} & \multicolumn{1}{c}{Adam} & \multicolumn{1}{c}{Certified} \\
    \midrule
    \multicolumn{9}{l}{\textit{Kodak images}} \\
    22 & 24/24 & 24/24 & 24/24 & 88\% & 12\% & 5.2 & 5.2 & 5.2 \\
    24 & 23/24 & 23/24 & 23/24 & 70\% & 30\% & 5.2 & 10.3 & 10.3 \\
    26 & 22/24 & 22/24 & 22/24 & 77\% & 23\% & 20.7 & 20.7 & 30.9 \\
    28 & 18/24 & 18/24 & 18/24 & 67\% & 33\% & 41.2 & 61.7 & 61.8 \\
    30 & 13/24 & 13/24 & 13/24 & 77\% & 23\% & 82.3 & 82.3 & 82.3 \\
    \midrule
    \multicolumn{9}{l}{\textit{Occupancy volumes}} \\
    19 & 5/5 & 5/5 & 5/5 & 20\% & 80\% & 3.9 & 7.7 & 14.1 \\
    20 & 5/5 & 5/5 & 5/5 & 0\% & 100\% & 4.8 & 15.5 & 31.0 \\
    21 & 5/5 & 5/5 & 5/5 & 0\% & 100\% & 7.8 & 31.1 & 62.2 \\
    22 & 5/5 & 5/5 & 5/5 & 0\% & 100\% & 14.1 & 62.2 & 124.4 \\
    23 & 5/5 & 4/5 & 5/5 & 0\% & 100\% & 15.6 & 124.4 & 237.6 \\
    \bottomrule
    \end{tabular}}
\end{table}

\paragraph{Cost.}
The bound gives the truncation, the ceiling and the certified FUTON at all six ranks of a $K$ in one call: on an A100, on average 0.02 to 0.17\,s per image and 0.05 to 3.0\,s per volume at $\kappa \le 1/2$, against 1.7 and 7.7\,s for one Adam run. Sizing a signal over every candidate setting, as above, takes 0.26\,s per image against 31\,s of training, and 3.4\,s per volume against 138\,s: 40 to 120 times less. The bound grows with $K$, as $\tns{Z}$ does, and stays below training: at $\kappa = 1$ and $2$, 0.8 to 1.0\,s per image, under a tenth of the six Adam runs, and 24\,s per volume, about half of them (\cref{fig:error_cost}), while the certified FUTON trails training further on the images there (\cref{tab:error_cost}, bottom).

\FloatBarrier
\section{Detailed Experimental Setup} \label{sec:detailed_experimental_setup}

Our implementation, the library Inro, with the configuration files and scripts behind every run, is available at \url{https://github.com/pashtari/inro}.

\paragraph{Data.}
\emph{Images.} The 24 Kodak images~\cite{kodak1993kodak}, $768 \times 512$ or $512 \times 768$, with RGB values scaled to $[-1, 1]$ and coordinates on $[-1, 1]^2$; every step uses the whole image. \emph{Occupancy volumes.} Five meshes from the Stanford 3D Scanning Repository~\cite{stanford3dscan}, normalized as in BACON~\cite{lindell2022bacon}. Each voxel center of a grid with 512 voxels per unit length is marked inside or outside the scan by a majority of five ray-parity tests, and the grid is cropped to the shape: $386 \times 460 \times 351$ (Armadillo), $461 \times 325 \times 206$ (Dragon), $190 \times 461 \times 189$ (Happy Buddha), $268 \times 154 \times 461$ (Lucy), and $274 \times 461 \times 236$ (Thai Statue), 17 to 62 million voxels. Targets are $\pm 1$; a step draws $2^{20}$ voxels at random, and an evaluation pass reads as many; IoU thresholds the output at zero over the whole grid. \emph{Radiance fields.} The eight Blender scenes~\cite{mildenhall2020nerf} under FINER's protocol~\cite{liu2024finer}, which follows torch-ngp~\cite{tang2022torchngp}: views downsampled by four to $200 \times 200$, every fourth of the 100 training views, every 25th validation view for learning curves, and all 200 test views for the final metrics. A radiance field combines the compared INR, the density network, which outputs a density and 15 geometry features, with a shared color network (two hidden layers of 64, view directions in spherical harmonics); rays are marched through an occupancy grid that skips empty space~\cite{muller2022instant,li2023nerfacc}, with at most 1,024 samples per ray and 4,096 rays per step.

\paragraph{Models and training.}
\Cref{tab:configs} lists every model's configuration, size, and learning rate per task. Each baseline is the smallest of its kind with at least 197,123 (images), 131,329 (volumes), or 74,131 (radiance fields, color network included) parameters, and FUTON takes the largest rank, a multiple of 5, that keeps it smaller than every baseline; on volumes, that cap was TensoRF's CP variant (131,585 parameters), which we ran but report, as in 3D throughout, in its vector--matrix form. FUTON normalizes each axis's features to unit length at every point, $\vct{\varphi}(x_c) / \norm[2]{\vct{\varphi}(x_c)}$, which keeps the separable structure and stabilizes training; the error analysis covers any basis (\cref{thm:error_general}), so it applies to these features too. FUTON's spectral resolution and head were chosen in a sweep of both on the 24 Kodak images, the five shapes (at 256 voxels per unit length), and the validation views of the eight scenes: the best quality at a training time and memory that stay low. Every model minimizes the mean squared error with PyTorch's default Adam~\cite{kingma2014adam}, its learning rate annealed by a cosine to a hundredth or, on radiance fields, decayed exponentially to a tenth, as in torch-ngp~\cite{tang2022torchngp}. Learning rates are the best of the grid $\{10^{-4}, 3 \cdot 10^{-4}, 10^{-3}, \ldots, 3 \cdot 10^{-1}\}$ on two Kodak images, on Thai Statue and Dragon (passing over rates that leave a shape empty), or on the validation views of Lego and Drums, the same grid for every model. Other hyperparameters follow the authors' settings for the task, except where they do not apply or break the budget: RFF's $\sigma$, MFN's input scale, $\alpha$, and $\beta$, and F-INR's first frequency were tuned for our coordinates and signal sizes; WIRE, in its real-valued form, uses $3 \cdot 10^{-3}$ on images and volumes, because the official $5 \cdot 10^{-3}$ diverges; Instant-NGP uses 20 levels on images and a 32-wide MLP on volumes to meet the budget, and learning rates of $3 \cdot 10^{-2}$ on images and radiance fields and $10^{-2}$ on volumes, against $10^{-2}$ and $10^{-4}$ in the references; F-INR's widths meet the budget; and radiance fields train for 25,000 steps instead of torch-ngp's 37,500. FreSh~\cite{kania2025fresh} selects SIREN's frequency scale for each signal with the authors' procedure, which compares the spectrum of the untrained model's output with the signal's; its search counts toward its training time. All models are compiled with \texttt{torch.compile}~\cite{ansel2024pytorch} and trained once, with seed 0.

\paragraph{Metrics and costs.}
PSNR, SSIM~\cite{wang2004image}, and LPIPS~\cite{zhang2018unreasonable} are computed on the whole image or test view, and IoU on the whole voxel grid. Training time is the wall-clock time of the training steps, without evaluation; memory is peak GPU memory in training and in inference; throughput is the signal's size over the median, over a run's evaluations, of its evaluation time, in megapixels or megavoxels per second, or frames per second for radiance fields. Costs are medians over the signals, since other work on a machine can only add time; qualities are means, with one standard error over the signals, computed within signals (Cousineau--Morey) for PSNR, SSIM, and LPIPS, pooled over all models, and plainly for IoU.

\paragraph{Compute.}
Every run uses one NVIDIA A100 (80~GB) with PyTorch~\cite{paszke2019pytorch} 2.14. The 985 distinct runs reported in the benchmarks and ablations train for 11.3 GPU-hours in total, 6.9 of them on radiance fields, and the error-bound study adds under an hour; learning-rate and configuration searches and preliminary experiments used additional compute.

\begin{table}[p]
    \caption{\textbf{Model configurations per task}: architecture, parameters, and learning rate. ``$3 \times 256$'' is three hidden layers of 256 units; $\omega_0$ and $\gamma_0$ are the activations' frequency and scale, $L_c$ is the signal's size along axis $c$, and $M$ the width of FUTON's MLP head. Radiance-field sizes include the shared color network.}
    \label{tab:configs}
    \centering
    \small
    \setlength{\tabcolsep}{4pt}
    \begin{tabular}{@{}l p{8.3cm} r r@{}}
        \toprule
        Model & Configuration & \#Params (k) & LR \\
        \midrule
        \multicolumn{4}{@{}l}{\textit{Images}} \\
        RFF & $3 \times 212$; 256 Fourier features, $\sigma = 10$ & 199.7 & $3 \cdot 10^{-3}$ \\
        PE-MLP & $4 \times 256$; positional encoding with 10 frequencies & 209.2 & $3 \cdot 10^{-3}$ \\
        MFN & $4 \times 256$; input scale 1024, $\alpha = 1.25$, $\beta = 0.125$ & 204.3 & $3 \cdot 10^{-2}$ \\
        SIREN & $4 \times 256$; $\omega_0 = 30$ & 198.9 & $10^{-3}$ \\
        FreSh (SIREN) & as SIREN, $\omega_0$ selected per image & 198.9 & $10^{-3}$ \\
        Gauss & $4 \times 256$; $\gamma_0 = 30$ & 198.9 & $3 \cdot 10^{-3}$ \\
        WIRE & $4 \times 256$, real Gabor; $\omega_0 = 20$, $\gamma_0 = 10$ & 199.3 & $3 \cdot 10^{-3}$ \\
        FINER & $4 \times 256$; $\omega_0 = 30$ & 198.9 & $10^{-3}$ \\
        Instant-NGP & 20 levels $\times$ 2 features, $2^{13}$ entries, resolutions 16 to $\max_c L_c / 2$; MLP $2 \times 64$ & 205.1 & $3 \cdot 10^{-2}$ \\
        TensoRF (CP) & rank 176, resolution 256; MLP $2 \times 256$ & 202.0 & $3 \cdot 10^{-2}$ \\
        GA-Planes & 26 features, resolutions 256 and 64; MLP $2 \times 256$ & 199.9 & $10^{-2}$ \\
        F-INR (SIREN) & CP rank 141; sub-networks $3 \times 140$, first $\omega_0 = 200$ & 198.8 & $10^{-3}$ \\
        FUTON (ours) & cosine, $K_c = L_c / 2$; CP, $R = 280$; MLP head, $M = 64$ & 197.4 & $10^{-1}$ \\
        \midrule
        \multicolumn{4}{@{}l}{\textit{Occupancy volumes}} \\
        RFF & $2 \times 188$; 256 Fourier features, $\sigma = 6$ & 132.2 & $3 \cdot 10^{-3}$ \\
        PE-MLP & $3 \times 241$; positional encoding with 10 frequencies & 132.3 & $10^{-3}$ \\
        MFN & $3 \times 256$; input scale 256, $\alpha = 6$, $\beta = 1$ & 138.0 & $3 \cdot 10^{-2}$ \\
        SIREN & $3 \times 256$; $\omega_0 = 30$ & 132.9 & $10^{-3}$ \\
        FreSh (SIREN) & as SIREN, $\omega_0$ selected per shape & 132.9 & $3 \cdot 10^{-3}$ \\
        Gauss & $3 \times 256$; $\gamma_0 = 30$ & 132.9 & $3 \cdot 10^{-3}$ \\
        WIRE & $3 \times 256$, real Gabor; $\omega_0 = 10$, $\gamma_0 = 40$ & 133.4 & $3 \cdot 10^{-3}$ \\
        FINER & $3 \times 256$; $\omega_0 = 30$ & 132.9 & $3 \cdot 10^{-3}$ \\
        Instant-NGP & 16 levels $\times$ 2 features, $2^{12}$ entries, resolutions 16 to 512; MLP $2 \times 32$ & 133.2 & $10^{-2}$ \\
        TensoRF (VM) & rank 4, resolution 97; MLP $2 \times 128$ & 132.4 & $10^{-2}$ \\
        GA-Planes & 25 features, resolutions 128, 32, and 8; MLP $2 \times 128$ & 132.0 & $3 \cdot 10^{-2}$ \\
        F-INR (SIREN) & CP rank 259; sub-networks $3 \times 96$ & 131.8 & $10^{-3}$ \\
        FUTON (ours) & cosine, $K_c = L_c / 2$; CP, $R = 215$ to $310$ by shape; linear head & 128.8--131.4 & $3 \cdot 10^{-2}$ \\
        \midrule
        \multicolumn{4}{@{}l}{\textit{Radiance fields (density network)}} \\
        RFF & $2 \times 180$; 90 Fourier features, $\sigma = 2$ & 74.5 & $3 \cdot 10^{-3}$ \\
        PE-MLP & $3 \times 166$; positional encoding with 10 frequencies & 75.1 & $10^{-2}$ \\
        MFN & $3 \times 176$; input scale 128, $\alpha = 2$, $\beta = 1$ & 75.8 & $10^{-2}$ \\
        SIREN & $3 \times 180$; $\omega_0 = 30$ & 75.2 & $3 \cdot 10^{-4}$ \\
        FreSh (SIREN) & as SIREN, $\omega_0$ selected per scene & 75.2 & $3 \cdot 10^{-4}$ \\
        Gauss & $3 \times 180$; $\gamma_0 = 30$ & 75.2 & $10^{-3}$ \\
        WIRE & $3 \times 180$, real Gabor; $\omega_0 = 20$, $\gamma_0 = 10$ & 74.5 & $10^{-3}$ \\
        FINER & $3 \times 180$; $\omega_0 = 30$ & 75.2 & $3 \cdot 10^{-4}$ \\
        Instant-NGP & 16 levels $\times$ 2 features, $2^{11}$ entries, resolutions 16 to 2048; MLP $1 \times 64$ & 75.0 & $3 \cdot 10^{-2}$ \\
        TensoRF (VM) & rank 14, resolution 32; MLP $2 \times 128$ & 74.8 & $3 \cdot 10^{-2}$ \\
        GA-Planes & 12 features, resolutions 128, 32, and 4; MLP $2 \times 128$ & 75.0 & $3 \cdot 10^{-2}$ \\
        F-INR (SIREN) & CP rank 39; sub-networks $3 \times 32$ & 74.7 & $10^{-3}$ \\
        FUTON (ours) & cosine, $K_c = 100$; CP, $R = 180$; MLP head, $M = 64$ & 73.0 & $10^{-2}$ \\
        \bottomrule
    \end{tabular}
\end{table}

\clearpage
\section{Full Results} \label{sec:full_results}

\begin{figure}[t]
    \centering
    {\setlength{\abovecaptionskip}{0pt}\setlength{\belowcaptionskip}{7pt}%
    \captionof{table}{\textbf{Time to target quality}, in seconds, read from each model's mean learning curve, computed as in \cref{fig:learning_speed} and interpolated linearly between evaluations, with the best bold and the second best underlined; ``--'' marks a target the mean curve never reaches, and ``$\le$'' one reached by the first evaluation. Radiance fields use the validation views.}\label{tab:time_to_target}}
    {\small\setlength{\tabcolsep}{4pt}%
\begin{tabular}{@{}l rrr rrr rrr@{}}
\toprule
& \multicolumn{3}{c}{Images, PSNR (dB)} & \multicolumn{3}{c}{Occupancy, IoU (\%)} & \multicolumn{3}{c}{Radiance fields, PSNR (dB)} \\
\cmidrule(lr){2-4}\cmidrule(lr){5-7}\cmidrule(l){8-10}
Model & 30 & 33 & 35 & 98 & 99 & 99.5 & 27 & 28 & 28.5 \\
\midrule
RFF & 13.7 & -- & -- & 11.8 & 23.1 & 37.1 & 49.3 & 117 & -- \\
PE-MLP & -- & -- & -- & -- & -- & -- & 63.5 & 172 & -- \\
MFN & $\le$2.9 & 6.7 & 14.6 & 79.1 & -- & -- & 42.5 & \underline{109} & 267 \\
SIREN & 12.5 & 22.9 & -- & 35.9 & -- & -- & 79.8 & 132 & 176 \\
FreSh (SIREN) & 9.8 & 18.5 & -- & 233 & 251 & -- & 78.8 & 141 & 203 \\
Gauss & 14.6 & -- & -- & 33.8 & 55.4 & -- & 139 & -- & -- \\
WIRE & 27.0 & -- & -- & 29.2 & 56.1 & -- & 104 & 169 & -- \\
FINER & 6.0 & 14.0 & 20.5 & 43.8 & 57.8 & -- & 74.7 & 127 & \underline{167} \\
Instant-NGP & \underline{0.3} & \underline{0.8} & \underline{1.6} & \underline{1.4} & \underline{4.0} & \underline{7.2} & -- & -- & -- \\
TensoRF & 2.8 & 7.2 & 15.0 & 3.7 & 11.1 & -- & 59.1 & 205 & -- \\
GA-Planes & 5.4 & -- & -- & 6.0 & 22.6 & 44.5 & \underline{37.3} & 121 & -- \\
F-INR (SIREN) & -- & -- & -- & -- & -- & -- & 152 & -- & -- \\
\midrule
FUTON (ours) & \textbf{$\le$0.2} & \textbf{0.4} & \textbf{0.9} & \textbf{0.5} & \textbf{2.0} & \textbf{3.2} & \textbf{$\le$16.1} & \textbf{22.0} & \textbf{33.3} \\
\bottomrule
\end{tabular}
}

    \vspace{12pt}
    \includegraphics{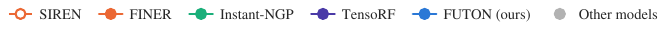}\\[1pt]
    \setlength{\tabcolsep}{0pt}
    \begin{tabular}{@{}c@{\hspace{0.09in}}c@{\hspace{0.09in}}c@{}}
        \footnotesize Images & \footnotesize Occupancy volumes & \footnotesize Radiance fields \\
        \includegraphics{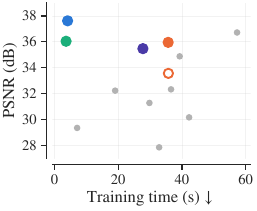} &
        \includegraphics{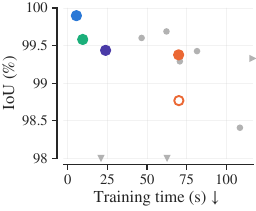} &
        \includegraphics{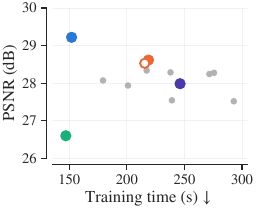} \\
        \includegraphics{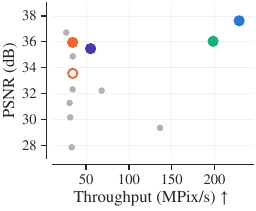} &
        \includegraphics{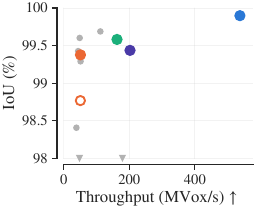} &
        \includegraphics{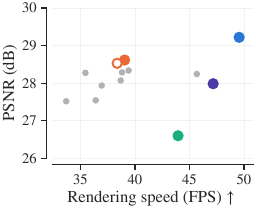}
    \end{tabular}
    \caption{\textbf{Quality against training time} (top) \textbf{and inference speed} (bottom) for every model; gray dots are the other baselines. On the volumes, the axes start at 98\% IoU, and models below sit on the axis line; a model past an axis's range sits at its end as a marker.}
    \label{fig:tradeoffs}
\end{figure}

\paragraph{Time to target.}
FUTON reaches every target first (\cref{tab:time_to_target}): 35\,dB on images in 0.9\,s, 99.5\% IoU on volumes in 3.2\,s, and 28.5\,dB on radiance fields in 33\,s. FUTON reaches the final quality of Instant-NGP, the fastest baseline, 2.7 times sooner on images (in 1.3\,s) and volumes (3.4\,s), and at least 9.1 times sooner on radiance fields, by its first evaluation (16\,s); and that of each task's most accurate baseline 34 times sooner on images (MFN, 1.7\,s against 57\,s), 16.5 times on volumes (GA-Planes), and 4.2 times on radiance fields (FINER).

\begin{figure}[t]
    \centering
    {\setlength{\abovecaptionskip}{0pt}\setlength{\belowcaptionskip}{7pt}%
    \captionof{table}{\textbf{Images}, means over the 24 Kodak images $\pm$ one standard error, and median costs, at about 197k parameters (\cref{tab:configs} lists each model's size). The best results are \textbf{bold}, and the second-best are \underline{underlined}.}\label{tab:image_full}}
    {\small\setlength{\tabcolsep}{3pt}%
\begin{tabular}{@{}lrrrrrrr@{}}
\toprule
\multicolumn{1}{c}{} & \multicolumn{2}{c}{Training} & \multicolumn{2}{c}{Inference} & \multicolumn{3}{c}{} \\
\cmidrule(lr){2-3}\cmidrule(lr){4-5}
Model & \multicolumn{1}{c}{Time} & \multicolumn{1}{c}{Memory} & \multicolumn{1}{c}{Throughput} & \multicolumn{1}{c}{Memory} & \multicolumn{1}{c}{PSNR} & \multicolumn{1}{c}{SSIM} & \multicolumn{1}{c}{LPIPS} \\
 & \multicolumn{1}{c}{(s) $\downarrow$} & \multicolumn{1}{c}{(GB) $\downarrow$} & \multicolumn{1}{c}{(MPix/s) $\uparrow$} & \multicolumn{1}{c}{(MB) $\downarrow$} & \multicolumn{1}{c}{(dB) $\uparrow$} & \multicolumn{1}{c}{$\uparrow$} & \multicolumn{1}{c}{$\downarrow$} \\
\midrule
RFF & 29.8 & 2.08 & 30.7 & 1208 & 31.28{\scriptsize$\pm$0.17} & 0.846{\scriptsize$\pm$0.007} & 0.271{\scriptsize$\pm$0.007} \\
PE-MLP & 32.8 & 2.11 & 33.2 & 805 & 27.86{\scriptsize$\pm$0.16} & 0.726{\scriptsize$\pm$0.015} & 0.431{\scriptsize$\pm$0.008} \\
MFN & 57.3 & 7.67 & 26.8 & 3224 & \underline{36.72}{\scriptsize$\pm$0.17} & \underline{0.938}{\scriptsize$\pm$0.003} & 0.129{\scriptsize$\pm$0.003} \\
SIREN & 35.7 & 3.25 & 34.4 & 805 & 33.57{\scriptsize$\pm$0.15} & 0.906{\scriptsize$\pm$0.003} & 0.192{\scriptsize$\pm$0.004} \\
FreSh (SIREN) & 39.3 & 3.25 & 34.5 & 805 & 34.88{\scriptsize$\pm$0.15} & 0.922{\scriptsize$\pm$0.003} & 0.150{\scriptsize$\pm$0.003} \\
Gauss & 36.6 & 3.64 & 34.3 & 805 & 32.33{\scriptsize$\pm$0.19} & 0.877{\scriptsize$\pm$0.005} & 0.213{\scriptsize$\pm$0.004} \\
WIRE & 42.2 & 4.00 & 31.5 & 856 & 30.17{\scriptsize$\pm$0.46} & 0.807{\scriptsize$\pm$0.017} & 0.260{\scriptsize$\pm$0.010} \\
FINER & 35.7 & 3.25 & 34.3 & 805 & 35.96{\scriptsize$\pm$0.14} & \underline{0.938}{\scriptsize$\pm$0.003} & \underline{0.120}{\scriptsize$\pm$0.002} \\
Instant-NGP & \textbf{3.6} & \underline{0.72} & \underline{198.3} & \underline{503} & 36.05{\scriptsize$\pm$0.20} & 0.932{\scriptsize$\pm$0.002} & 0.142{\scriptsize$\pm$0.004} \\
TensoRF & 27.8 & 1.79 & 55.2 & 805 & 35.48{\scriptsize$\pm$0.13} & 0.920{\scriptsize$\pm$0.002} & 0.161{\scriptsize$\pm$0.003} \\
GA-Planes & 19.0 & 1.32 & 68.1 & 805 & 32.23{\scriptsize$\pm$0.20} & 0.878{\scriptsize$\pm$0.004} & 0.227{\scriptsize$\pm$0.005} \\
F-INR (SIREN) & 7.1 & 2.70 & 136.3 & 1338 & 29.35{\scriptsize$\pm$0.13} & 0.786{\scriptsize$\pm$0.007} & 0.343{\scriptsize$\pm$0.005} \\
\midrule
FUTON (ours) & \underline{4.1} & \textbf{0.36} & \textbf{228.7} & \textbf{201} & \textbf{37.63}{\scriptsize$\pm$0.55} & \textbf{0.955}{\scriptsize$\pm$0.004} & \textbf{0.098}{\scriptsize$\pm$0.004} \\
\bottomrule
\end{tabular}
}

    \vspace{12pt}
    \includegraphics{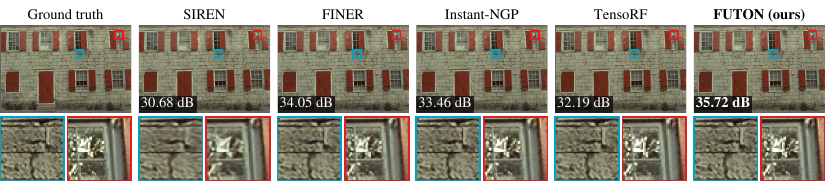}\\[3pt]
    \includegraphics[trim={0 0 0 11bp},clip]{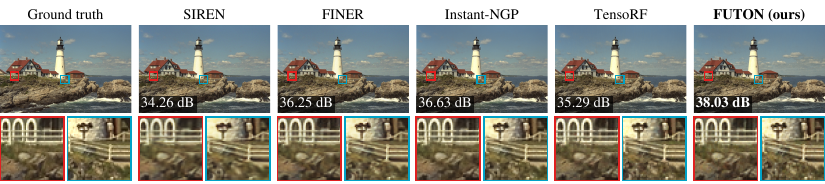}\\[3pt]
    \includegraphics[trim={0 0 0 11bp},clip]{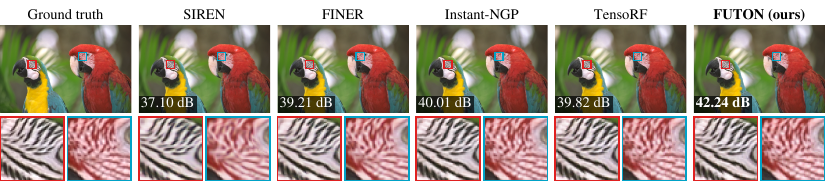}
    \caption{\textbf{Image fitting} on kodim01, kodim21, and kodim23, as in \cref{fig:qualitative_image}.}
    \label{fig:qualitative_kodim01}
\end{figure}

\begin{figure}[t]
    \centering
    {\setlength{\abovecaptionskip}{0pt}\setlength{\belowcaptionskip}{7pt}%
    \captionof{table}{\textbf{Occupancy volumes}: IoU on every shape and its mean $\pm$ one standard error, and median costs, at about 131k parameters. The best results are \textbf{bold}, and the second-best are \underline{underlined}.}\label{tab:occupancy_full}}
    {\footnotesize\setlength{\tabcolsep}{3.2pt}%
\begin{tabular}{@{}lrrrrrrrrrr@{}}
\toprule
 & \multicolumn{2}{c}{Training} & \multicolumn{2}{c}{Inference} & \multicolumn{6}{c}{IoU (\%) $\uparrow$} \\
\cmidrule(lr){2-3}\cmidrule(lr){4-5}\cmidrule(l){6-11}
Model & \multicolumn{1}{c}{Time} & \multicolumn{1}{c}{Memory} & \multicolumn{1}{c}{Throughput} & \multicolumn{1}{c}{Memory} & \multicolumn{1}{c}{Armadillo} & \multicolumn{1}{c}{Dragon} & \multicolumn{1}{c}{Happy} & \multicolumn{1}{c}{Lucy} & \multicolumn{1}{c}{Thai} & \multicolumn{1}{c}{Mean} \\
 & \multicolumn{1}{c}{(s) $\downarrow$} & \multicolumn{1}{c}{(GB) $\downarrow$} & \multicolumn{1}{c}{(MVox/s) $\uparrow$} & \multicolumn{1}{c}{(GB) $\downarrow$} & & & \multicolumn{1}{c}{Buddha} & & \multicolumn{1}{c}{Statue} & \\
\midrule
RFF & 46.5 & 4.15 & 48.9 & 3.33 & \underline{99.71} & 99.65 & 99.59 & 99.78 & 99.28 & 99.60{\scriptsize$\pm$0.09} \\
PE-MLP & 62.6 & 4.09 & 48.3 & 2.14 & 96.96 & 96.92 & 97.54 & 91.96 & 95.71 & 95.82{\scriptsize$\pm$1.01} \\
MFN & 108.5 & 15.07 & 38.9 & 5.49 & 99.23 & 97.47 & 97.50 & 99.10 & 98.74 & 98.41{\scriptsize$\pm$0.39} \\
SIREN & 70.0 & 6.48 & 51.0 & 2.26 & 98.78 & 99.18 & 99.41 & 98.53 & 97.94 & 98.77{\scriptsize$\pm$0.26} \\
FreSh (SIREN) & 263.6 & 6.48 & 51.3 & 2.26 & 99.34 & 99.67 & 99.65 & 99.15 & 98.83 & 99.33{\scriptsize$\pm$0.16} \\
Gauss & 70.7 & 6.48 & 51.5 & 2.26 & 99.52 & 99.59 & 99.70 & 99.07 & 98.57 & 99.29{\scriptsize$\pm$0.21} \\
WIRE & 81.5 & 7.62 & 44.2 & 2.39 & 99.52 & 99.59 & 99.74 & 99.43 & 98.84 & 99.43{\scriptsize$\pm$0.15} \\
FINER & 69.9 & 6.48 & 51.3 & 2.26 & 99.45 & 99.59 & 99.70 & 99.29 & 98.85 & 99.38{\scriptsize$\pm$0.15} \\
Instant-NGP & \underline{9.4} & 2.11 & 162.6 & 2.93 & 99.50 & 99.59 & 99.69 & 99.78 & \underline{99.35} & 99.58{\scriptsize$\pm$0.08} \\
TensoRF & 23.8 & \underline{1.75} & \underline{202.3} & \underline{1.19} & 99.53 & 99.71 & 99.78 & 99.34 & 98.82 & 99.44{\scriptsize$\pm$0.17} \\
GA-Planes & 62.2 & 2.31 & 111.9 & 1.37 & 99.58 & \underline{99.79} & \underline{99.93} & \underline{99.79} & 99.34 & \underline{99.69}{\scriptsize$\pm$0.10} \\
F-INR (SIREN) & 20.9 & 6.59 & 179.2 & 3.39 & 97.75 & 98.25 & 98.08 & 96.27 & 96.42 & 97.35{\scriptsize$\pm$0.42} \\
\midrule
FUTON (ours) & \textbf{5.4} & \textbf{0.28} & \textbf{537.4} & \textbf{0.61} & \textbf{99.86} & \textbf{99.94} & \textbf{99.99} & \textbf{99.96} & \textbf{99.74} & \textbf{99.90}{\scriptsize$\pm$0.04} \\
\bottomrule
\end{tabular}
}

    \vspace{12pt}
    \includegraphics{figures/results/legend.pdf}\\[1pt]
    \setlength{\tabcolsep}{0pt}
    \begin{tabular}{@{}c@{\hspace{0.09in}}c@{\hspace{0.09in}}c@{}}
        \footnotesize Images & \footnotesize Occupancy volumes & \footnotesize Radiance fields \\
        \includegraphics{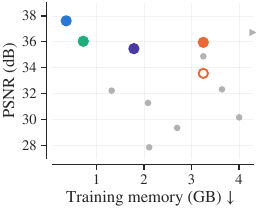} &
        \includegraphics{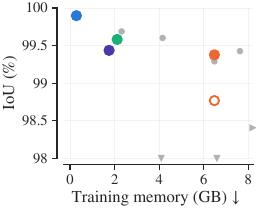} &
        \includegraphics{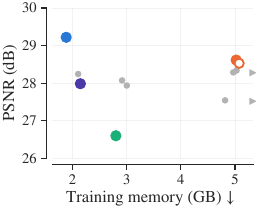} \\
        \includegraphics{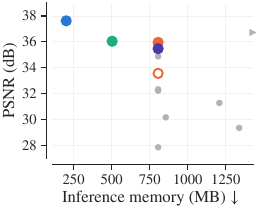} &
        \includegraphics{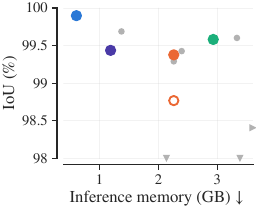} &
        \includegraphics{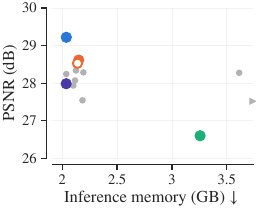}
    \end{tabular}
    \caption{\textbf{Quality against training memory} (top) \textbf{and inference memory} (bottom), as in \cref{fig:tradeoffs}.}
    \label{fig:tradeoffs_memory}
\end{figure}

\begin{figure}[t]
    \centering
    \includegraphics{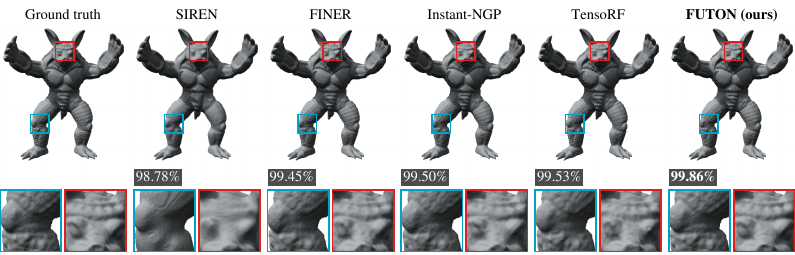}\\[3pt]
    \includegraphics[trim={0 0 0 11bp},clip]{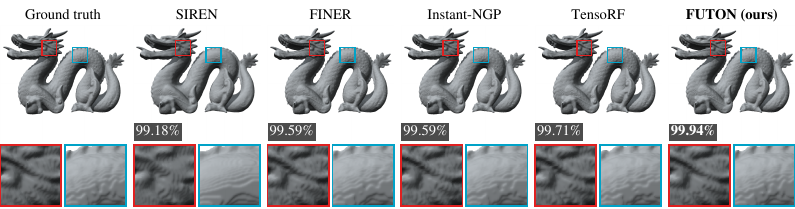}\\[3pt]
    \includegraphics[trim={0 0 0 11bp},clip]{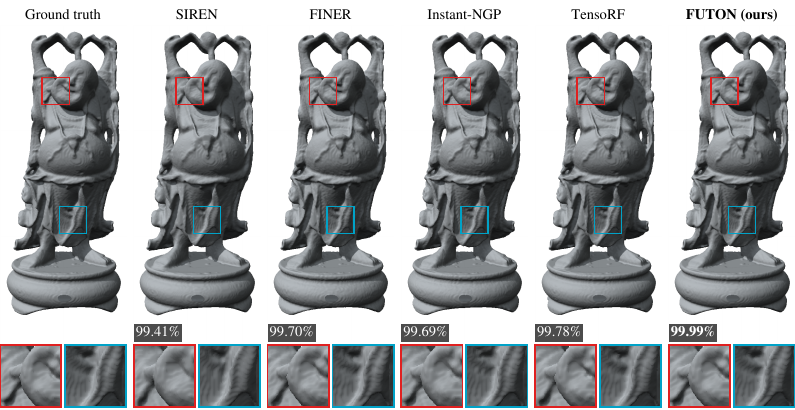}\\[3pt]
    \includegraphics[trim={0 0 0 11bp},clip]{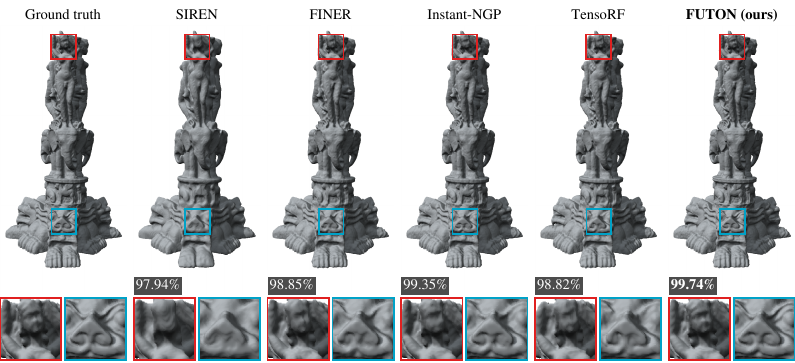}
    \caption{\textbf{Occupancy volumes} on Armadillo, Dragon, Happy Buddha, and Thai Statue, as in \cref{fig:qualitative_occupancy}.}
    \label{fig:qualitative_volumes_a}\label{fig:qualitative_volumes_b}
\end{figure}

\begin{table}[t]
    \caption{\textbf{Radiance fields} at about 74k parameters: median costs and means over the eight Blender scenes $\pm$ one standard error (top); per scene, the PSNR of the baselines \cref{tab:nerf} omits and every model's SSIM and LPIPS (bottom), marked among all models; 200 test views per scene.}
    \label{tab:nerf_means}\label{tab:nerf_scenes}
    \centering
    {\footnotesize\setlength{\tabcolsep}{3pt}%
\begin{tabular}{@{}lrrrrrrr@{}}
\toprule
\multicolumn{1}{c}{} & \multicolumn{2}{c}{Training} & \multicolumn{2}{c}{Inference} & \multicolumn{3}{c}{Mean over the scenes} \\
\cmidrule(lr){2-3}\cmidrule(lr){4-5}\cmidrule(l){6-8}
Model & \multicolumn{1}{c}{Time} & \multicolumn{1}{c}{Memory} & \multicolumn{1}{c}{Speed} & \multicolumn{1}{c}{Memory} & \multicolumn{1}{c}{PSNR} & \multicolumn{1}{c}{SSIM} & \multicolumn{1}{c}{LPIPS} \\
 & \multicolumn{1}{c}{(s) $\downarrow$} & \multicolumn{1}{c}{(GB) $\downarrow$} & \multicolumn{1}{c}{(FPS) $\uparrow$} & \multicolumn{1}{c}{(GB) $\downarrow$} & \multicolumn{1}{c}{(dB) $\uparrow$} & \multicolumn{1}{c}{$\uparrow$} & \multicolumn{1}{c}{$\downarrow$} \\
\midrule
RFF & 179.3 & 2.91 & 38.7 & 2.12 & 28.07{\scriptsize$\pm$0.30} & 0.928{\scriptsize$\pm$0.004} & 0.082{\scriptsize$\pm$0.008} \\
PE-MLP & 201.0 & 3.00 & 37.0 & 2.10 & 27.94{\scriptsize$\pm$0.22} & 0.931{\scriptsize$\pm$0.003} & 0.079{\scriptsize$\pm$0.006} \\
MFN & 275.6 & 9.97 & 35.5 & 3.61 & 28.28{\scriptsize$\pm$0.16} & 0.932{\scriptsize$\pm$0.002} & 0.078{\scriptsize$\pm$0.006} \\
SIREN & 215.5 & 5.08 & 38.4 & 2.14 & 28.53{\scriptsize$\pm$0.16} & 0.935{\scriptsize$\pm$0.002} & 0.106{\scriptsize$\pm$0.007} \\
FreSh (SIREN) & 217.2 & 5.03 & 39.4 & 2.13 & 28.34{\scriptsize$\pm$0.23} & 0.928{\scriptsize$\pm$0.004} & 0.146{\scriptsize$\pm$0.016} \\
Gauss & 239.1 & 4.82 & 36.4 & 2.18 & 27.54{\scriptsize$\pm$0.21} & 0.922{\scriptsize$\pm$0.003} & 0.105{\scriptsize$\pm$0.008} \\
WIRE & 237.9 & 4.97 & 38.8 & 2.19 & 28.29{\scriptsize$\pm$0.19} & 0.928{\scriptsize$\pm$0.003} & 0.126{\scriptsize$\pm$0.007} \\
FINER & 218.8 & 5.02 & 39.1 & 2.15 & \underline{28.62}{\scriptsize$\pm$0.19} & 0.934{\scriptsize$\pm$0.002} & 0.120{\scriptsize$\pm$0.009} \\
Instant-NGP & \textbf{147.0} & 2.80 & 44.0 & 3.26 & 26.60{\scriptsize$\pm$0.37} & 0.908{\scriptsize$\pm$0.009} & 0.119{\scriptsize$\pm$0.011} \\
TensoRF & 246.2 & 2.14 & \underline{47.2} & \textbf{2.04} & 27.99{\scriptsize$\pm$0.18} & 0.931{\scriptsize$\pm$0.003} & 0.076{\scriptsize$\pm$0.007} \\
GA-Planes & 271.6 & \underline{2.10} & 45.7 & \textbf{2.04} & 28.24{\scriptsize$\pm$0.19} & \underline{0.936}{\scriptsize$\pm$0.003} & \underline{0.071}{\scriptsize$\pm$0.005} \\
F-INR (SIREN) & 292.8 & 16.18 & 33.7 & 6.69 & 27.52{\scriptsize$\pm$0.47} & 0.911{\scriptsize$\pm$0.012} & 0.127{\scriptsize$\pm$0.020} \\
\midrule
FUTON (ours) & \underline{152.0} & \textbf{1.88} & \textbf{49.6} & \textbf{2.04} & \textbf{29.22}{\scriptsize$\pm$0.24} & \textbf{0.943}{\scriptsize$\pm$0.004} & \textbf{0.065}{\scriptsize$\pm$0.007} \\
\bottomrule
\end{tabular}
}

    \medskip
    {\footnotesize\setlength{\tabcolsep}{3.5pt}%
\begin{tabular}{@{}lrrrrrrrr@{}}
\toprule
Model & \multicolumn{1}{c}{Chair} & \multicolumn{1}{c}{Drums} & \multicolumn{1}{c}{Ficus} & \multicolumn{1}{c}{Hotdog} & \multicolumn{1}{c}{Lego} & \multicolumn{1}{c}{Materials} & \multicolumn{1}{c}{Mic} & \multicolumn{1}{c}{Ship} \\
\midrule
\multicolumn{9}{@{}l}{\textit{PSNR (dB) $\uparrow$, the baselines \cref{tab:nerf} omits}} \\
PE-MLP & 32.09 & 22.84 & 25.75 & 32.06 & 28.55 & 26.64 & 32.44 & \underline{23.12} \\
Gauss & 30.92 & 23.69 & 26.41 & 32.07 & 27.66 & 25.21 & 32.46 & 21.94 \\
WIRE & 32.63 & 24.18 & 27.54 & 32.00 & 28.44 & 26.14 & 33.23 & 22.13 \\
\midrule
\multicolumn{9}{@{}l}{\textit{SSIM $\uparrow$}} \\
RFF & \underline{0.975} & 0.901 & \underline{0.959} & \textbf{0.967} & 0.935 & 0.922 & 0.982 & 0.783 \\
PE-MLP & 0.970 & 0.896 & 0.946 & 0.960 & 0.943 & 0.934 & 0.980 & 0.816 \\
MFN & 0.973 & 0.908 & 0.957 & 0.964 & 0.943 & 0.920 & 0.980 & 0.813 \\
SIREN & 0.972 & \textbf{0.916} & 0.955 & 0.965 & 0.943 & 0.930 & \textbf{0.983} & 0.817 \\
FreSh (SIREN) & 0.970 & 0.901 & 0.952 & 0.964 & 0.952 & 0.897 & 0.979 & 0.806 \\
Gauss & 0.961 & 0.902 & 0.950 & 0.958 & 0.927 & 0.913 & 0.980 & 0.785 \\
WIRE & 0.970 & 0.911 & 0.957 & 0.957 & 0.933 & 0.921 & 0.982 & 0.793 \\
FINER & 0.972 & \textbf{0.916} & 0.956 & 0.962 & 0.952 & 0.922 & \textbf{0.983} & 0.809 \\
Instant-NGP & 0.967 & 0.894 & 0.865 & 0.953 & 0.925 & 0.907 & 0.978 & 0.779 \\
TensoRF & 0.968 & 0.908 & 0.942 & \textbf{0.967} & 0.935 & 0.933 & 0.980 & 0.818 \\
GA-Planes & 0.973 & 0.910 & 0.944 & \textbf{0.967} & \underline{0.956} & 0.931 & 0.979 & \underline{0.826} \\
F-INR (SIREN) & 0.973 & 0.850 & 0.904 & 0.960 & 0.952 & \underline{0.941} & 0.976 & 0.731 \\
FUTON (ours) & \textbf{0.976} & 0.913 & \textbf{0.960} & 0.966 & \textbf{0.970} & \textbf{0.948} & 0.982 & \textbf{0.827} \\
\midrule
\multicolumn{9}{@{}l}{\textit{LPIPS $\downarrow$}} \\
RFF & 0.036 & 0.109 & \textbf{0.046} & 0.045 & 0.084 & 0.086 & \textbf{0.026} & 0.227 \\
PE-MLP & 0.039 & 0.122 & 0.059 & 0.053 & 0.060 & 0.072 & \underline{0.029} & 0.195 \\
MFN & 0.040 & 0.102 & 0.056 & 0.044 & 0.070 & 0.084 & 0.033 & 0.192 \\
SIREN & 0.052 & 0.130 & 0.163 & 0.052 & 0.091 & 0.116 & 0.043 & 0.205 \\
FreSh (SIREN) & 0.076 & 0.191 & 0.210 & 0.056 & 0.086 & 0.230 & 0.087 & 0.228 \\
Gauss & 0.059 & 0.129 & 0.071 & 0.065 & 0.088 & 0.143 & 0.043 & 0.241 \\
WIRE & 0.058 & 0.146 & 0.132 & 0.079 & 0.133 & 0.158 & 0.051 & 0.249 \\
FINER & 0.062 & 0.146 & 0.187 & 0.073 & 0.075 & 0.149 & 0.040 & 0.228 \\
Instant-NGP & 0.047 & 0.130 & 0.196 & 0.069 & 0.085 & 0.132 & 0.041 & 0.253 \\
TensoRF & 0.041 & \textbf{0.094} & 0.065 & \textbf{0.042} & 0.075 & 0.070 & 0.032 & 0.185 \\
GA-Planes & \underline{0.034} & 0.097 & 0.063 & \textbf{0.042} & \underline{0.045} & 0.077 & 0.031 & \textbf{0.179} \\
F-INR (SIREN) & 0.052 & 0.226 & 0.160 & 0.062 & 0.061 & \underline{0.063} & 0.063 & 0.331 \\
FUTON (ours) & \textbf{0.033} & \textbf{0.094} & \underline{0.052} & 0.044 & \textbf{0.030} & \textbf{0.057} & 0.030 & \underline{0.182} \\
\bottomrule
\end{tabular}
}

\end{table}

\begin{figure}[t]
    \centering
    \includegraphics{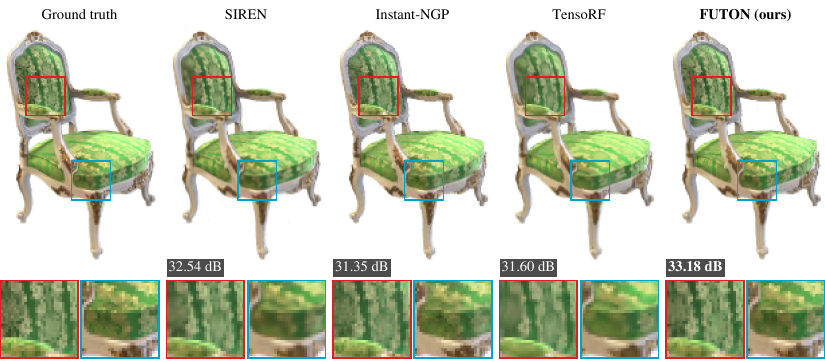}\\[3pt]
    \includegraphics[trim={0 0 0 11bp},clip]{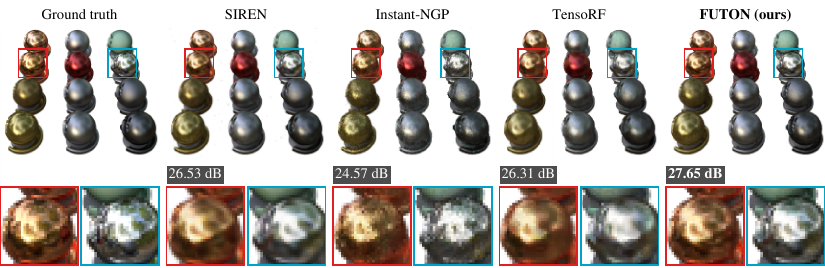}\\[3pt]
    \includegraphics[trim={0 0 0 11bp},clip]{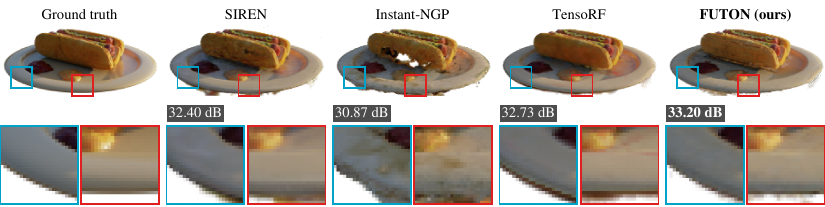}\\[3pt]
    \includegraphics[trim={0 0 0 11bp},clip]{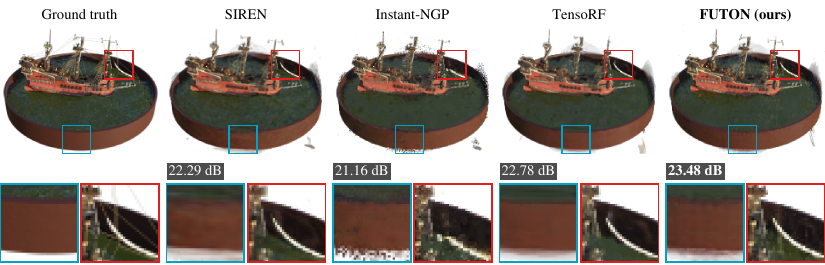}
    \caption{\textbf{Novel view synthesis} on Chair, Materials, Hotdog, and Ship, as in \cref{fig:qualitative_nerf}.}
    \label{fig:qualitative_nerf_more}
\end{figure}

\clearpage
\section{Ablation Studies} \label{sec:ablation_details}

\begin{figure}[t]
    \centering
    \begin{subfigure}[b]{0.32\linewidth}\centering\includegraphics{figures/results/ablation/bandwidth.pdf}\caption{Relative bandwidth}\label{fig:ablation_bandwidth}\end{subfigure}\hspace{0.02\linewidth}%
    \begin{subfigure}[b]{0.32\linewidth}\centering\includegraphics{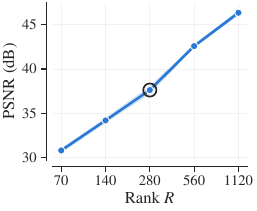}\caption{Rank}\label{fig:ablation_rank}\end{subfigure}\\[6pt]
    \begin{subfigure}[b]{0.32\linewidth}\centering\includegraphics{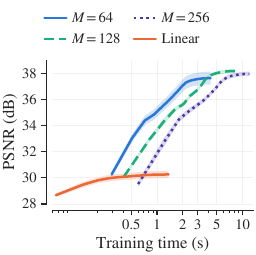}\caption{Head}\label{fig:ablation_head}\end{subfigure}\hfill
    \begin{subfigure}[b]{0.32\linewidth}\centering\includegraphics{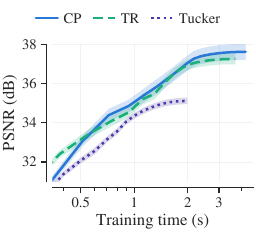}\caption{Combiner}\label{fig:ablation_combiner}\end{subfigure}\hfill
    \begin{subfigure}[b]{0.32\linewidth}\centering\includegraphics{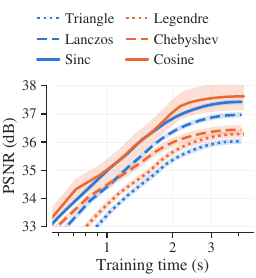}\caption{Basis}\label{fig:ablation_basis}\end{subfigure}
    \caption{\textbf{Ablations on Kodak}, mean PSNR $\pm$ one standard error, as in \cref{tab:ablation_bandwidth}: (a)~relative bandwidth and (b)~rank, the defaults circled; (c--e)~PSNR during training for four heads, three combiners, and six bases.}
    \label{fig:ablations}
\end{figure}

We ablate FUTON's main design choices on the 24 Kodak images, changing one at a time from the image benchmark's FUTON: cosine basis, $\kappa = 1/2$, CP combiner with $R = 280$, and an MLP head of width $M = 64$, 197k parameters, trained as in \cref{sec:detailed_experimental_setup}. Except in the rank study, every variant keeps the size by adjusting the rank. \Cref{tab:ablation_bandwidth} reports the results, and \cref{fig:ablations} the curves.

\paragraph{Relative bandwidth.}
At a fixed size, a larger $\kappa$ resolves higher frequencies but leaves a smaller rank (\cref{tab:ablation_bandwidth}). The best PSNR is at $\kappa = 1/2$, and $\kappa = 1/4$ and $1$ stay within 1.1\,dB; $\kappa = 1/8$ truncates the images (4.2\,dB lower), and $\kappa = 2$ leaves rank 75 (5.4\,dB lower). Training time falls as $\kappa$ grows, with the rank.

\paragraph{Rank.}
At $\kappa = 1/2$, PSNR grows by 3.4, 3.4, 5.0, and 3.8\,dB per doubling of $R$ from 70 to 1,120, with no plateau at these sizes, while training time grows from 2.2 to 10.9\,s (\cref{tab:ablation_rank}).

\paragraph{Head.}
A linear head of the same size, with $R = 305$, scores 30.26\,dB, 7.4\,dB below the MLP head (\cref{tab:ablation_head}). Wider MLP heads at equal size trade rank for width: $M = 128$ gains 0.5\,dB on average and $M = 256$ 0.3\,dB, at about two and three times the training time and three times the memory, so we keep $M = 64$.

\paragraph{Combiner.} \label{sec:ablation_combiner}
At about equal size (the ring holds 180k parameters, as its next rank exceeds the budget), the CP combiner leads a tensor ring, the cyclic form of a tensor train~\cite{oseledets2011tensor}, by 0.4\,dB and Tucker~\cite{tucker1966some} by 2.5\,dB (\cref{tab:ablation_combiner}). On a grid, both alternatives train faster, Tucker in half the time, but neither reaches CP's quality.

\paragraph{Basis.}
Among six bases at equal size, cosines are the most accurate (37.63\,dB), ahead of sinc (37.43\,dB) and Lanczos (36.97\,dB), shifted kernels on a uniform grid, and of the Chebyshev (36.44\,dB) and Legendre (36.29\,dB) polynomials and the triangle (hat) functions (36.03\,dB). All train in about the same time (\cref{tab:ablation_basis}).

\begin{table}[t]
    \caption{\textbf{Ablations on Kodak}: means over the 24 images $\pm$ one standard error, and median costs, with the best and second best marked within each block. Top: the relative bandwidth $\kappa$ at about 197k parameters, the rank adjusted to keep the size, and the rank $R$ at $\kappa = 1/2$, from 49.5k to 789k parameters. Bottom: at equal size, MLP heads of width $M = 64$, 128, and 256 and a linear head, each with the largest rank that keeps the size; CP, a tensor ring (TR, 180k parameters), and Tucker, at about equal size; and six bases.}
    \label{tab:ablation_bandwidth}\label{tab:ablation_rank}\label{tab:ablation_head}\label{tab:ablation_combiner}\label{tab:ablation_basis}
    \centering
    {\small\setlength{\tabcolsep}{3pt}%
\begin{tabular}{@{}llrrrrrrr@{}}
\toprule
\multicolumn{2}{c}{} & \multicolumn{2}{c}{Training} & \multicolumn{2}{c}{Inference} & \multicolumn{3}{c}{} \\
\cmidrule(lr){3-4}\cmidrule(lr){5-6}
$\kappa$ & $R$ & \multicolumn{1}{c}{Time} & \multicolumn{1}{c}{Memory} & \multicolumn{1}{c}{Throughput} & \multicolumn{1}{c}{Memory} & \multicolumn{1}{c}{PSNR} & \multicolumn{1}{c}{SSIM} & \multicolumn{1}{c}{LPIPS} \\
 &  & \multicolumn{1}{c}{(s) $\downarrow$} & \multicolumn{1}{c}{(MB) $\downarrow$} & \multicolumn{1}{c}{(MPix/s) $\uparrow$} & \multicolumn{1}{c}{(MB) $\downarrow$} & \multicolumn{1}{c}{(dB) $\uparrow$} & \multicolumn{1}{c}{$\uparrow$} & \multicolumn{1}{c}{$\downarrow$} \\
\midrule
\multicolumn{9}{@{}l}{\textit{Relative bandwidth, at about 197k parameters}} \\
1/8 & 885 & 9.0 & 442 & 111.9 & 219 & 33.41{\scriptsize$\pm$0.15} & 0.914{\scriptsize$\pm$0.003} & 0.167{\scriptsize$\pm$0.004} \\
1/4 & 515 & 6.1 & 392 & 156.9 & \textbf{201} & \underline{36.95}{\scriptsize$\pm$0.18} & \underline{0.948}{\scriptsize$\pm$0.003} & \underline{0.118}{\scriptsize$\pm$0.004} \\
1/2 & 280 & 4.1 & 359 & 228.7 & \textbf{201} & \textbf{37.63}{\scriptsize$\pm$0.51} & \textbf{0.955}{\scriptsize$\pm$0.003} & \textbf{0.098}{\scriptsize$\pm$0.003} \\
1 & 145 & \underline{2.8} & \underline{341} & \underline{254.1} & \textbf{201} & 36.56{\scriptsize$\pm$0.19} & 0.936{\scriptsize$\pm$0.002} & 0.140{\scriptsize$\pm$0.003} \\
2 & 75 & \textbf{2.2} & \textbf{331} & \textbf{335.5} & \textbf{201} & 32.26{\scriptsize$\pm$0.26} & 0.860{\scriptsize$\pm$0.007} & 0.271{\scriptsize$\pm$0.008} \\
\midrule
\multicolumn{9}{@{}l}{\textit{Rank, at $\kappa = 1/2$}} \\
1/2 & 70 & \textbf{2.2} & \textbf{330} & \textbf{305.8} & \textbf{201} & 30.81{\scriptsize$\pm$0.28} & 0.826{\scriptsize$\pm$0.010} & 0.304{\scriptsize$\pm$0.008} \\
1/2 & 140 & \underline{2.7} & \underline{340} & \underline{249.8} & \textbf{201} & 34.20{\scriptsize$\pm$0.29} & 0.903{\scriptsize$\pm$0.003} & 0.190{\scriptsize$\pm$0.004} \\
1/2 & 280 & 4.1 & 359 & 228.7 & \textbf{201} & 37.63{\scriptsize$\pm$0.46} & 0.955{\scriptsize$\pm$0.003} & 0.098{\scriptsize$\pm$0.003} \\
1/2 & 560 & 6.2 & 398 & 152.0 & \textbf{201} & \underline{42.58}{\scriptsize$\pm$0.20} & \underline{0.983}{\scriptsize$\pm$0.005} & \underline{0.040}{\scriptsize$\pm$0.004} \\
1/2 & 1120 & 10.9 & 473 & 95.2 & 251 & \textbf{46.33}{\scriptsize$\pm$0.30} & \textbf{0.992}{\scriptsize$\pm$0.005} & \textbf{0.016}{\scriptsize$\pm$0.005} \\
\bottomrule
\end{tabular}
}

    \medskip
    {\small\setlength{\tabcolsep}{3pt}%
\begin{tabular}{@{}lrrrrrrr@{}}
\toprule
\multicolumn{1}{c}{} & \multicolumn{2}{c}{Training} & \multicolumn{2}{c}{Inference} & \multicolumn{3}{c}{} \\
\cmidrule(lr){2-3}\cmidrule(lr){4-5}
Variant & \multicolumn{1}{c}{Time} & \multicolumn{1}{c}{Memory} & \multicolumn{1}{c}{Throughput} & \multicolumn{1}{c}{Memory} & \multicolumn{1}{c}{PSNR} & \multicolumn{1}{c}{SSIM} & \multicolumn{1}{c}{LPIPS} \\
 & \multicolumn{1}{c}{(s) $\downarrow$} & \multicolumn{1}{c}{(MB) $\downarrow$} & \multicolumn{1}{c}{(MPix/s) $\uparrow$} & \multicolumn{1}{c}{(MB) $\downarrow$} & \multicolumn{1}{c}{(dB) $\uparrow$} & \multicolumn{1}{c}{$\uparrow$} & \multicolumn{1}{c}{$\downarrow$} \\
\midrule
\multicolumn{8}{@{}l}{\textit{Head}} \\
$M{=}64$ ($R{=}280$) & \underline{4.2} & \underline{359} & \underline{248.6} & \underline{201} & 37.62{\scriptsize$\pm$0.49} & \textbf{0.955}{\scriptsize$\pm$0.002} & \textbf{0.098}{\scriptsize$\pm$0.003} \\
$M{=}128$ ($R{=}255$) & 8.1 & 1024 & 131.7 & 602 & \textbf{38.16}{\scriptsize$\pm$0.15} & \underline{0.954}{\scriptsize$\pm$0.002} & \underline{0.099}{\scriptsize$\pm$0.001} \\
$M{=}256$ ($R{=}220$) & 12.0 & 1176 & 90.4 & 749 & \underline{37.95}{\scriptsize$\pm$0.17} & 0.951{\scriptsize$\pm$0.002} & 0.106{\scriptsize$\pm$0.002} \\
Linear ($R{=}305$) & \textbf{1.4} & \textbf{37} & \textbf{346.1} & \textbf{19} & 30.26{\scriptsize$\pm$0.27} & 0.874{\scriptsize$\pm$0.006} & 0.201{\scriptsize$\pm$0.005} \\
\midrule
\multicolumn{8}{@{}l}{\textit{Combiner}} \\
CP & 4.2 & 359 & 203.5 & \textbf{201} & \textbf{37.62}{\scriptsize$\pm$0.41} & \textbf{0.955}{\scriptsize$\pm$0.001} & \textbf{0.098}{\scriptsize$\pm$0.002} \\
TR & \underline{3.7} & \underline{356} & \underline{216.9} & \textbf{201} & \underline{37.26}{\scriptsize$\pm$0.27} & \underline{0.947}{\scriptsize$\pm$0.001} & \underline{0.112}{\scriptsize$\pm$0.002} \\
Tucker & \textbf{2.0} & \textbf{330} & \textbf{325.3} & \textbf{201} & 35.12{\scriptsize$\pm$0.16} & 0.915{\scriptsize$\pm$0.002} & 0.170{\scriptsize$\pm$0.003} \\
\midrule
\multicolumn{8}{@{}l}{\textit{Basis}} \\
Cosine & 4.2 & \textbf{359} & 230.4 & \textbf{201} & \textbf{37.63}{\scriptsize$\pm$0.49} & \textbf{0.955}{\scriptsize$\pm$0.001} & \textbf{0.098}{\scriptsize$\pm$0.002} \\
Chebyshev & \textbf{4.1} & \textbf{359} & 230.3 & \textbf{201} & 36.44{\scriptsize$\pm$0.16} & 0.939{\scriptsize$\pm$0.001} & 0.124{\scriptsize$\pm$0.002} \\
Legendre & 4.2 & 360 & 210.4 & \textbf{201} & 36.29{\scriptsize$\pm$0.12} & 0.937{\scriptsize$\pm$0.001} & 0.129{\scriptsize$\pm$0.001} \\
Triangle & \textbf{4.1} & \textbf{359} & \underline{242.6} & \textbf{201} & 36.03{\scriptsize$\pm$0.10} & 0.934{\scriptsize$\pm$0.001} & 0.126{\scriptsize$\pm$0.001} \\
Lanczos & \textbf{4.1} & \textbf{359} & 223.4 & \textbf{201} & 36.97{\scriptsize$\pm$0.09} & 0.944{\scriptsize$\pm$0.001} & 0.112{\scriptsize$\pm$0.001} \\
Sinc & \textbf{4.1} & \textbf{359} & \textbf{248.4} & \textbf{201} & \underline{37.43}{\scriptsize$\pm$0.09} & \underline{0.949}{\scriptsize$\pm$0.001} & \underline{0.108}{\scriptsize$\pm$0.001} \\
\bottomrule
\end{tabular}
}

\end{table}

\FloatBarrier
\section{Limitations} \label{sec:limitations}

While FUTON demonstrates strong performance across the tasks considered, we identify several limitations and directions for improvement:

\begin{enumerate}
    \item \textbf{Separability assumption.} FUTON's basis functions are constructed as separable products of one-dimensional functions~\eqref{eq:separable_basis}. While this is theoretically complete (\cref{thm:separable_basis_orthonormality}) and practically effective for the tasks we evaluate, signals with strong cross-dimensional correlations (\eg, oriented textures at arbitrary angles) may be less efficiently represented than with non-separable bases. The CP rank $R$ compensates for this by mixing dimensions, but at increased parameter cost{: an oblique edge needs a rank that grows with its sharpness (\cref{tab:edge_test})}.

    \item \textbf{Low-rank assumption.} FUTON assumes that the coefficient tensor $\tns{W}$ admits an effective low-rank CP approximation. For signals whose Fourier spectra lack low-rank structure (\eg, white noise or highly irregular textures), the required rank $R$ may become large, diminishing the computational advantage over the full-rank GFS. \Cref{prop:error} measures the rank a signal needs, and \cref{prop:typical_rank} shows that no signal-independent bound can do better than \cref{thm:futon_expressiveness}'s order, so the assumption must be checked on the data at hand.

    \item \textbf{Scaling to high dimensions.} While FUTON's complexity is linear in $C$ for fixed $R$ and {$K_c$}, the sufficient rank for universal approximation grows as $R \le K^{C-1}\min\{K, D\}$ (\cref{thm:futon_expressiveness}). For very high-dimensional inputs ($C \gg 3$), the practical rank required for accurate representation remains an open question. Our experiments cover $C \in \{2, 3\}$; validation on higher-dimensional tasks (\eg, video, 4D light fields) is left for future work.

    \item \textbf{Single-scene optimization.} Like other coordinate-based INRs, FUTON is optimized per scene and does not generalize across scenes without retraining. Meta-learning or hypernetwork-based extensions could address this but are beyond the current scope.

    \item \textbf{Evaluation scope.} Our experiments cover images, occupancy volumes, and radiance fields against twelve baselines, with one seed per run. Broader evaluation on additional signal modalities (\eg, audio, signed distance fields) and on inverse problems would provide a more comprehensive picture of FUTON's relative strengths.

    \item \textbf{Scope of the error analysis.} \Cref{prop:error} covers the linear head on a fully sampled signal: the coefficient tensor is computed from every sample, so radiance fields and inverse problems, where the signal is observed through a renderer or an operator, are outside it, and so are the MLP head, to which only the upper bound transfers, and the $\tanh$ output, which \cref{rem:nonlinear} covers only for targets inside $(-1, 1)$, unlike the benchmarks'. On a grid, its norm says nothing off the grid. For three or more axes, the lower bound falls to the truncation once $R$ passes the largest $K_c$, so the bracket is wide on volumes; a tighter lower bound for that regime is open, and a search that improves on the certified FUTON's terms would tighten the upper one.

    \item \textbf{Implementation-dependent speed.} Wall-clock comparisons depend on implementations. All models here run in PyTorch, compiled; hash-grid and factorized models with fused CUDA kernels, as in their authors' code, can train faster than in our implementation.
\end{enumerate}

\end{document}